\documentclass[times,twocolumn,final,authoryear]{elsarticle}

\usepackage{jasr}
\usepackage{framed,multirow}
\usepackage{threeparttable}

\usepackage{amssymb}
\usepackage{amsmath}
\usepackage{bm}
\usepackage{latexsym}
\usepackage{xspace}

\usepackage{lineno}

\usepackage{xcolor}
\definecolor{newcolor}{rgb}{1.0,.349,.1}
\definecolor{red}{rgb}{0.0,0.0,0.0}

\usepackage[colorlinks,linkcolor=magenta,citecolor=blue,filecolor=cyan,urlcolor=purple]{hyperref}

\journal{Advances in Space Research}

\begin{document}

\verso{Benjamin Lynch \textit{et al.}}

\begin{frontmatter}


\title{On the utility of flux rope models for CME magnetic structure below 30\,$R_{\odot}$}

\author[1]{Benjamin~J.~\snm{Lynch}\corref{cor1}}
\cortext[cor1]{Corresponding author}
  \ead{blynch@berkeley.edu}
\author[2]{Nada~\snm{Al-Haddad}}
\author[2]{Wenyuan~\snm{Yu}}
\author[3]{Erika~\snm{Palmerio}}
\author[2]{No\'{e}~\snm{Lugaz}}

\address[1]{Space Sciences Laboratory, University of California--Berkeley, Berkeley, CA 94720, USA}
\address[2]{Space Science Center, University of New Hampshire, Durham, NH 03824, USA}
\address[3]{Predictive Science Inc., San Diego, CA 92121, USA}

\received{04/12/2021}
\finalform{03/04/2022}
\accepted{03/05/2022}
\availableonline{}

\begin{abstract}

We present a comprehensive analysis of the three-dimensional magnetic flux rope structure generated during the Lynch et al. (2019, ApJ 880:97) magnetohydrodynamic (MHD) simulation of a global-scale, $360^{\circ}$-wide streamer blowout coronal mass ejection (CME) eruption. We create both fixed and moving synthetic spacecraft to generate time series of the MHD variables through different regions of the flux rope CME. Our moving spacecraft trajectories are derived from the spatial coordinates of Parker Solar Probe's past encounters 7 and 9 and future encounter 23. Each synthetic time series through the simulation flux rope ejecta is fit with three different in-situ flux rope models commonly used to characterize the large-scale, coherent magnetic field rotations observed in a significant fraction of interplanetary CMEs (ICMEs). \textcolor{red}{We present each of the in-situ flux rope model fits to the simulation data and discuss the similarities and differences between the model fits and the MHD simulation's flux rope spatial orientations, field strengths and rotations, expansion profiles, and magnetic flux content.} We compare in-situ model properties to those calculated with the MHD data for both classic bipolar and unipolar ICME flux rope configurations as well as more problematic profiles such as those with a significant radial component to the flux rope axis orientation or profiles obtained with large impact parameters. We find general agreement among the in-situ flux rope fitting results for the classic profiles and much more variation among results for the problematic profiles. \textcolor{red}{We also examine the force-free assumption for a subset of the flux rope models and quantify properties of the Lorentz force within MHD ejecta intervals.} We conclude that the in-situ flux rope models are generally a decent approximation to the field structure, but all the caveats associated with in-situ flux rope models will still apply (and perhaps moreso) at distances below $30\,R_\odot$. We discuss our results in the context of future PSP observations of CMEs in the extended corona.

\end{abstract}

\begin{keyword}
\KWD Magnetohydrodynamics (MHD)\sep Solar corona\sep Coronal mass ejection (CME)\sep Magnetic flux rope\sep Parker Solar Probe (PSP)
\end{keyword}

\end{frontmatter}



\section{Introduction} \label{sec:intro}

Interplanetary coronal mass ejections (ICMEs) that contain a magnetic flux-rope structure are commonly referred to as magnetic clouds \citep[e.g.][]{Burlaga1981, Marubashi1986, Lepping1990}. Magnetic cloud ICMEs are characterized by enhanced magnetic field magnitudes, a smooth rotation of the magnetic field, low proton temperature, and low plasma beta, corresponding to large-scale magnetic flux ropes with typical durations at 1~AU of tens of hours \citep{Burlaga1988}. The properties of ICMEs have been investigated mainly at ${\sim}1$~AU, where the solar wind and interplanetary magnetic field have been measured continuously for several decades \citep[e.g.][]{Richardson2010, Li2014, Jian2018, Nieves-Chinchilla2019}. In particular, magnetic cloud ICMEs have been analyzed extensively from the Sun--Earth L1 point, in both statistical \citep[e.g.][]{Lynch2005, Li2011, Janvier2014b, Wood2017} and detailed case studies \citep[e.g.][]{Moestl2008, Moestl2009b, Kilpua2009a, Palmerio2017}. The heliospheric evolution of ICMEs has been recently reviewed by \citet{Manchester2017}, \citet{Kilpua2017}, and \citet{Luhmann2020}.

With the launch of the Parker Solar Probe \citep[PSP;][]{Fox2016}, we now have the opportunity for unprecedented coordinated multi-spacecraft observations, including complementary remote-sensing and in-situ observations of the same solar wind and transient structures, as well as in-situ measurements over a range of angular separations and radial distances \citep[e.g.][]{Velli2020,Hadid2021,Moestl2022}. There have already been several well-observed slow, streamer blowout CME events in the PSP data that exhibit flux rope morphology in remote-sensing observations \citep[e.g.][]{HowardR2019, Hess2020, Wood2020, Liewer2021} as well as in-situ measurements \citep[e.g.][]{Korreck2020, Lario2020, Nieves-Chinchilla2020, Palmerio2021c}. In fact, a large percentage of coronal mass ejections (CMEs) observed by PSP thus far (either remotely or in situ) has been of the slow, streamer blowout variety \citep{Vourlidas2018}, which tend to originate high in the corona above polarity inversion lines of essentially quiet-Sun magnetic field distributions and typically erupt via the evolutionary processes described by \citet{lynch2016b}.

\citet{Al-Haddad2019b} examined the inner heliospheric evolution of two simulated CMEs and compared the radial evolution of their ejecta size, field strength, and velocity profiles to various empirical scaling relations obtained from earlier statistical analyses of CMEs observed in situ by Helios, MESSENGER, STEREO, and the Wind and ACE spacecraft at L1. Here we employ a similar methodology---analysis of magnetohydrodynamic (MHD) simulation data---but with a specific focus on the three-dimensional (3D) magnetic structure that would be seen by synthetic spacecraft observers with radial distances ${<}30\,R_\odot$ that are either stationary or have time-dependent, PSP-like trajectories. 

In order to compare the set of simulation time series with future PSP observations of CMEs in the extended solar corona, we apply several common in-situ flux rope models and their fitting techniques \citep[e.g.\ see][and references therein]{Al-Haddad2013,Al-Haddad2018}. There are well-known limitations to in-situ flux rope models that have been previously discussed, including statistical descriptions of model parameter uncertainties \citep{Lepping2003a, Lynch2005}, the inability of these models to correctly estimate the shape of the flux rope cross-section \citep{Riley2004, Owens2006a, Owens2008}, and the inability to distinguish between the large-scale magnetic field rotations of twisted versus writhed field lines \citep{Al-Haddad2011,Al-Haddad2019a}. Different flux rope models have been shown to give different flux rope orientations and other parameters (e.g. impact parameter, field strength, etc.) when applied to the same observational data. This led \citet{Al-Haddad2013} to conclude that ``having multiple methods able to successfully fit or reconstruct the same event gives more reliable results regarding the orientation of the ICME axis'' when two or more of the different models give similar answers.

\textcolor{red}{In this paper we examine the application of idealized, in-situ flux rope models to the 3D magnetic structure of the \citet{Lynch2019} MHD simulation of a global streamer blowout eruption. We generate a set of time series by placing synthetic spacecraft throughout the computational domain and each time series is fit with three common cylindrical flux rope models that are used to analyze in-situ observations of magnetic cloud ICMEs. In Section~\ref{sec:sim}, we briefly review the MHD simulation results. In Section~\ref{sec:orbits}, we describe the construction of the synthetic spacecraft time series. In Section~\ref{sec:insitu}, we present our results for the in-situ flux rope models applied to each of the 16 synthetic time series. In Section~\ref{subsec:types}, we detail a subset of the time series that include classic bipolar and unipolar flux rope orientations as well as more problematic orientations, such as sampling along the flux rope axis or having a large impact parameter due to a glancing trajectory. In Section~\ref{sec:MHD-FR-comp}, we compare the in-situ flux rope model fits to each other and to the MHD results, including the cylinder axis orientations (\ref{subsec:geom}), the hodogram signatures of the magnetic field rotation (\ref{subsec:hodogram}), inferred CME expansion profiles (\ref{subsec:exp}), the force-free nature of the ejecta (\ref{subsec:ffness}), and the derived flux content (\ref{subsec:fluxes}). Lastly, in Section~\ref{sec:conclusions}, we summarize our results and discuss the implications for PSP observations of CME--ICME observations in the extended corona.}


\section{Overview of the Numerical Simulation} \label{sec:sim}

The MHD simulation was run with the Adaptively Refined MHD Solver \citep[ARMS;][]{DeVore2008}, which solves the equations of ideal MHD with a finite-volume, multidimensional flux-corrected transport techniques \citep{DeVore1991}. The ARMS code utilizes the adaptive-mesh toolkit PARAMESH \citep{MacNeice2000} to enable efficient multiprocessor parallelization and dynamic, solution-adaptive grid refinement. ARMS has been used to model a variety of dynamic phenomena in the solar atmosphere, including CME initiation in 2.5D and 3D spherical geometries, the initiation of coronal jets and their propagation into the solar wind, and interchange reconnection processes and their resulting time-dependent solar wind outflow. One of the major advantages of the ARMS code is its conservation of magnetic helicity under a wide variety of magnetic field geometries and complex reconnection scenarios \citep{pariat2015,knizhnik2015,knizhnik2017}.

In this work, we perform a detailed analysis of the \citet{Lynch2019} MHD simulation of an idealized, global-scale eruption of the entire coronal helmet streamer belt, i.e.\ a $360^\circ$-wide streamer blowout CME. While this particular simulation did not model a specific solar (or stellar) CME event, magnetic reconnection during the eruption process creates a large-scale flux rope ejecta that contains significant longitudinal variation in its orientation, reflecting the underlying structure of the polarity inversion line (PIL) of the streamer-belt flux distribution. Given the common formation and eruption mechanism for bipolar streamer-blowout eruptions \citep[e.g.][]{lynch2016b,lynch2021,Vourlidas2018}, despite the exaggerated size of the \citet{Lynch2019} flux rope ejecta, we will show that the simulation results contain a variety of \emph{local} orientations that we can sample with synthetic spacecraft trajectories.

\begin{figure*}[th]
    \centering
    \includegraphics[width=0.95\textwidth]{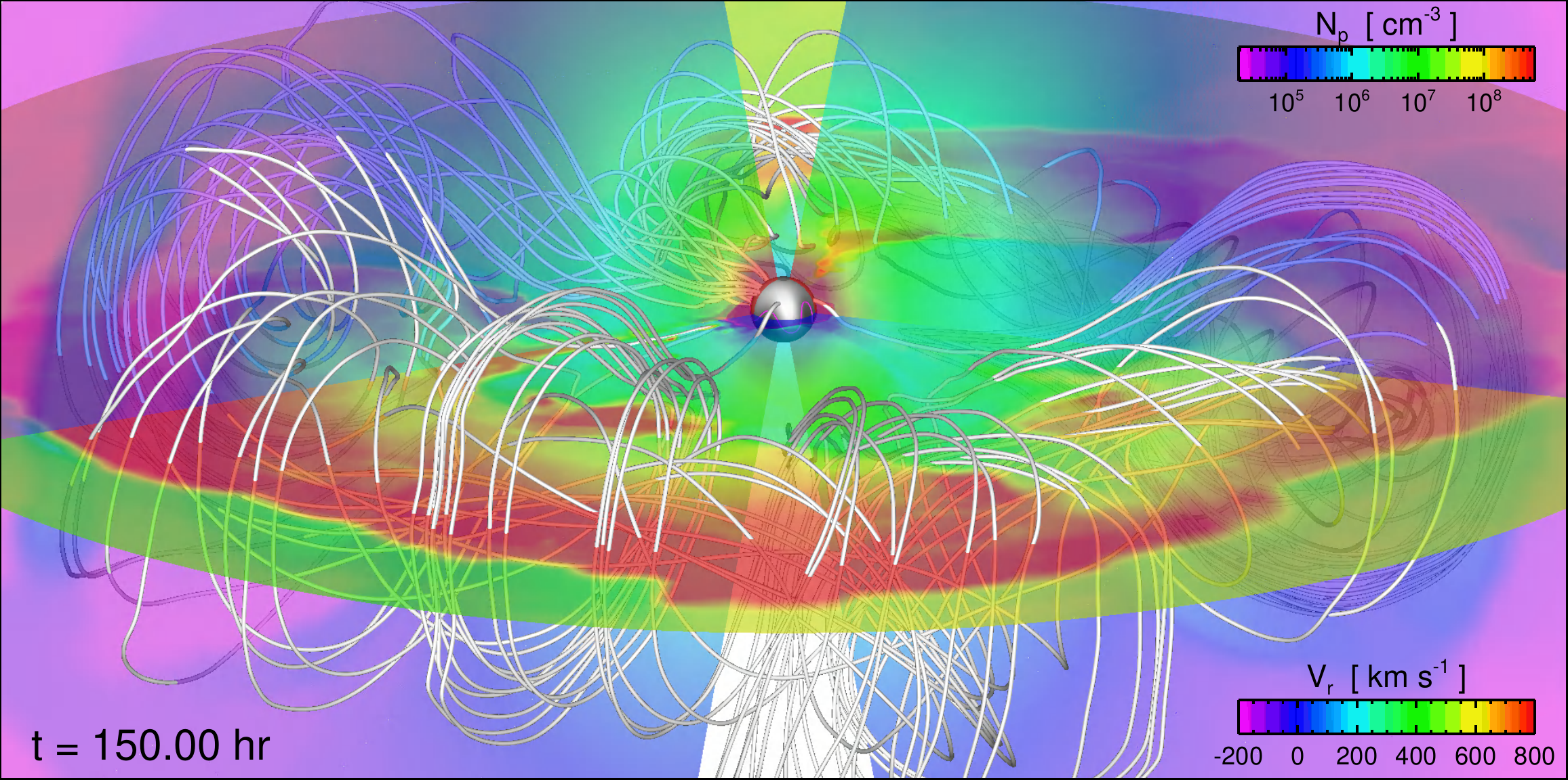}
    \caption{Snapshot of the \citet{Lynch2019} global-scale streamer blowout eruption at $t = 150$~hr. Representative magnetic field lines of the 3D flux rope ejecta are shown in light gray. The semi-transparent equatorial plane shows the radial velocity $V_r$. The meridional planes at longitudes $\phi = -130^{\circ}$ and $+4^{\circ}$ show the proton number density $N_p$. The viewing perspective is centered at longitude $\phi=-67^{\circ}$ and here the CME flux rope axis lies just below the equatorial plane.}   
    \label{fig:recap}
\end{figure*}

Figure~\ref{fig:recap} presents a 3D visualization of the global-scale, MHD flux rope CME ejecta at $t=150$~hr. Representative magnetic field lines are plotted in light gray. The semi-transparent equatorial plane shows the radial velocity $V_r$ and the leading edge of the flux rope structure at ${\gtrsim}800$~km/s. The semi-transparent meridional planes show the proton number density (on a logarithmic scale), which highlights the density structure of the ejecta flux rope cross-section---the standard three-part CME structure of an approximately circular, leading edge enhancement surrounding a depleted cavity region and followed by a dense, central core \citep{Illing1985,Vourlidas2013}. We note that in this particular quadrant of the simulation the axis of the flux rope is slightly below the equatorial plane.

\section{Synthetic Spacecraft Trajectories} \label{sec:orbits}

\begin{figure*}[t!]
    \centering
    \includegraphics[width=0.98\textwidth]{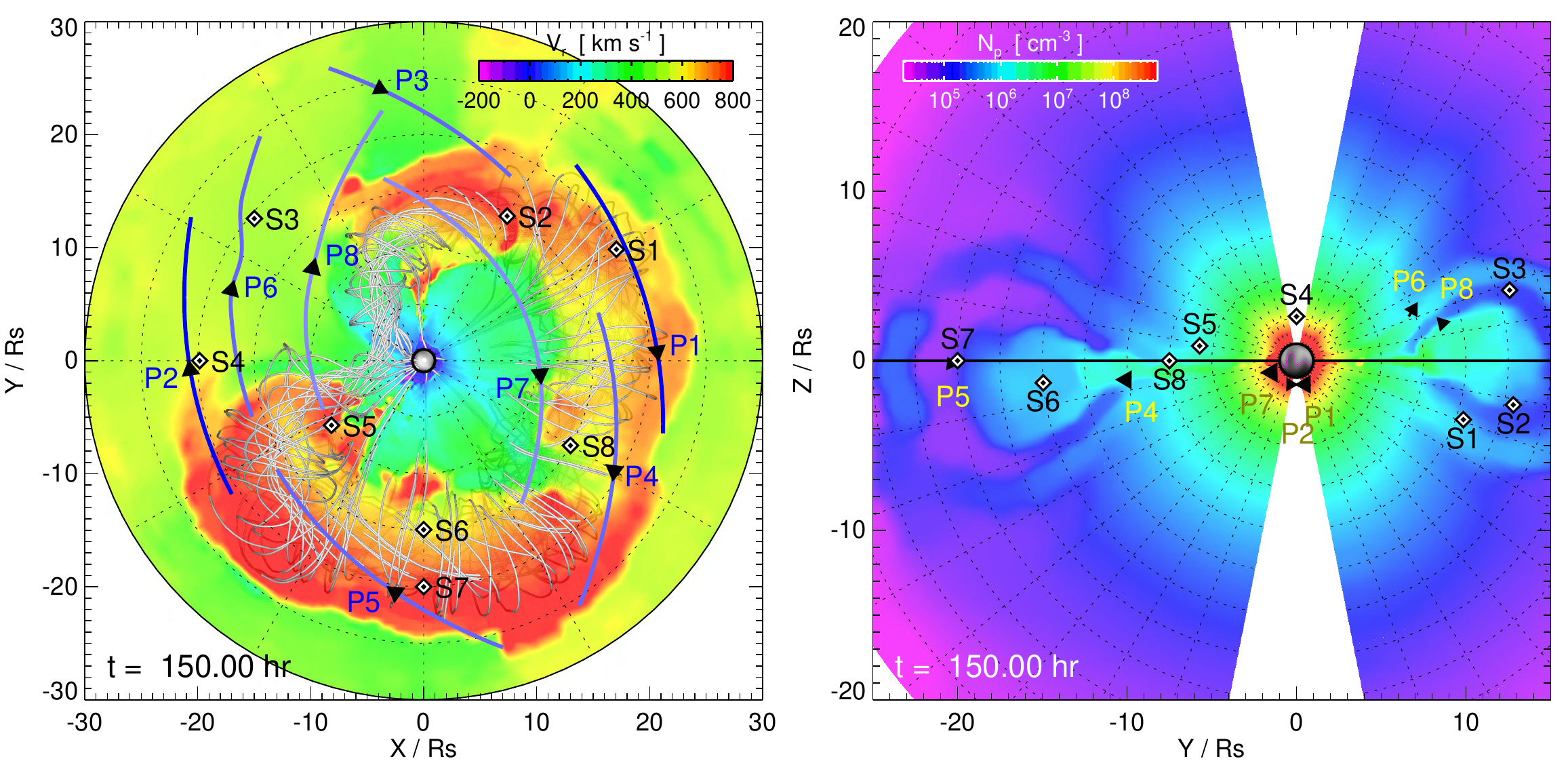}
    \caption{Location of the 16 synthetic spacecraft observers at $t=150$~hr. 
    The left panel shows the radial velocity $V_r$ (with representative CME flux rope field lines) from a top-down view of the ecliptic plane and the right panel shows $N_p$ in the meridional plane-of-the-sky from longitude $\phi=0^{\circ}$.
    The position of the stationary observers (S1--8) are denoted by diamonds, whereas the instantaneous position of the moving observers (P1--8) are shown as the triangular arrow heads along the $x$--$y$ plane projections of their respective trajectories during the numerical simulation.}
    \label{fig:orbits}
\end{figure*}

In the following sections, we have constructed a set of time series corresponding to data that would be measured by 16 different synthetic spacecraft observers, \textcolor{red}{as illustrated in Figure~\ref{fig:orbits}.} Eight synthetic spacecraft are stationary in space (labeled S1--S8) and the other eight are based, in part, on previous and future PSP trajectories near perihelion (labeled P1--P8). \textcolor{red}{While each observer generates their own times series of the global MHD eruption passing by, in this study we treat each observers' encounter as isolated and independent, i.e., the 16 resulting time series should be considered as 16 different events that are grouped by certain characteristics of their encounters.} 

The synthetic time series are obtained from physical quantities in the native ARMS spherical coordinates. We take each synthetic observing spacecraft and create a new magnetic field time series, $\boldsymbol{B}_{\mathrm{RTN}}(t)$, where $(\, B_R, \, B_T,\, B_N \,)$ are obtained from the ARMS $(\, B_r,\, B_\theta, \, B_\phi \,)$ values via straightforward linear transformations. For eleven of the 16 synthetic spacecraft, the RTN transform is given by

\begin{equation}
\boldsymbol{B}_{\mathrm{RTN}}
= \left[
\begin{array}{ccc}
1 & 0 & 0 \\
0 & 0 & 1 \\
0 & -1 & 0
\end{array} \right] \left[
\begin{array}{r}
B_r\\
B_\theta\\
B_\phi
\end{array} \right] \; .
\end{equation}
For the remaining five synthetic spacecraft (S2, S6, S8, P3, and P7), we have rotated the MHD quantities by $90^{\circ}$ counterclockwise around $\boldsymbol{\hat{r}}$ to obtain the RTN time series via the transform
$\boldsymbol{B}_{\mathrm{RTN}} = B_r\boldsymbol{\hat{r}} + B_\theta \boldsymbol{\hat{t}} + B_\phi \boldsymbol{\hat{n}}$.

In the synthetic spacecraft RTN coordinates, the inclination angle $\boldsymbol{B}_{\mathrm{RTN}}$ makes with respect to the $R$--$T$ plane is given by

\begin{equation} \label{eqdelt}
    \delta = \sin^{-1}\left[ \frac{ B_N }{\sqrt{ B_R^2 + B_T^2 + B_N^2}} \right] 
\end{equation}
and the azimuthal angle within the $R$--$T$ plane is given by the usual

\begin{equation} \label{eqlamb}
    \lambda = \left\{ \begin{array}{ll}
                \sin^{-1}\psi & \mathrm{for} \;\; B_R \ge 0, \; B_T \ge 0 \\
                2\pi + \sin^{-1}\psi & \mathrm{for} \;\; B_R \ge 0, \; B_T < 0 \\
                \pi - \sin^{-1}\psi & \mathrm{for} \;\; B_R < 0, \; B_T \ge 0 \\
                \pi + \sin^{-1}\psi & \mathrm{for} \;\; B_R < 0, \; B_T < 0 
              \end{array} \right.
\end{equation}
where $\psi \equiv B_T/\sqrt{B_R^2 + B_T^2}$. Equations~\ref{eqdelt} and \ref{eqlamb} yield the standard convention for the angular ranges, $\delta \in [-90^\circ,+90^\circ]$ and $\lambda \in [\, 0^\circ,360^\circ]$.

We use volume-weighted interpolation to obtain an estimate of the MHD variables at each synthetic spacecraft's position from the surrounding 8 grid points. The temporal cadence of the simulation output files are one every two hours for $t \le 140$~hr, one per hour for $140 < t \le 145$~hr, and one every 5~minutes for $t > 145$~hr. While actual in-situ plasma and field time series can be of much higher cadence (i.e. seconds resolution), historically ICME flux rope magnetic fields have been analyzed at 1~hr resolution, so our synthetic time series temporal resolution of 5~minutes is more than adequate. 

\begin{table*}
 \centering
 \begin{threeparttable}[h]
 \caption{Synthetic spacecraft observer positions/orbits.}
 \label{tab:spc}
 \begin{tabular}{|c|rrr||c|cccr|}
  \hline
  Observer & $r$ [$R_\odot$] & Lat.\tnote{a} \; [$^\circ$] & Long.\tnote{a} \; [$^\circ$] & Observer & PSP Enc.\# & PSP Orbit  $t_0$ & Sim. $t_0$ [hr] & $r_{\rm min}$ [$R_\odot$] \\ 
  \hline
  \hline
  {S1} & 20 & $-10.0$ & 30.0 & 	{P1} & 7 & 2021-01-16, 21:35 &  120.583 & 20.36 \\
  {S2}\tnote{b} & 15 & $-10.0$ & 60.0 & 	{P2} & 7\tnote{c} & 2021-01-17, 07:40 &  130.667 & 20.36 \\
  {S3} & 20 & 12.0 & 140.0 & 	{P3}\tnote{b} & 9 & 2021-08-07, 20:20 &  140.333 & 16.36 \\
  {S4} & 20 & 7.5 & 180.0 & 	{P4} & 9 & 2021-08-09, 20:55 &  128.917 & 16.00 \\
  {S5} & 10 & 5.0 & $-145.0$ & 	{P5} & 9\tnote{c}& 2021-08-07, 22:15 &  122.250 & 16.18\\
  {S6}\tnote{b} & 15 & $-5.0$ & $-90.0$ & 	{P6} & 9\tnote{c,d} & 2021-08-09, 19:25 &  132.417 & 15.98 \\
  {S7} & 20 & 0.0 & $-90.0$ & 	{P7}\tnote{b} & 23 & 2025-03-22, 00:15 &  123.250 & 9.86 \\
  {S8}\tnote{b} & 15 & 0.0 & $-30.0$ & 	{P8} & 23\tnote{c,d} & 2025-03-22, 12:05 &  135.083 & 9.86 \\
  \hline
  \end{tabular}
  \begin{tablenotes}
   \item[a] \textcolor{red}{Latitude and longitude are defined with respect to and within the ecliptic plane, respectively.}
   \item[b] The MHD simulation data has been rotated by $90^{\circ}$ about $\boldsymbol{\hat{r}}$.
   \item[c] The PSP orbit coordinates for this encounter have been rotated by $180^{\circ}$ in longitude.
   \item[d] The PSP orbit latitude includes a time-dependent offset. See Section~\ref{sec:orbits:psp} for details.
  \end{tablenotes}
 \end{threeparttable}
\end{table*}

\begin{figure*}[!t]
    \centering
    \includegraphics[width=1.00\textwidth]{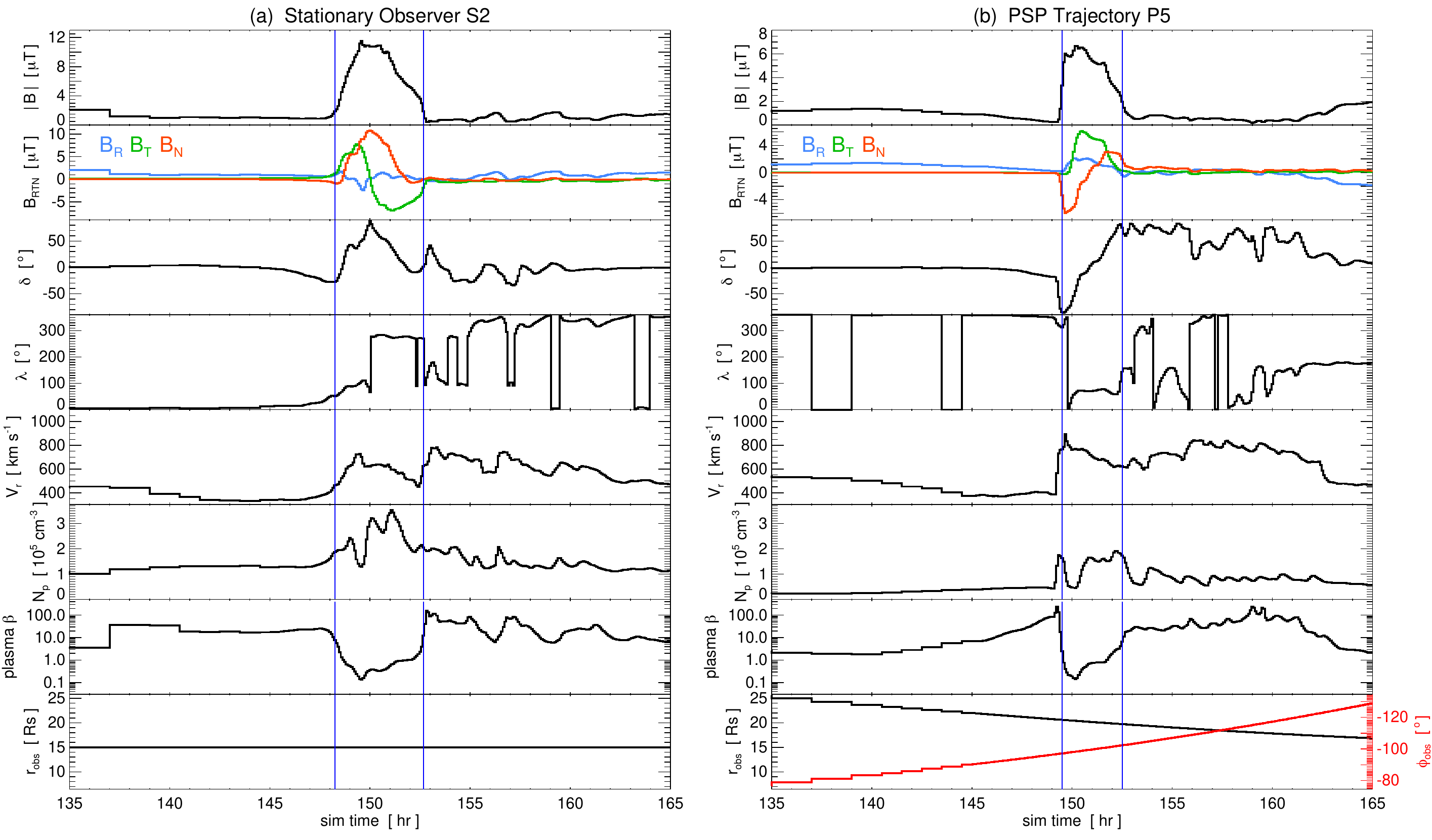}
    \caption{Representative in-situ time series obtained by synthetic spacecraft observers, S2 (left) and P5 (right). Each column plots magnetic field magnitude, the $\boldsymbol{B}_{\mathrm{RTN}}$ components, the latitude ($\delta$) and longitude ($\lambda$) of the vector field orientation, the radial velocity $V_r$, the proton number density $N_p$, plasma $\beta$ and the observer's radial distance $r_{\mathrm{obs}}$ (in the P5 case, we also show the observer's longitudinal position, $\phi_{\mathrm{obs}}$, in red). The vertical blue lines in each column indicate the boundaries of the flux rope CME seen by each spacecraft.}
    \label{fig:insitu}
\end{figure*}

\subsection{Stationary Observers} \label{sec:orbits:stationary}

The eight stationary synthetic spacecraft are spread around the computational domain at radial distances between $10$--$20\,R_\odot$ to sample different regions of the simulation's global-scale flux rope CME. Due to the 3D structure of the MHD ejecta, many of the synthetic spacecraft are slightly above or slightly below the ecliptic plane. \textcolor{red}{We note that these positions include latitudes that exceed PSP's maximum out-of-the-ecliptic excursion of ${\sim}3^{\circ}$ but, for the purposes of this study, the locations were chosen to sample different portions of the MHD ejecta.} Each of the stationary observers' location in radial distance, latitude \textcolor{red}{(with respect to the ecliptic plane)}, and longitude \textcolor{red}{(within the ecliptic plane)} is given in the first three columns of Table~\ref{tab:spc} and shown visually in Figure~\ref{fig:orbits}. The left panel of Figure~\ref{fig:orbits} shows these positions projected into the equatorial plane and right panel shows them against the plane of the sky from a longitude of 0$^{\circ}$. Each contour plot shows the spatial distribution of the radial velocity at $t=150$~hr during the eruption from their respective viewpoints (the same time as shown in Figure~\ref{fig:recap}). \textcolor{red}{We note that the synthetic spacecraft in the upper left quadrant (S3, S4, P2, P3, P6, and P8) of the ecliptic view in Figure~\ref{fig:orbits} have not yet encountered the CME by $t=150$~hr.}

Figure~\ref{fig:insitu}(a) plots a representative time series of the MHD quantities sampled by stationary observer S2. From top to bottom we have displayed the magnetic field magnitude $|B|$, the field components $B_R$ (blue), $B_T$ (green), and $B_N$ (red), the magnetic field vector's elevation angle $\delta$, its azimuthal angle $\lambda$, the radial velocity $V_r$, the proton number density $N_p$, the plasma $\beta = 8\pi P / B^2$, and finally the radial distance of the observer $r_{\mathrm{obs}}$. The CME ejecta impacts observer S2 at a simulation time of $t=148.25$~hr \textcolor{red}{(approximately 3.25 hours after the CME erupts from the Sun at $t=145$~hr)} and we estimate it has passed over the spacecraft by $t=152.67$~hr. The magnetic flux rope interval is bounded by the two vertical lines. The major identifying characteristics of a magnetic cloud flux rope ejecta are immediately visible. The magnetic field magnitude is enhanced, the $B_T$ and $B_N$ field components show a smooth, coherent rotation through a large angle, and while there is a significant density enhancement during the ejecta interval compared to the upstream solar wind values, the ejecta itself is magnetically dominated, with a sharp transition to low $\beta$ (${<}1$) for the duration of the flux rope. As will be discussed later, by visual inspection, the magnetic cloud has a clear unipolar, West--North--East (WNE) orientation.

\subsection{Parker Solar Probe Trajectories} \label{sec:orbits:psp}

The spatial coordinates of the eight moving synthetic spacecraft are derived from PSP encounter orbits 7, 9, and 23. The encounter~7 perihelion was at a radial distance of $20.36\,R_\odot$ and occurred on 17 January 2021 at 17:35~UT. The encounter~9 perihelion was at $15.98\,R_\odot$ on 9 August 2021 at 19:10~UT. The encounter~23 perihelion is currently scheduled to reach a distance of $9.86\,R_\odot$ on 22 March 2025 at 21:55~UT. These particular orbits were chosen such that their perihelia roughly correspond to the radial distances of the stationary observers ($10\,R_\odot$, $15\,R_\odot$, and $20\,R_\odot$).

In order to increase the number of PSP-like spacecraft time series, we have duplicated the three PSP encounter orbits and added an additional 180$^{\circ}$ shift in longitude, thus obtaining two sets of coordinate trajectories for each PSP encounter. Synthetic spacecraft P1 and P2 are derived from encounter~7. P3 and P4 occur on the inbound and outbound legs of encounter~9, while P5 and P6 follow the inbound and outbound legs of the phase-shifted encounter~9 duplicate. P7 and P8 follow portions of encounter~23 and its duplicate. Again, due to the 3D structure of the MHD ejecta, we have constructed a latitudinal offset for P6 and P8 trajectories so that they intersect more of the CME. 
The latitude in the P6 trajectory is defined as

\begin{equation}
    \theta_{\rm obs}(t) = \theta_{\rm P6}^{\rm \, PSP}(t) + 20.3098 \, \left( \frac{1}{2} - \frac{1}{2}\cos{\left[ 2\pi\frac{(t-154.5)}{20} \right]} \right)
\end{equation}
in units of degrees and for the duration of the interval $t \in [144.5,164.5]$~hr.
Likewise, the P8 trajectory latitude is

\begin{equation}
    \theta_{\rm obs}(t) = \theta_{\rm P8}^{\rm \, PSP}(t) + 13.2843 \, \left( \frac{1}{2} - \frac{1}{2}\cos{\left[ 2\pi\frac{(t-151)}{20} \right]} \right)
\end{equation}
over the interval $t \in [141,161]$~hr.

Table~\ref{tab:spc} also summarizes properties of the P1--P8 observers. The column labeled `PSP Orbit $t_0$' lists the date and time during the PSP Encounter \# that we have assigned to the simulation time, `Sim.\ $t_0$.' This temporal mapping lines up each set of PSP trajectory coordinates with their respective sampling intervals during the MHD simulation. Each of the P1--P8 time-dependent trajectories are shown in Figure~\ref{fig:orbits} as lines and the arrowheads (and labels) indicate the synthetic spacecraft's instantaneous positions at $t=150$~hr.

Figure~\ref{fig:insitu}(b) plots the corresponding representative time series of the MHD quantities sampled by the PSP-like observer P5. Here the synthetic spacecraft encounters the leading edge of the CME at $t_0=149.5$~hr \textcolor{red}{(approximately 4.5 hours after the CME erupts from the Sun at $t=145$~hr)} and the magnetic flux rope portion of the ejecta appears to end at $t_1=152.5$~hr. The characteristics of a coherent flux rope ejecta are immediately visible, just as in Figure~\ref{fig:insitu}(a), and the P5 synthetic observations show a clear bipolar, South--West--North (SWN) type of orientation. \textcolor{red}{At the beginning of the CME encounter P5 is at $\boldsymbol{r}_{\rm P5}(t_0) = ( \, 20.610\, R_\odot, \, -0.531^{\circ}, \, -97.144^{\circ} \, )$ and by the end it has reached $\boldsymbol{r}_{\rm P5}(t_1) = ( \, 19.738\,R_\odot, \, -0.872^{\circ}, \, -102.294^{\circ} \, )$.} Therefore, during the 3-hour CME encounter, P5 transverses a radial distance of $\Delta r_{\rm obs} = 0.872\,R_{\odot}$, a change in longitude of $\Delta \phi_{\rm obs} = 5.15^{\circ}$, and a change in latitude of $\Delta \theta_{\rm obs} = 0.34^{\circ}$ (which we will neglect below). The total distance traveled will be the arc length $\Delta S$, which can be estimated via $\Delta S=\int r_{\rm obs}(t) \, d\phi$ over the encounter interval corresponding to $[\, \phi_{\rm obs}(t_0),\, \phi_{\rm obs}(t_1)\, ]$. This yields $\Delta S = 1.86\,R_\odot$ and an average observer velocity of $v_{\rm obs} \sim \Delta S/\Delta t = 120$~km/s through the CME. \textcolor{red}{At $t=151$~hr, the location of P5 corresponds to a computational grid cell size of $\Delta r = 0.375\,R_\odot$, $ r \Delta \theta = 0.286\,R_\odot$, and $ r \sin{\theta} \Delta \phi = 0.330\,R_\odot$. Thus, the synthetic spacecraft P5 travels ${\sim}2.3$ grid cells in the radial direction and ${\sim}5.5$ grid cells in longitude during the passage of the MHD ejecta. }


%
%

\section{Simulation Flux Rope Profiles and Fitting Results} \label{sec:insitu}

Our synthetic time series are grouped into four types representing different flux rope CME orientations and/or CME--spacecraft configurations: Type~1 encounters are \emph{classic bipolar} magnetic flux rope orientations, which correspond to the synthetic observers S1, S7, P1, and P5; Type~2 encounters are \emph{classic unipolar} magnetic flux rope orientations, corresponding to S2, S6, P3, and P7; we call the Type~3 profiles \emph{problematic orientation} encounters because the CME--spacecraft configurations for S3, S5, P6, and P8 were designed to intersect the CME flux rope axis at a significant angle in the $R$--$T$ plane; and the Type~4 profiles represent \emph{problematic impact parameter} encounters, where the S4, S8, P2, and P4 CME--spacecraft configurations only intersect the outer regions/periphery of the flux rope CME.

The three in-situ flux rope models we have fit to the simulation data are described in detail in \ref{sec:frmodels}. They are the \citet{Lundquist1950} constant-$\alpha$, linear force-free cylinder model (LFF), the \citet{Gold1960} uniform-twist model (GH), and a non-force free circular cross-section model \citep[CCS; e.g.][]{Hidalgo2000,Nieves-Chinchilla2016}. In the following sections we present our synthetic spacecraft time series of the vector magnetic field as the MHD CME intersects the various spacecraft positions and evaluate the performance of the different in-situ flux rope model fits to the simulation data. \textcolor{red}{For each MHD time series, we selected the flux rope ejecta boundaries by visual inspection of the magnetic field and plasma profiles using a combination of the traditional \citet{Burlaga1988} magnetic cloud criteria (enhanced $|\boldsymbol{B}|$, smooth rotation in $\delta$, $\lambda$, and low plasma $\beta$) and co-temporal changes in the density and radial velocity profiles (as illustrated in Figure~\ref{fig:insitu}). These fixed boundaries were used in each of the in-situ flux rope model reconstructions. While the flux rope boundaries were clear in our synthetic profiles, we should mention this is a nontrivial procedure with observational data and modifying the boundaries can, under certain circumstances, strongly affect the resulting flux rope model fits \citep[e.g., see][]{Lynch2003,Riley2004,Ruffenach2012,Al-Haddad2013}.}

\subsection{Model Fits to Synthetic Time Series}
\label{subsec:types}

\subsubsection{Type 1 -- Classic Bipolar Profiles}

\begin{figure*}[!t]
    \centering
    \includegraphics[width=0.85\textwidth]{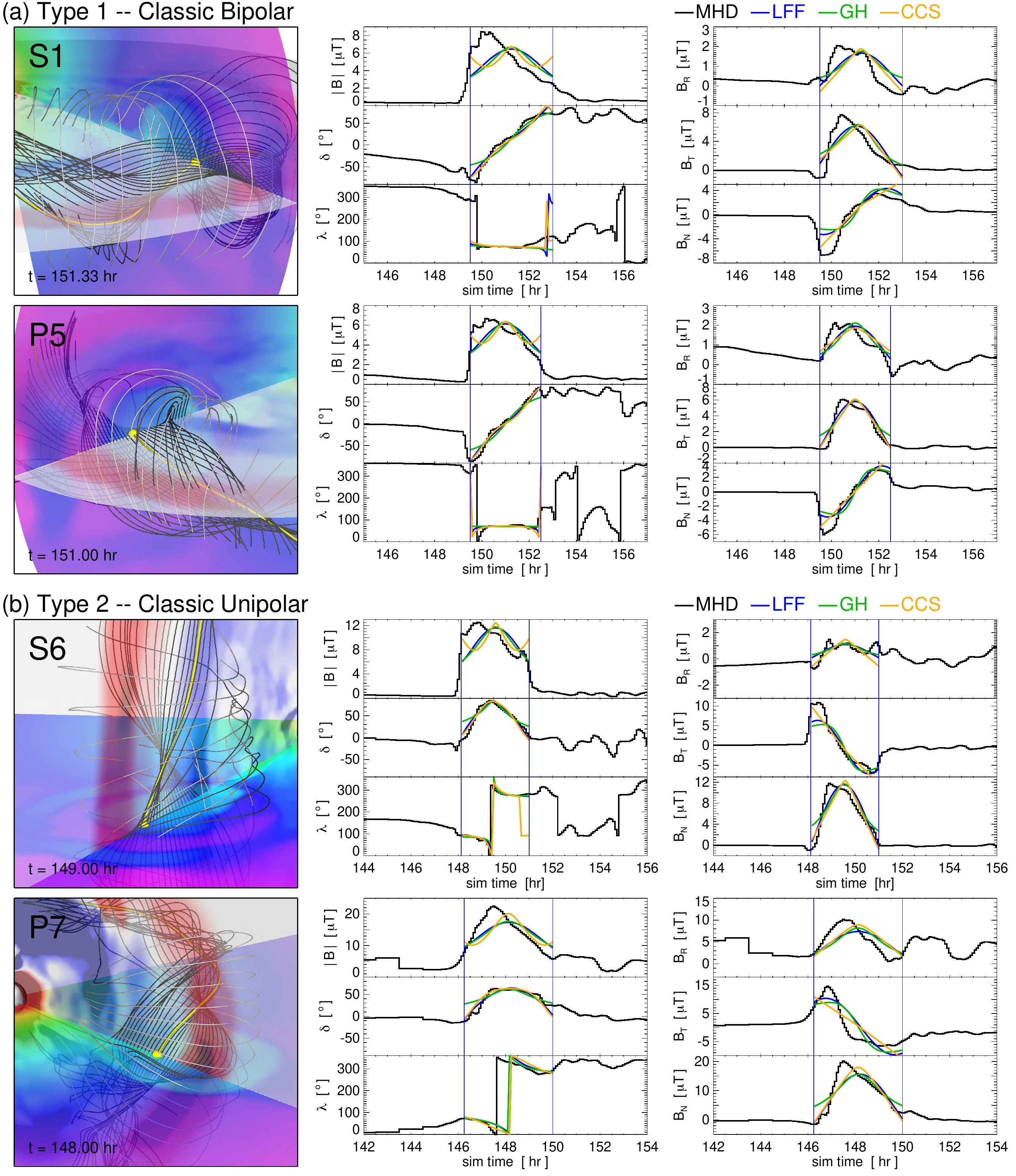}
    \caption{(a) Type 1 magnetic field profiles representing classic bipolar MC/ICME orientations from synthetic observers S1 and P5. 
    (b) Type 2 magnetic field profiles representing classic unipolar MC/ICME orientations from synthetic observers S6 and P7.
    The planar cuts through the flux rope cross-sections show the logarithim of mass density while the planar cuts along the axis shows the twist component of the flux rope magnetic fields.
    The remaining Type 1 and Type 2 encounters are shown in Figure~\ref{fig:fit2}.}
    \label{fig:fit0}
\end{figure*}

The Type~1 simulated CME magnetic field profiles have a local flux rope geometry that corresponds to the classic, bipolar flux rope/magnetic cloud orientation \citep[also defined as a `low-inclination flux rope' in][]{Palmerio2018}. A bipolar flux rope axis orientation is essentially parallel to the ecliptic plane (or to the $R$--$T$ plane in RTN coordinates) and points in more-or-less the $\pm\boldsymbol{\hat{t}}$ direction. Thus, the normal field component either starts positive and smoothly transitions to negative (i.e. North-to-South) over the course of the flux rope transient or vice-versa (South-to-North). The tangential field component is approximately zero at the leading edge flux rope boundary but smoothly increases to its maximum magnitude at the center of the flux rope with positive (negative) values pointing West (East) and then smoothly decreases back to approximately zero at the trailing edge boundary. These ``horizontal'' flux rope orientations result in the large-scale, coherent magnetic field rotations that can be classified as NWS, NES, SWN, or SEN using the \citet{Bothmer1998} and \citet{Mulligan1998} convention.

Figure~\ref{fig:fit0}(a) shows two of the four Type~1 synthetic spacecraft CME encounters, S1 and P5, that represent these classic bipolar flux rope profiles. Each synthetic spacecraft time series is shown on its own row. From left to right, first we show a 3D visualization of the spacecraft position (yellow sphere currently sampling the yellow field line),  a set of adjacent field lines (dark gray) traced at $0.25\,R_\odot$ intervals in the radial direction either side of the spacecraft point, and a set of light gray field lines indicating the approximate MHD flux rope boundary. The next two columns are the time series of the vector magnetic field profiles as $( \, |\boldsymbol{B}|, \, \delta, \, \lambda \, )$ in the center and $( \, B_R, \, B_T, \, B_N \, )$ on the right. In each of the time series panels, we have indicated the flux rope ejecta boundaries as vertical lines and plotted each of the in-situ flux rope model fits: LFF--blue, GH--green, and CCS--orange. The parameter values for each models' fit to S1 and P5 are given in Table~\ref{tab:fits12}.

The Type~1 flux rope profiles ought to be the simplest and most straightforward to fit with the in-situ flux rope models. In general, each of the flux rope models captures some or even most of the overall trend in the coherent field rotation, but there are aspects of the MHD profiles that certain flux rope models cannot reproduce. Specifically, the asymmetry in the $|\boldsymbol{B}|$ profile---where the peak/maximum value is more toward the front of the ejecta than the center. The LFF, GH, and CCS models are all symmetric, by construction. This aspect is also seen in the RTN components. All of the models underestimate the amplitude of the initial negative $B_N$ components because their $B_N$ profiles are symmetric with respect to the center of the spacecraft crossing. This is a well known shortcoming of models with cylindrical symmetry.

The other feature worth mentioning is that the MHD profiles, even in the simplest bipolar flux rope orientations, have a non-zero $B_R$ component. While this does not significantly affect the overall structure of the large-scale field rotation, it does indicate that the idealized, in-situ flux rope models will not result in a perfect fit. In \ref{sec:frmodels}, the formulas for each model's field components are given and every model has a radial component of zero (in their local flux-rope frame). Thus, to generate a non-zero $B_R$ profile, it must originate from a tilt or angle of the flux rope symmetry axis with respect to the RTN coordinates and/or from a trajectory that passes either above or below the flux rope cylinder symmetry axis. In other words, an orientation of $\phi_0 = 90^{\circ}$, $\theta_0 = 0^{\circ}$, which points the $z_{\rm FR}$ axis parallel to the RTN $\pm\boldsymbol{\hat{t}}$ direction will always give $B_R = 0$ for an impact parameter of $p_0 = 0$. This is also true for a completely vertical flux rope orientation of $\theta_0 = \pm90^{\circ}$, which we discuss below.

\subsubsection{Type 2 -- Classic Unipolar Profiles}

The Type~2 magnetic field profiles correspond to a local unipolar flux rope/magnetic cloud orientation \citep[also defined as a `high-inclination flux rope' in][]{Palmerio2018}. In a unipolar flux rope event, the axis is essentially perpendicular to the $R$--$T$ plane, aligned with the $\pm\boldsymbol{\hat{n}}$ direction---or at least it makes a significant angle so the axis is highly inclined. In these cases the normal field component is predominately North or South, while the bipolar rotation signature of positive-to-negative (negative-to-positive) is in the West-to-East (East-to-West) direction. These ``vertical'' flux rope orientations result in the large-scale, coherent magnetic field rotations that can be classified as WNE, WSE, ENW, or ESW in the \citet{Bothmer1998} and \citet{Mulligan1998} convention. We note that, in general, unipolar South magnetic clouds tend to drive the most intense geomagnetic storms \citep[e.g.][]{Zhang2004}, however our particular MHD coordinate transform results in unipolar North configurations.

Figure~\ref{fig:fit0}(b) shows two examples of the Type~2 synthetic spacecraft CME encounters, S6 and P7, that represent classic unipolar flux rope orientations in the same format as Figure~\ref{fig:fit0}(a). The parameter values for each models' fit to S6 and P7 are also given in Table~\ref{tab:fits12}. Visually, the quality of the in-situ flux rope model fits to the unipolar MHD time series seem comparable to those for the classic bipolar events in Figure~\ref{fig:fit0}(a). In general, the large-scale magnetic structure of the coherent field rotations are reasonably well captured in the flux rope models, with some components matching better than others (e.g. the $B_T$ profiles of S6 look better than those of P7 but the $B_N$ profiles look similar).

The fit parameters for the P7 encounter are essentially identical across models (given the typical parameter uncertainties) indicating a fairly robust reconstruction. For example, the axial tilt $\theta_0$ values are all within $70^{\circ}$--$80^{\circ}$, the impact parameters are all ${\sim}20$\%, and each model determines the flux rope size to be $R_c \approx 0.02$~AU.  The S6 encounter fits are also pretty consistent in certain parameters (e.g. $\theta_0$, $R_c$), however the LFF and CCS models give impact parameters $p_0/R_c$ of only a few \% whereas GH gives 23\%. The P7 fits are apparently a bit better than the S6 fits, in that there is less variation between models in the orientation parameters. This is likely due to the S6 time series having more $|\boldsymbol{B}|$ asymmetry than P7. However, as we will see in Section~\ref{subsec:hodogram}, taking the variation between different models as an assessment of fit quality may not encompass the whole picture.

\subsubsection{Type 3 -- Problematic Orientation Profiles}

\begin{figure*}[!t]
    \centering
     \includegraphics[width=0.85\textwidth]{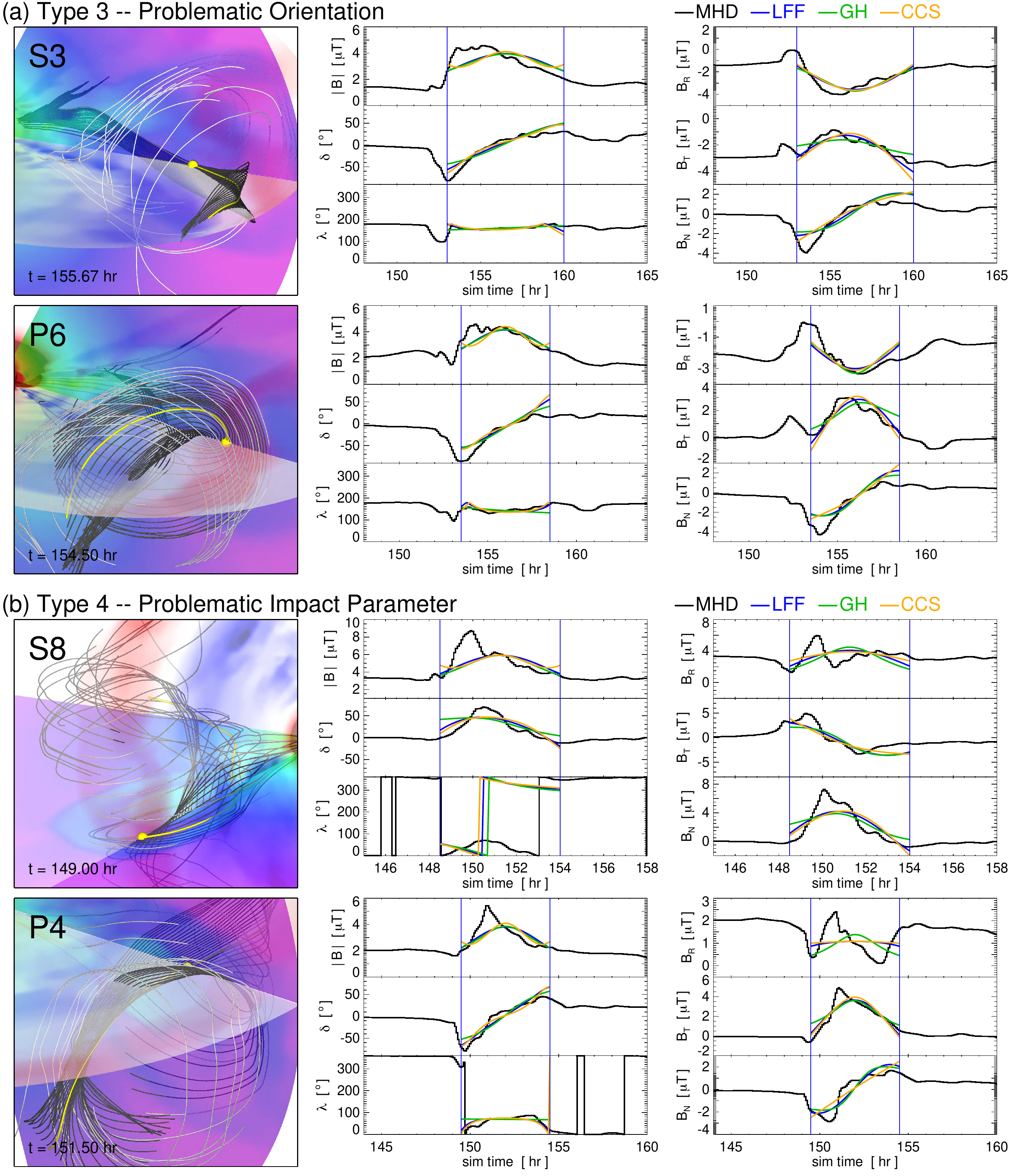}
    \caption{(a) Type 3 magnetic field profiles representing problematic orientation MC/ICME events from synthetic observers S3 and P6. 
    (b) Type 4 magnetic field profiles representing problematic impact parameter MC/ICME events from synthetic observers S8 and P4.
    The remaining Type 3 and Type 4 encounters are shown in Figure~\ref{fig:fit3}.}
    \label{fig:fit1}
\end{figure*}

The next category of flux rope encounter we are investigating in the MHD simulation and the flux rope model fits are profiles that we have described as having a ``problematic orientation.'' What this means is that the flux rope axis makes a significant angle \emph{within} the spacecraft's $R$--$T$ plane but is still has a relatively low inclination, i.e. the flux rope parameter angle representing tilt out of the $R$--$T$ plane, $\theta_0$, is small, but the azimuthal angle, $\phi_0$, has a large departure from 90$^{\circ}$ or 270$^{\circ}$. This results in a spacecraft trajectory that includes a significant component along or parallel to the magnetic flux rope axis. This type of flux rope orientation has been discussed by \citet{Marubashi1997}, who sketched the geometry (see their Figure~3) and presented two examples of magnetic cloud observations consistent with this interpretation. Since then, there has been a number of well-observed flux rope ICME events that have similar morphology and interpretation \citep{Owens2012}. This scenario has also been recently considered in the context of PSP CME encounters \citep{Moestl2020}.

Figure~\ref{fig:fit1}(a) shows the Type~3 problematic orientation events seen by observers S3 and P6. In these examples, the significant angle the (local) flux rope axis makes with respect to the $\pm\boldsymbol{\hat{t}}$ direction can be seen in the constant-latitude plane rendering of the MHD simulation $B_\theta$ values (in the red-to-white-to-blue color scheme). Here the red (blue) colors represent positive (negative) values, illustrating the bipolar or twist component of the flux rope. The local flux rope axis, therefore, can be considered as the $B_\theta=0$ point of the sign change, i.e. essentially the midpoint of the central white stripe between the red-to-blue transition. The intersection of the meridional plane ($R$--$N$) and the latitudinal plane ($R$--$T$) shows the $\boldsymbol{\hat{r}}$ direction. Since the angle between the flux rope axis (white stripe) and $\boldsymbol{\hat{r}}$ is not 90$^{\circ}$, there will be some contribution of both the axial and azimuthal fields of the flux rope to the spacecraft's $B_R$ time series.

The flux rope parameter values for S3 and P6 are given in Table~\ref{tab:fits34}. Every in-situ model fit for S3 gives $\phi_0$ values with anywhere from 20$^{\circ}$--70$^{\circ}$ departure from the classic bipolar orientation of $\phi_0 = 90^{\circ}$, while the model fits to P6 give $\phi_0$ values with a 5$^{\circ}$--60$^{\circ}$ departure. The model profiles shown in the central and right columns of Figure~\ref{fig:fit1}(a) do not give the impression of especially good or especially bad flux rope fits, even when compared to the classic Types~1 and 2 of Figure~\ref{fig:fit0}. However, there is much more variation in the fit parameters (and derived quantities) between models for these cases than in those previous events. For example, in the S3 (P6) fits, the flux rope radius, $R_c$, ranges from 0.013--0.045~AU (0.018--0.037~AU), largely due to the variation in the impact parameter. In the LFF and CCS models, the impact parameter $p_0/R_c$ is found to be on the order of 50\% for both S3 and P6, but the visual representation of the spacecraft in Figure~\ref{fig:fit1}(a) shows this is clearly not the case. Thus, despite the more-or-less adequate fits to the MHD time series components, these in-situ flux rope fits should be considered \emph{more uncertain} than those of the classic types.

\subsubsection{Type 4 -- Problematic Impact Parameter Profiles}

Our final category of synthetic observer flux rope encounters is the problematic impact parameter type. These events correspond to spacecraft trajectories that pass a substantial distance from the central symmetry axis of the ICME flux rope. In terms of in-situ flux rope model parameters, the Type 4 events are usually characterized by normalized impact parameters of the order $p_0/R_c \gtrsim 0.50$. The effects of the large impact parameter on the flux rope magnetic field signatures is typically to decrease the magnitude of every component as well as to reduce the overall rotation that the $\boldsymbol{B}$ vector makes during the ICME interval. It is well known that the quality of the in-situ flux rope model fits also suffers when the impact parameter becomes large. For example, in one of the first papers to compare different flux rope model fits to MHD simulation data, \citet{Riley2004} showed that when the impact parameter was small, essentially every model gave similar and consistent fitting results, whereas for a large impact parameter there was much more variation between model fits and none of the models did particularly well at reproducing the actual distorted, elliptical shape of the MHD ejecta cross-section.

Figure~\ref{fig:fit1}(b) shows the Type~4 events, S8 and P4. The S8 encounter is rotated in the same manner as the classic unipolar cases. Here we can see the $\delta$ profile has a single, central peak at $\sim$70$^{\circ}$ corresponding to the unipolar $B_N$ component profile. In both of the MHD visualizations, the large impact parameter puts the S8 and P4 spacecraft trajectories through their flux ropes much closer to the outer layers. Hence, the dark gray field lines are seen to wrap \emph{around} the flux rope axis much more than in Figure~\ref{fig:fit0} or \ref{fig:fit1}(a) where they were essentially tracing field lines \emph{through} the central axis.

The in-situ flux rope model fit parameters are fairly consistent for S8, i.e. each give an axis inclination angle of $\sim$67$^{\circ}$, impact parameters between 50--60\% of $R_c$, and $R_c$ values of 0.036~AU.  The flux rope model fit parameters for the P4 synthetic trajectory yield consistent values for some aspects of the eject (e.g. $\theta_0$, $R_c$) but considerably more variation in others (e.g. $\phi_0$, $p_0$). Qualitatively, the S8 time series has a larger field strength and more of a coherent rotation in the angular profiles and/or component profiles. The P4 time series shows almost no rotation in the $(\delta, \lambda)$ angles.

An important feature of both the Type 3 and Type 4 problematic encounter profiles is that the set of $\chi^2$ error norms---defined in Equation~\ref{eq:chi2} and used in the optimization of the model fit parameters through its minimization---are not quantitatively worse than those of Types 1 and 2. It is also not immediately obvious that a given set of model fits to a Type 3 or 4 time series are, by visual inspection, qualitatively worse than the Type 1 or 2 fits. In a certain sense, each set of model fits for the Type 3 and Type 4 encounters \emph{are} worse than the corresponding set of model fits for each Type 1 and Type 2 encounter, but we would have to use a ``quality of fit'' metric that is includes the variation of the models' best-fit parameters within each set.

\subsection{Comparisons of MHD and In-situ Model CME Properties}
\label{sec:MHD-FR-comp}
\subsubsection{\textcolor{red}{Flux Rope Geometry and Orientation}}
\label{subsec:geom}

\textcolor{red}{One of the primary uses of the in-situ flux rope models is to estimate the large-scale ICME flux rope geometric properties and orientation. The parameters describing the flux rope model cylinder axis direction are the two orientation angles: the azimuthal angle $\phi_0$ in the $R$--$T$ plane and the elevation angle $\theta_0$. The third parameter required to fully specify the 3D orientation with respect to the observing spacecraft is the impact parameter $p_0$, which describes how close the spaceraft trajectory passes to the flux rope cylinder axis. The radial size of the cylindrical cross-section, $R_c$, can then be calculated as a function of the three model orientation parameters and the observed radial velocity time series---specifically, the ejecta duration and an average $V_r$ during the interval (e.g.\ see equation A4 in \citealt{Lynch2005}).}

\textcolor{red}{In order to compare the in-situ flux rope model fit parameters with the MHD simulation ``ground truth,'' we need to estimate equivalent cylinder geometry parameters from the MHD data cubes. Our procedure for estimating the MHD version of $\{\, \phi_0, \, \theta_0, \, p_0/R_c, \, R_c \, \}$ is as follows. The flux rope size, $R_c$, is the most straightforward; we choose left ($L$) and right ($R$) boundary features in the $r$ and $\theta$ directions based on the mass density and current density structure of the flux rope to obtain
\begin{equation}
    R_c^{\rm MHD} \approx \frac{1}{2}\left[ \, \frac{\left( r_R - r_L\right)}{2} + \left(\frac{ r_L + r_R}{2} \right) \left( \frac{| \, \theta_R - \theta_L \, |}{2} \right) \, \right] \; . 
\end{equation}
The normalized impact parameter can then be obtained by estimating the center of the MHD ejecta cross-section as the mid-point of the $L$, $R$ boundary features, ($r_m$, $\theta_m$), where $r_m = (r_L+r_R)/2$, $\theta_m = (\theta_L + \theta_R)/2$, and using the coordinates of the synthetic observing spacecraft. This yields
\begin{equation}
    \left( \frac{p_{0}}{R_c} \right)^{\rm MHD} \approx \frac{r_m \, \left( \, \theta_m - \theta_{\rm obs}\left(t^{*}\right) \, \right)}{R_c} \; ,
\end{equation}
where $\theta_{\rm obs}(t^*)$ is the observer latitude at the times shown in each of the MHD visualization panels of Figures~\ref{fig:fit0}, \ref{fig:fit1}, \ref{fig:fit2}, and \ref{fig:fit3}.
}

\textcolor{red}{The cylinder axis orientation parameters, ($\phi_0^{\rm MHD}$, $\theta_0^{\rm MHD}$), are estimated from two planar cuts centered on the estimated midpoint. First, the azimuthal angle, $\phi_0^{\rm MHD}$, is determined from the spatial orientation of the $B_\theta = 0$ contour in the $r$--$\phi$ plane at the latitude midpoint, $\theta_m$. The $\phi_0^{\rm MHD}$ angle is defined with respect to the $+\boldsymbol{\hat{r}}$ direction and the $\pm$180$^{\circ}$ ambiguity is resolved by choosing the direction of the positive axial field. Similarly, the elevation angle, $\theta_0^{\rm MHD}$, is determined from the spatial orientation of the $B_r = 0$ contour on a spherical wedge at the radial midpoint $r_m$ and defined as elevation above or below the $r$--$\phi$ plane at $\theta_m$ (again, in the direction of positive axial field). For both angles, we fit a straight line to the spatial points of the $\{\, B_\theta, \, B_r \,\} = 0$ contours on their respective 2D planes, using the position $(r_m, \, \theta_m, \, \phi_{\rm obs})$ as the origin of a local RTN Cartesian coordinate system and calculate ($\phi_0^{\rm MHD}$, $\theta_0^{\rm MHD}$) from these linear fits. For the rotated cases (S2, S6, S8, P3, P7), we have applied the same procedure as in the non-rotated MHD data and then converted these angles into the corresponding cylinder axis orientation values for the local (rotated) coordinate system.}

\begin{figure*}[!t]
    \centering
    \includegraphics[width=0.98\textwidth]{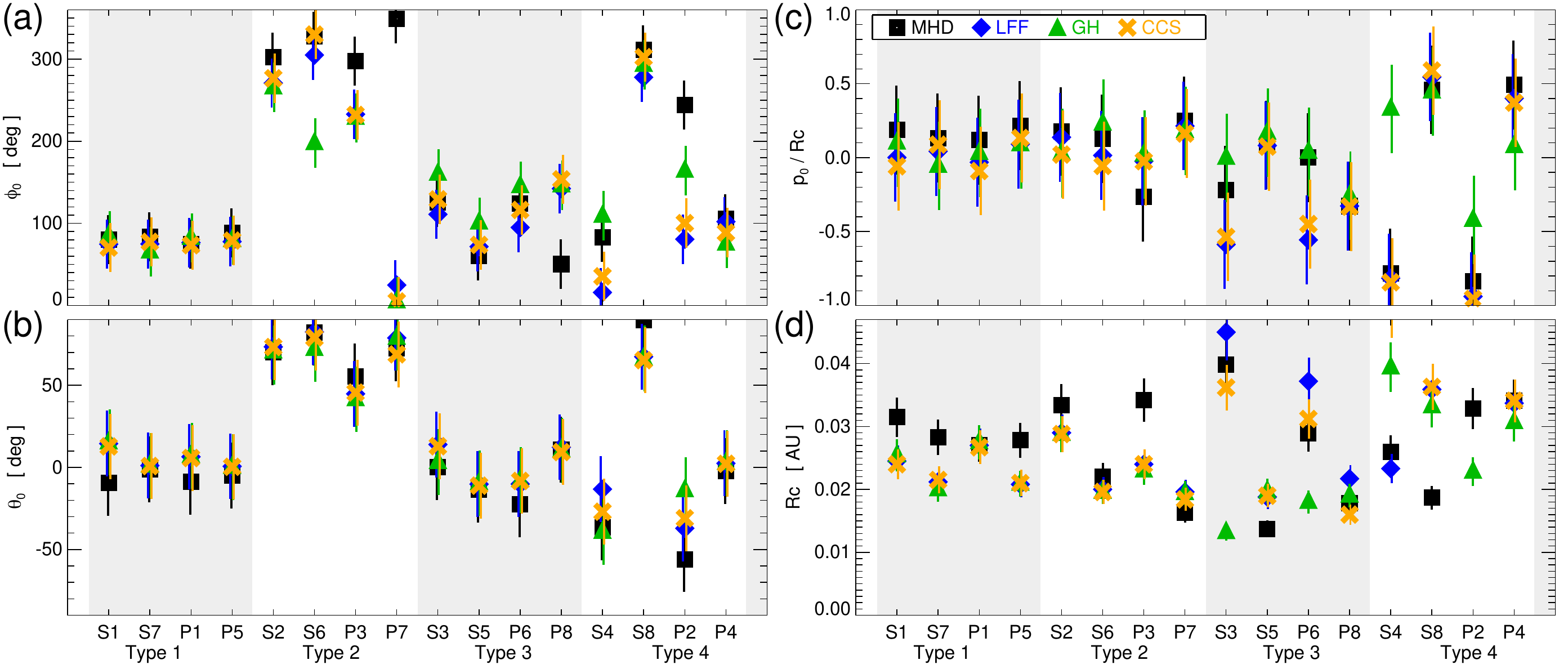}
    \caption{\textcolor{red}{Comparison between the in-situ flux rope model geometric parameters and estimates of their corresponding MHD CME values. (a) Azimuthal angle $\phi_0$ in $R$--$T$ plane. (b) Elevation angle $\theta_0$ out of the $R$--$T$ plane. (c) Normalized impact parameter $p_0/R_c$. (d) Flux rope radius $R_c$. In each panel the plot symbols are: MHD--black full squares; LFF--blue diamonds; GH--green triangles; CCS--orange $\times$'s.} }
    \label{fig:param}
\end{figure*}

\textcolor{red}{Figure~\ref{fig:param} shows the differences between the best-fit in-situ flux rope model parameters $\{\, \phi_0, \, \theta_0, \, p_0/R_c, \, R_c \, \}$ and the corresponding estimates of their MHD equivalent. The three in-situ models and the MHD values are plotted as different symbols: MHD--black full squares; LFF--blue diamonds; GH--green triangles; CCS--orange $\times$'s. The error bars are representative 1-$\sigma$ parameter uncertainties derived from the LFF model \citep[e.g., see][]{Lepping2003a,Lynch2005}. We have applied these to every model and the MHD values. The parameter uncertainties are taken to be $\sigma_\phi = 30^{\circ}$, $\sigma_\theta = 20^{\circ}$, $\sigma_{(p0/Rc)} = 0.30$, and $\sigma_{Rc} = 0.10\, R_c$. The MHD version of the flux rope cylinder parameters for each synthetic spacecraft's encounter with the simulation ejecta are also listed in Tables~\ref{tab:fits12} and \ref{tab:fits34}. }

\textcolor{red}{The cylinder axis orientation angles ($\phi_0$, $\theta_0$) are well approximated by every in-situ flux rope model for both the Type 1 classic bipolar profiles and the Type 2 unipolar profiles. While it looks like there is large disagreement in the $\phi_0$ values for the Type 2 events, recall that for highly-inclined flux ropes ($\theta_0 = \pm90^{\circ}$) the azimuthal angle becomes essentially degenerate. The Type 3 problematic orientation profiles show more agreement between the in-situ flux rope models and the MHD estimates for $\theta_0$ than $\phi_0$. While the overall scatter is the greatest for the Type 4 problematic impact parameter profiles, the flux rope elevation angles for the S8 and P4 profiles are the closest to the MHD values. }

\textcolor{red}{The (normalized) impact parameters ($p_0/R_c$) are reasonably consistent between the in-situ flux rope models and the MHD estimates for both the Type 1 and 2 events, although we note that the flux rope model's impact parameter uncertainty of 30\% is the largest of the parameters examined here. For the problematic Type 3 and 4 profiles, the impact parameters are statistically well matched for some events (S5, P8, S8) and not for others (S3, P6, P2, P4).  The flux rope cross-section radius, $R_c$, shows the greatest disagreement between the MHD estimates and the in-situ flux rope estimates. Five of the eight flux rope model fits for the classic Type 1 and Type 2 profiles systematically underestimate the MHD flux rope size (S1, S7, P5, S2, P3), while the other three agree reasonably well (P1, S6, P7). Similar to the impact parameter performance, the flux rope model $R_c$ estimates for the problematic Type 3 and 4 profiles show more variation and certain fits return obviously incorrect answers (e.g.\ GH for S3, P6, S4; CCS for S4; every model for P2, etc). }

\subsubsection{Hodogram Signatures of the Flux Rope Field Rotations}
\label{subsec:hodogram}

Another method of determining how much coherent rotation the magnetic field vector experiences during a particular event interval is to examine the hodograms of each of the different components, i.e. $B_R$--$B_T$, $B_R$--$B_N$, and $B_T$--$B_N$. Hodograms have been used since the first generation of interplanetary magnetic field measurements \citep{Klein1982,Berchem1982}, but they are now being used fairly regularly in the analysis of ICME flux rope ejecta \citep[e.g.][]{Bothmer1998,Nieves-Chinchilla2018a}. 
The general form of the hodograms for magnetic flux ropes are that two of the three component-pair plots show little-to-no coherent structure while the third shows a large, relatively smooth rotation. In the local flux rope frame coordinate, it is obvious that the axial and azimuthal fields will be the pair of components with the coherent rotation while neither will vary with the radial component (defined as zero in the mathematical formulations of \ref{sec:frmodels}).
This can also be understood as relating to the directions of minimum, intermediate, and maximum variance in the magnetic field over the ejecta interval \citep[specificially, the eigenvectors of the variance space matrix one obtains with Minimum Variance Analysis, e.g.\ see][and references therein]{RosaOliveira2020, RosaOliveira2021}.
Herein, we will consider only the $B_T$--$B_N$ hodograms as these are expected to be the directions of maximum and intermediate variance, respectively (or vice versa for the rotated, unipolar cases).

\begin{figure*}[!t]
    {\includegraphics[width=0.48\textwidth]{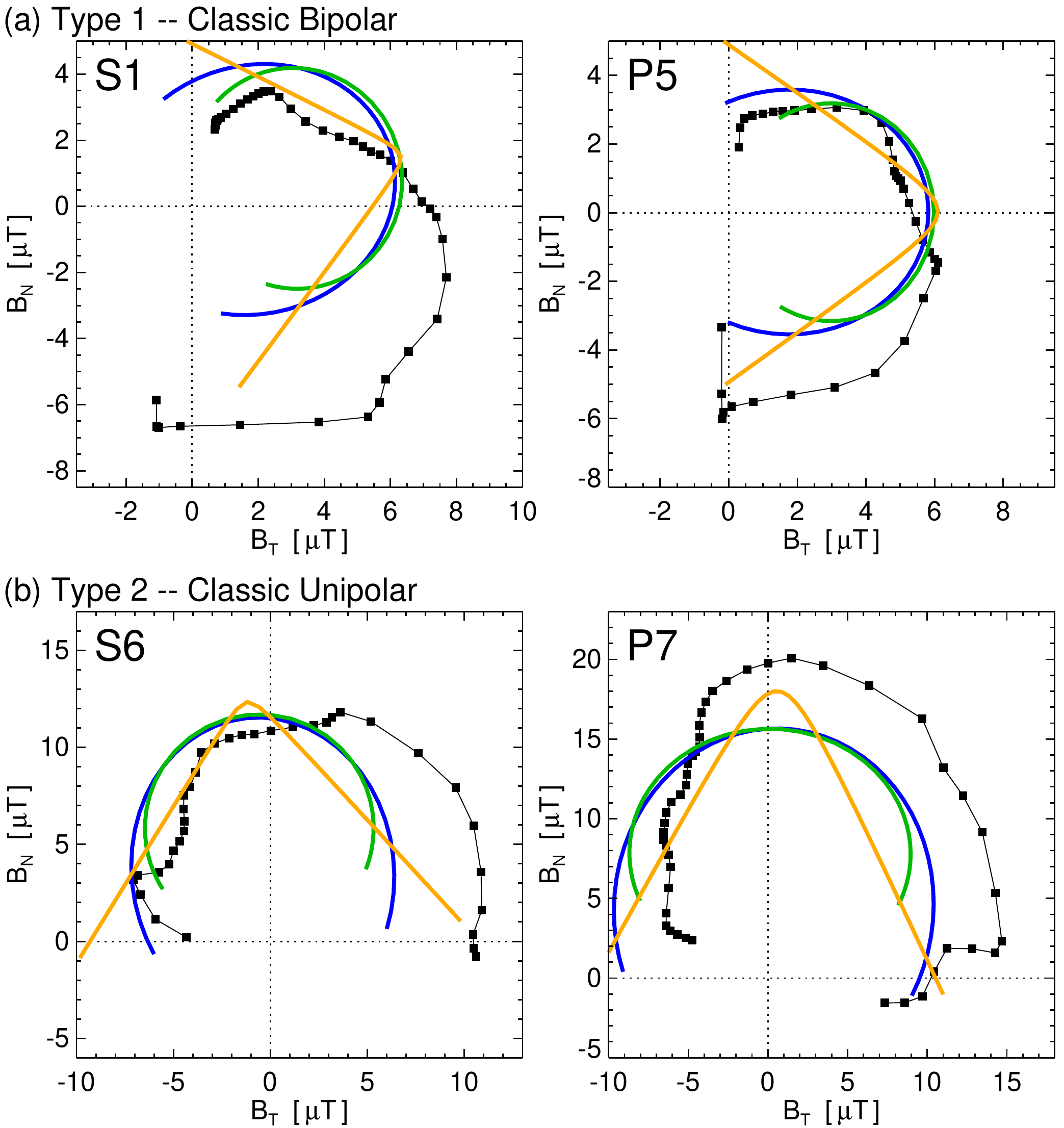}}
    \xspace\xspace
    {\includegraphics[width=0.48\textwidth]{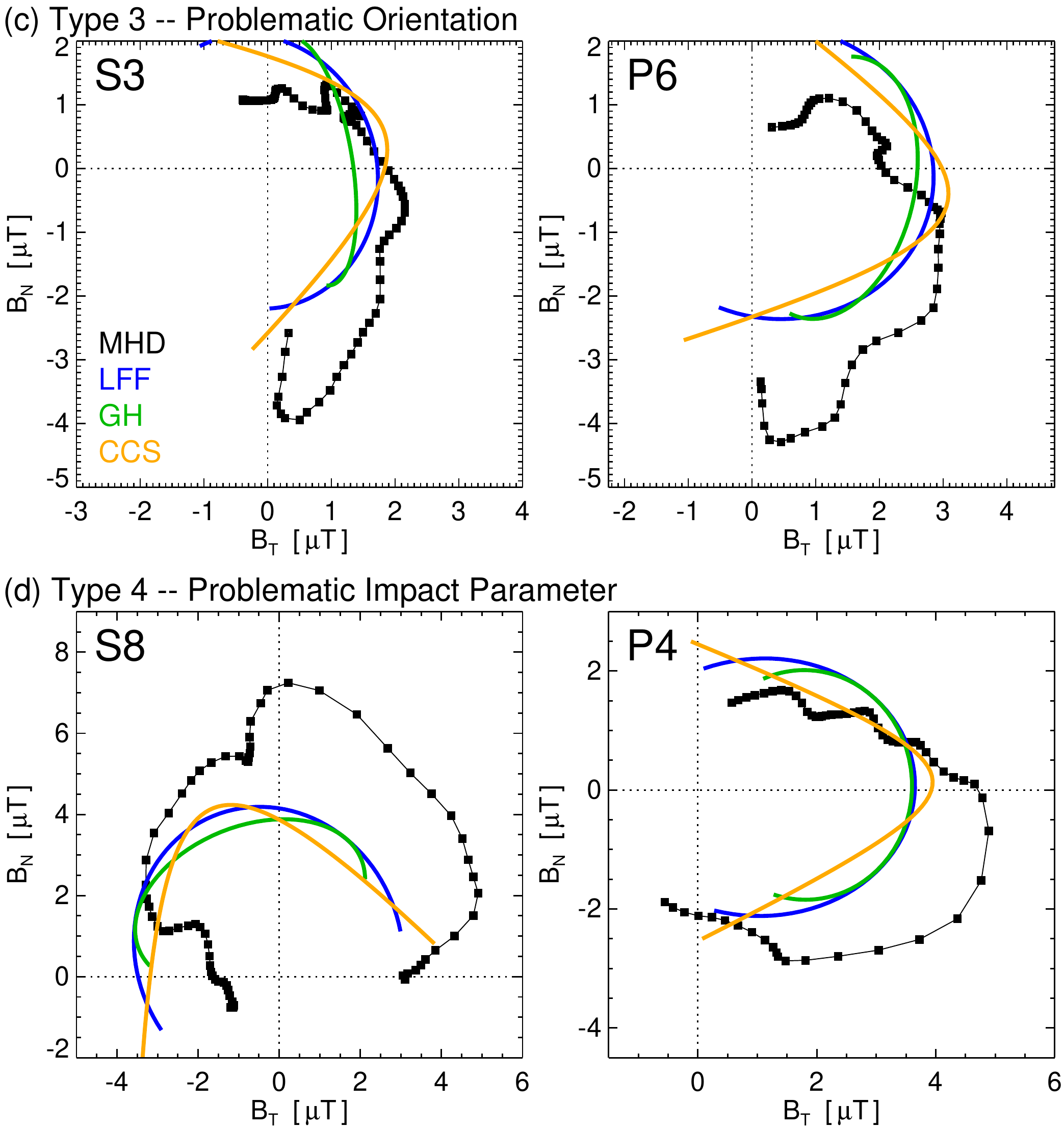}}
    \caption{Hodograms of the $B_T$--$B_N$ rotations during each observers' CME ejecta encounter organized by profile type. (a) Type~1: S1, P5; (b) Type~2: S6, P7; (c) Type~3: S3, P6; (d) Type~4: S8, P4. The rotations of the in-situ flux rope model fits are shown in the same color scheme as Figures~\ref{fig:fit0}, \ref{fig:fit1} (MHD--black, LFF--blue, GH--green, CCS--orange).
    The remaining synthetic observer hodograms can be found in Figure~\ref{fig:hodor2}.}
    \label{fig:hodor}
\end{figure*}

Figure~\ref{fig:hodor} shows the $B_T$--$B_N$ hodogram plots for each of the synthetic spacecraft samples of the 3D MHD flux rope ejecta shown in Figures~\ref{fig:fit0} and \ref{fig:fit1}, organized by profile type: \ref{fig:hodor}(a) Type 1 bipolar profiles (S1, P5), \ref{fig:hodor}(b) Type 2 unipolar profiles (S6, P7), \ref{fig:hodor}(c) Type 3 problematic orientation profiles (S3, P6), and \ref{fig:hodor}(d) Type 4 problematic impact parameter profiles (S8, P4). In each panel the MHD simulation data are shown with black squares and each of the in-situ flux rope models are shown in their respective color-scheme: LFF (blue), GH (green), and CCS (orange). The remaining eight observers' hodogram plots are shown in Figure~\ref{fig:hodor2}.
There are a number of interesting features in Figure~\ref{fig:hodor}: first, the hodogram representation of the field rotations in the MHD simulation data on their own; second, the hodogram representation of each of the in-situ flux rope models; and third, in the comparison of the two.

Considering the MHD simulation data, we see that the Type 1 and Type 2 curves make a more continuous circular arc than the Type 3 and Type 4 curves. In each panel the $B_T$:$B_N$ aspect ratio is 1:1 so that the relative shapes between panels can be more easily compared. The Type 3 and 4 hodogram curves tend to be less circular and have considerably more small-scale bumps, wiggles, and/or sharp discontinuities, i.e.\ the curves are less smooth. However, the Type 1 and 2 hodograms are not completely smooth either, as there is still some substructure along the large-scale arcs.

The hodograms for the in-situ flux rope models are generally smoother than the simulation data, as one might expect from analytic expressions. In every case, the LFF and GH models give continuous circular arcs, often with a significant amount of overlap, although the GH starting and ending points tend to not extend as far as the LFF ones. The CCS hodograms tend to make more of a V-shape, which makes sense given the linear relationship of both the azimuthal and axial field components with radius from the cylinder center ($B_\varphi \, , B_z \propto \rho$; see Equation~\ref{eq:ccs}) and these components map to $B_T$ and $B_N$ in the syntheric observer's RTN coordinates.

The $B_T$--$B_N$ hodogram plots are a complementary way of evaluating the quality of the in-situ flux rope model fits, as well as how close the observational (simulation) data themselves are to the form of an idealized flux rope structure. While there was some indication in the Figure~\ref{fig:fit0}, \ref{fig:fit1} time series that encounter Types 1 and 2 were ``better fit'' than Types 3 and 4, the hodogram vizualizations of the coherent field rotations show more difference in the sizes, shapes, and positions of the in-situ model fits with respect to the simulation data and more structural differences between the classic and problematic events. The same trends are also seen in the hodograms of Figure~\ref{fig:hodor2}.

\subsubsection{\textcolor{red}{CME Flux Rope Expansion Profiles}}
\label{subsec:exp}

\textcolor{red}{Given the somewhat unexpected disparity between the MHD and in-situ flux rope models' estimates for the flux rope sizes, $R_c$, we have examined the time evolution of each model's $R_c$ values. It may be the case that, despite a consistent offset between the MHD and flux rope model sizes, the estimates of the flux rope expansion rate(s) are similar. Our previous analyses have treated each synthetic spacecraft trajectory as an independent CME event encounter, but here we now treat each event encounter as a a measurement of the same CME, just at different times. }

\begin{figure}[!t]
    \centering
    \includegraphics[width=0.48\textwidth]{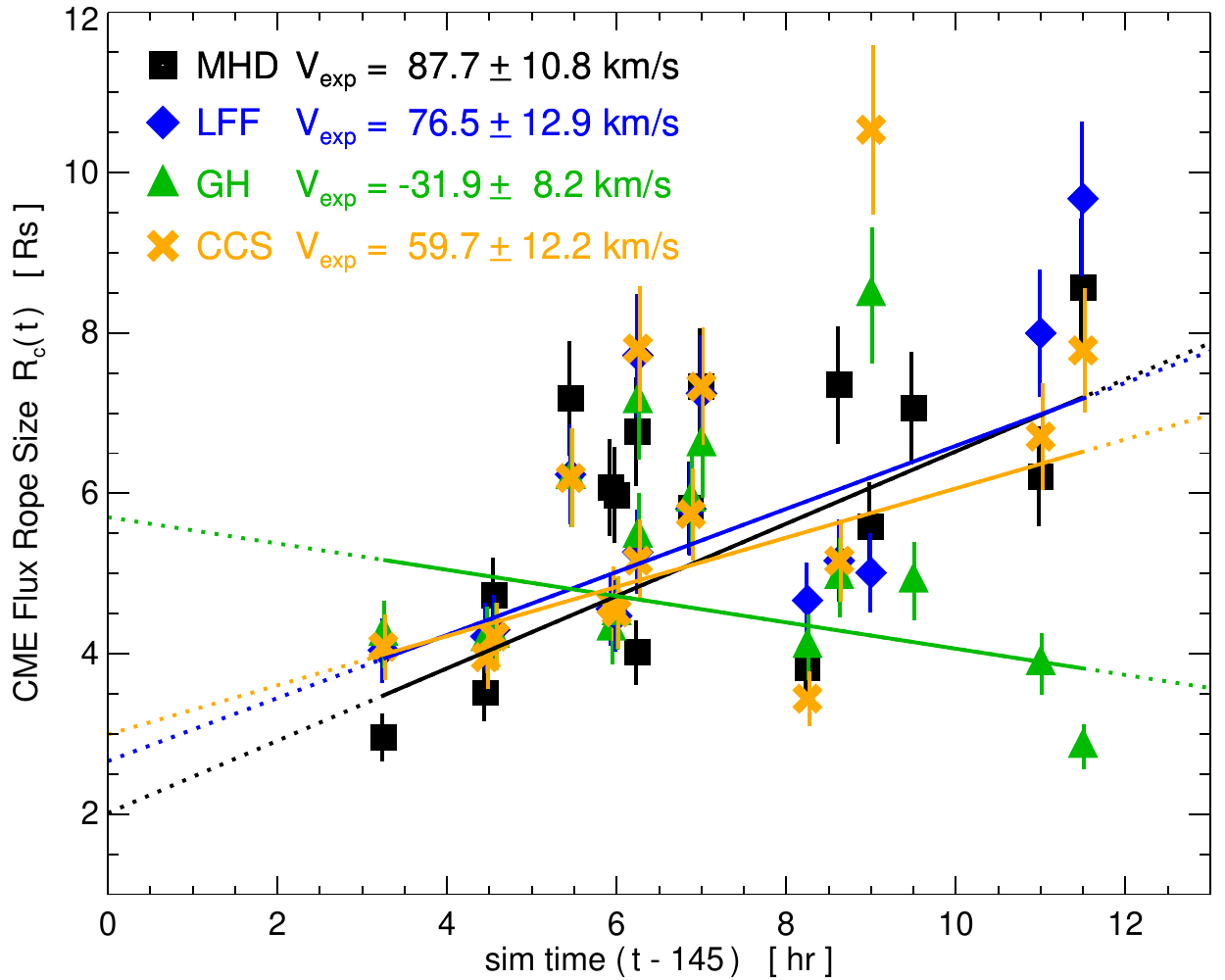}
    \caption{\textcolor{red}{Estimates of the CME flux rope expansion profiles $R_c(t)$ and their associated expansion velocities $V_{\rm exp}$. The points for each flux rope model and the MHD values follow our usual convention (MHD--black full squares; LFF--blue diamonds; GH--green triangles; CCS--orange $\times$'s) and the linear fit to each flux rope model and the MHD expansion profiles are shown as the thin solid lines in their respective colors.}}
    \label{fig:revol}
\end{figure}

\textcolor{red}{Figure~\ref{fig:revol} plots the $R_c$ values from each of our trajectories (from Figure~\ref{fig:param}(d)) with respect to their temporal midpoints, $t_m = (t_s + t_e)/2$, to yield a series of $R_c(t_m)$ samples. The points for each flux rope model and the MHD values follow our usual convention (MHD--black full squares; LFF--blue diamonds; GH--green triangles; CCS--orange $\times$'s). We use the same uncertainty estimate of $\sigma_{Rc} / R_c = 10\%$. The linear fit to each flux rope model and the MHD values are shown as the solid lines in the associated color. We define the expansion velocity as 
$V_{\rm exp} = \partial R_c / \partial t$ and convert the slope of each linear fit from $R_\odot$/hr to km/s. These values, along with their parameter uncertainties, are listed in the Figure~\ref{fig:revol} legend. }

\textcolor{red}{There are two points to highlight in these results. First, the MHD, LFF, and CCS models all yield remarkably similar $V_{\rm exp}$ values on the order of 60--90~km/s. The exception is the GH model $R_c(t)$ values, which give a linear fit with a negative slope---largely biased by the two problematic orientation points (S3, P6) at $(t-145) \gtrsim 11$~hr. The second point is that the mean $V_{\rm exp}$ consensus for the MHD, LFF, and CCS models is consistent with the observed expansion speeds of flux rope CMEs in the STEREO/COR2 coronagraph observations during solar minimum. The multi-viewpoint STEREO/COR2 CME catalog \citep{Vourlidas2017} shows the average CME expansion velocity is $V_{\rm exp} \sim 100$~km/s for the three years of solar minimum (2007--2009). This period is when the majority of streamer blowout CMEs have an easily identifiable flux rope morphology \citep{Vourlidas2018}. Given that our MHD simulation is essentially one giant streamer-blowout eruption, the agreement with the observed $V_{\rm exp}$ is encouraging. Additionally, our $V_{\rm exp}$ values are also entirely consistent with the in-situ velocity profiles within flux rope ICMEs measured throughout the inner heliosphere \citep[e.g.][and references therein]{Lugaz2020}. This also implies that the LFF and CCS flux rope models are ``good enough'' to be able to capture a significant portion of the MHD ejecta's expansion, despite the fact these are both \emph{static} models and there is apparently some systematic disagreement in the estimated CME flux rope sizes. }

\subsubsection{\textcolor{red}{Evaluating the Force-Free Assumption}}
\label{subsec:ffness}

\textcolor{red}{An important implication of flux rope expansion in the extended solar corona is that the magnetic structure must not be in equilibrium, i.e.\ it is not force-free. 
The force-free in-situ flux rope models (LFF, GH) have, by definition, $\boldsymbol{j} \parallel \boldsymbol{B}$ so the Lorentz force, $\boldsymbol{j} \times \boldsymbol{B}$, is identically zero. It is instructive, therefore, to investigate how appropriate this assumption is in our MHD simulation's time series, and evaluate just how non-force free the CCS model fits end up being. }

\textcolor{red}{A number of authors have examined this issue via different methodologies.
One way of quantifying the departure from a force-free state is to examine the ratio of perpendicular-to-parallel currents $|\, \boldsymbol{j}_{\perp}\, | / | \, \boldsymbol{j}_{\parallel} \, |$ \citep[e.g.][]{Moestl2009b} or perpendicular-to-total current $|\, \boldsymbol{j}_{\perp}\, | / | \, \boldsymbol{j}_{\rm total} \, |$ \citep[e.g.][]{Nieves-Chinchilla2016}. This current density decomposition, $\boldsymbol{j}_{\rm total} = \boldsymbol{j}_{\perp} + \boldsymbol{j}_{\parallel}$, is defined with respect to the magnetic field direction and is therefore obtained via
\begin{equation}
\boldsymbol{j}_{\perp} = \frac{\boldsymbol{j} \times \boldsymbol{B}}{| \, \boldsymbol{B} \, |} \; , \;\;\;\;\; \boldsymbol{j}_{\parallel} = \frac{\left( \boldsymbol{j} \cdot \boldsymbol{B} \right) \boldsymbol{B} }{| \, \boldsymbol{B} \, |^2} \; .
\end{equation}
Given that the force-free component of the current density ($\boldsymbol{j}_{\parallel}$) is parallel to $\boldsymbol{B}$, these current ratios are simply the equivalent of examining the misalignment angle between the $\boldsymbol{j}$ and $\boldsymbol{B}$ vectors, defined as $\sin {\vartheta} = | \, \boldsymbol{j} \times \boldsymbol{B} \, | / ( \, |\, \boldsymbol{j}\,|\,|\,\boldsymbol{B} \,| \, )$. }

\textcolor{red}{Herein, we use the \citet{Nieves-Chinchilla2016} ratio of the perpendicular-to-total current densities because $|\, \boldsymbol{j}_{\perp}\, | / | \, \boldsymbol{j}_{\rm total} \, | = \sin \vartheta$ whereas the \citet{Moestl2009b} ratio gives $|\, \boldsymbol{j}_{\perp}\, | / | \, \boldsymbol{j}_{\parallel} \, | = \tan \vartheta$. In practice, the force-free threshold chosen by \citet{Moestl2009b} of $\tan \vartheta < 0.30$ can be applied to the $|\, \boldsymbol{j}_{\perp}\, | / | \, \boldsymbol{j}_{\rm total} \, |$ ratio with almost no discernible difference, i.e., a maximum of $0.75^{\circ}$ in the resulting values of $\sin^{-1}(0.30) \approx \tan^{-1}(0.30) \approx 17^{\circ}$.}

\begin{figure}[!t]
    \centering
    \includegraphics[width=0.48\textwidth]{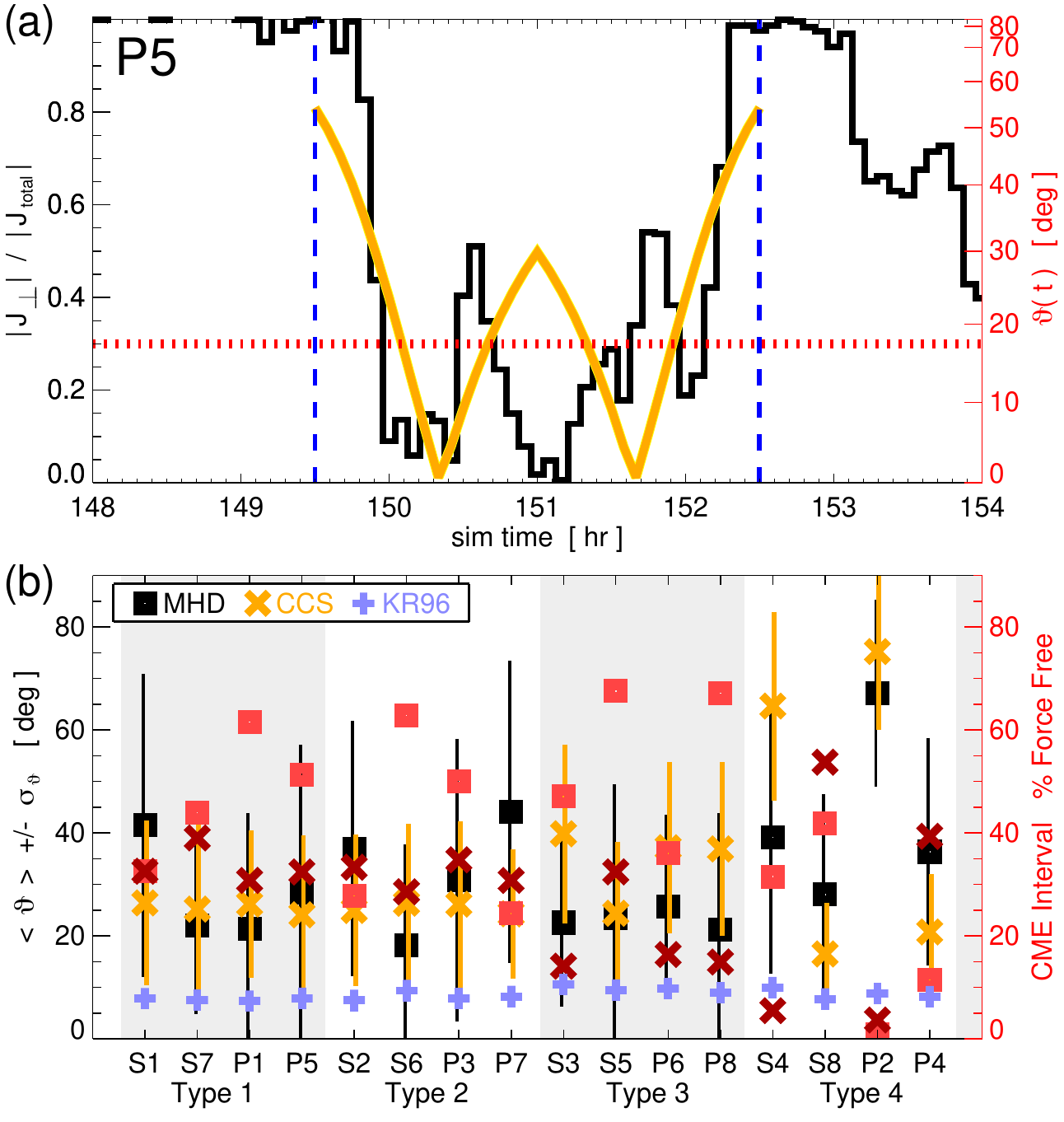}
    \caption{\textcolor{red}{Properties of the Lorentz force during the synthetic spacecraft time series. (a) Representative example of the ejecta profiles for the P5 observer. The $| \, \boldsymbol{j}_{\perp} \, | / | \, \boldsymbol{j}_{\rm total} \,|$ profile is shown in black and corresponding misalignment angle, $\vartheta(t)$, is shown on the right axis. The best-fit CCS model profile, $\sin \vartheta(t)^{\rm CCS}$, is plotted as the orange W-shaped curve. The magnetic flux rope boundaries are indicated with the blue vertical dashed lines. The red dotted line at $\sin \vartheta = 0.30$ ($\vartheta \approx 17.5^{\circ}$) indicates the \citet{Moestl2009b}, \citet{Nieves-Chinchilla2016} ``approximately force free'' threshold. (b) Comparison between the mean misalignment angles in each of the MHD profiles (black squares) and their CCS model fits (orange $\times$'s). The \citet{Kumar1996} misalignment angles are shown as purple~+ symbols. The red square and $\times$ symbols indicate the percentage of each flux rope interval that can be considered approximately force free for the MHD and CCS models, respectively.}}
    \label{fig:nff}
\end{figure}

\textcolor{red}{Figure~\ref{fig:nff}(a) shows a representative example of the $|\, \boldsymbol{j}_{\perp}\, | / | \, \boldsymbol{j}_{\rm total} \, |$ ratio (and its equivalent misalignment angle $\vartheta$) derived from the MHD simulation data as a function of time seen by the P5 synthetic observer. The orange W-shaped profile shows the misalignment angle of the non-force free CCS flux rope model, $\sin \vartheta(t)^{\, \rm CCS}$, determined by the best-fit model parameters (see \ref{sec:frmodels} and Equation~\ref{eq:angle}). The red dotted horizontal line denotes the misalignment angle threshold $\sin(0.30) = 17.45^{\circ}$ and the blue dashed vertical lines indicate the flux rope boundaries at start time $t_s$ and end time $t_e$.}

\textcolor{red}{The largest non-force free regions within the CME flux rope are clearly at the boundaries, i.e. at the CME--solar wind interface, but the misalignment angle is not uniformly distributed throughout the magnetic ejecta. Using the \citet{Moestl2009b}
and \citet{Nieves-Chinchilla2016} thresholding, we find that  51\% (19/37 data points) of the P5 ejecta interval identified in Figure~\ref{fig:nff}(a) can be considered approximately force-free, whereas the CCS flux rope model fit over the same interval yields 32\% (12/37 data points).}

\textcolor{red}{We have also calculated the mean and standard deviation of the misalignment angle from the MHD $\boldsymbol{j}(t) \times \boldsymbol{B}(t)$ time series directly over each of the flux rope ejecta intervals. Figure~\ref{fig:nff}(b) plots $\langle \, \vartheta \, \rangle \pm \sigma_\vartheta$ for each of the CME flux rope intervals in our set of synthetic time series. The MHD values are shown as the black squares and the analogous values derived from the CCS model profiles are shown as the orange $\times$'s. The large $\sigma_\vartheta$ values are a consequence of the misalignment angle structure at the flux rope boundaries (as seen in Figure~\ref{fig:nff}(a)).}

\textcolor{red}{A somewhat complementary approach has been employed by \citet{Subramanian2014} in their study of the self-similar expansion of CMEs observed in STEREO/SECCHI coronagraph data and how that relates to the overall Lorentz ``self-force''  as a driver of the CME eruption and/or propagation. 
\citet{Subramanian2014} used an expression for the $\boldsymbol{j} \times \boldsymbol{B}$ misalignment angle  
\begin{equation}
	\sin \vartheta = \kappa / x_{01} \approx \kappa / 2.405 \; ,
\end{equation}
to evaluate the range of values for $\vartheta$ inferred by the observed CME expansion. Here, $x_{01}$ is the first zero of the Bessel function $J_0(x)$ and $\kappa$ is the observed self-similar expansion parameter, defined as the ratio of the flux rope's minor radius to its major radius. 
This expression for the misalignment angle was derived by \citet{Kumar1996} under the assumption that the flux rope's magnetic structure has only a small departure from the LFF Bessel function solution.}

\textcolor{red}{We have performed the same calculation, defining the self-similar parameter for each of our MHD CME samples as $\kappa = R_c / r_m$, using the $R_c$ values from Figure~\ref{fig:revol} and the midpoint distances $r_m$ (from Section~\ref{subsec:geom}). We obtain an average self-similar parameter of $\langle \kappa \rangle = 0.357 \pm 0.042$, which is in excellent agreement with the \citet{Subramanian2014} observational values (e.g.\ see their Table 1). This self-similar parameter yields an average misalignment angle of $\langle \,\vartheta \, \rangle^{\rm KR96} = 8.5^{\circ} \pm 1.0^{\circ}$ over our set of 16 encounters and each of the individual $\vartheta^{\rm KR96}$ values are also shown in Figure~\ref{fig:nff}(b) as the purple + symbols.}

\textcolor{red}{For most of the synthetic encounters, the MHD and CCS values are consistent within the statistical spread, however it is worth highlighting that both of these are always greater than the corresponding \citet{Kumar1996} values. Similar to our previous findings, the CCS flux rope model results are more consistent---with each other and with the MHD values---over the subset of Type 1 and 2 events. The other important takeaway from Figure~\ref{fig:nff}(b) is that the percentage of each MHD flux rope interval that can be considered force-free is greater than the corresponding percentage derived from the CCS model fits. In other words, for 12 of our 16 profiles, the MHD intervals have more points below the ``force-free threshold'' than the CCS fits to the same intervals. }

\textcolor{red}{There is some ambiguity in exactly what range of misalignment angles can be considered ``nearly force-free.'' \citet{Subramanian2014} obtained a similar range of $\langle \, \vartheta \, \rangle^{\rm KR96}$ angles (5$^{\circ}$--10$^{\circ}$) and suggested these values were sufficiently large to demonstrate the ``nearly force-free'' assumption of \citet{Kumar1996} was inconsistent with coronagraph observations---despite being comfortably below the \citet{Moestl2009b} and \citet{Nieves-Chinchilla2016} threshold (${\sim}17^{\circ}$) used in the in-situ flux rope analyses. Overall, we conclude that the MHD flux rope profiles obtained by our synthetic observers, while obviously not force-free, are in some sense, \emph{more} force free than the in-situ CCS flux rope model fits to the same MHD profiles and \emph{less} force free than the \citet{Kumar1996} assumption of a small deviation from the LFF model structure.}

\subsubsection{CME Flux Content}
\label{subsec:fluxes}

There is a direct quantitative relationship between the reconnection flux in the corona and the magnetic flux swept by the flare ribbons \citep{Forbes1983, Qiu2007, Kazachenko2017}. 
Magnetic reconnection rapidly adds flux to the erupting magnetic structure(s) during the eruption \citep{Welsch2018} resulting in the formation of coherent, twisted flux rope structures in agreement with remote-sensing and in-situ CME observations \citep[e.g.][]{Bothmer1998, Vourlidas2013, Hu2014}.
In addition to the relationship between eruptive flare reconnection and CME acceleration \citep{Jing2005, Qiu2010}, properties of the magnetic field---handedness, orientation, and magnetic flux content---are important quantities for making the CME--ICME connection \citep[e.g.][]{Qiu2007, Demoulin2008, Hu2014, Palmerio2017, Gopalswamy2017, Gopalswamy2018}.

\begin{figure*}[!t]
    \centering
    \includegraphics[width=1.0\textwidth]{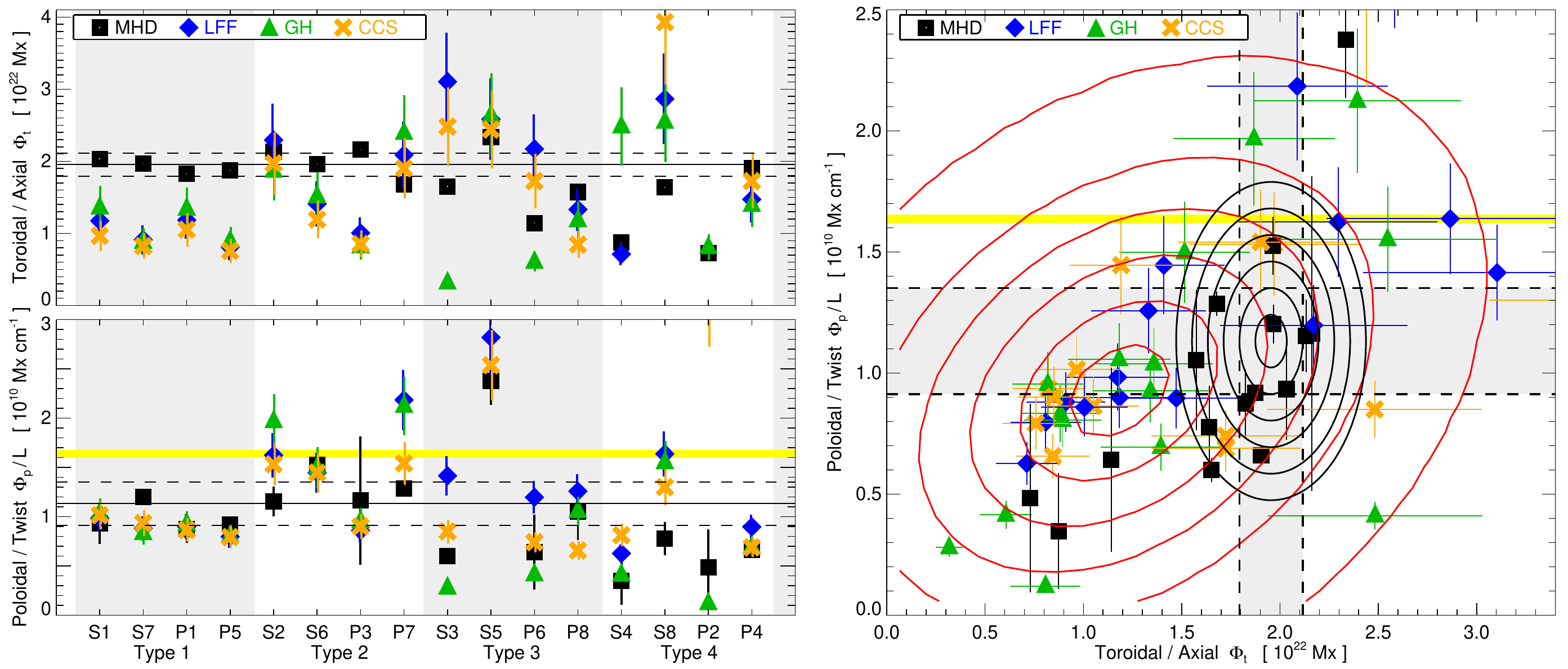}
    \caption{Left panels: Estimates of toroidal/axial flux and poloidal/twist flux from the CME cross-sections in the MHD simulation results and for each observers' in-situ model fit. The color scheme is the same as previous figures: MHD--black full squares; LFF--blue diamonds; GH--green triangles; CCS--orange $\times$'s. The thin black lines are the mean MHD value from the Type 1 and Type 2 events and the dashed lines show plus and minus the standard deviation. The thick yellow line indicates the \citet{Lynch2019} estimate of the reconnection flux. Right panel: Distribution of $(\, \Phi_t, \Phi_p/L \, )$ points showing overlap between a 2D Gaussian model fit to the  set of all in-situ flux rope model fits (red contours) and MHD Type 1 and 2 values (black contours). The contours are at 0.5-$\sigma$ intervals between 0.5$\sigma$ and 2.5$\sigma$. See text for details.}
    \label{fig:flux}
\end{figure*}

The toroidal/axial flux $\Phi_t$ and poloidal/twist flux $\Phi_p$ of the in-situ flux rope models are given by

\begin{equation}
\Phi_t = \int \boldsymbol{B} \cdot \boldsymbol{\hat{z}} \, \rho \, d\rho \, d\varphi = \int_{0}^{2\pi} d\varphi \int_{0}^{R_c(\varphi)} \rho \, d\rho \, B_a(\rho) \; ,
\end{equation}
\begin{equation} 
\Phi_p = \int \boldsymbol{B} \cdot \boldsymbol{\hat{\varphi}} \, d\rho \, dz = \int_{0}^{L} dz \int_0^{R_c} \, d\rho \, B_\varphi(\rho) \;\; ,
\end{equation}
in cylindrical flux rope coordinates $( \rho, \, \varphi, \, z )$. In these models with circular cross-sections, $R_c(\varphi) = R_c$ and the area is simply $\pi R_c^2$. The poloidal/twist flux calculation involves integrating $dz$ along the flux rope axis over the entire length of the ICME. Since the axial length $L$ of interplanetary flux ropes are generally unknown, it is common to either estimate their value as 2--2.5 times the radial distance of the observation \citep[e.g.][]{Leamon2004,Pal2021} or just calculate the poloidal/twist flux per unit length, $\Phi_p/L$.

Figure~\ref{fig:flux} plots $\Phi_t$ and $\Phi_p/L$ calculated for each in-situ flux rope model. In terms of the model parameters (or quantities derived from the model fit parameters), the toroidal/axial fluxes are calculated as:

\begin{equation}
    \Phi_t^{\mathrm{LFF}} = \left( 2 \, J_1(x_{01})/x_{01} \right) B_{0} \, \pi R_{c}^2 \;,
\end{equation}
\begin{equation}
    \Phi_t^{\mathrm{GH}} = \ln{\left[1 + \tau^2R_c^2\right]} \, B_{0} \, \pi \left(1/\tau\right)^{2} \;,
\end{equation}
\begin{equation}
    \Phi_t^{\mathrm{CCS}} =  \left( 1/3 \right) B_0 \, \pi R_c^2 \;.
\end{equation}
We refer the reader to \ref{sec:frmodels} for the precise definitions of each of the flux rope model parameters but note here the general form of $\Phi_t \sim$ constant~$\times \, (B \, \pi R^2)$. 
The poloidal/twist fluxes (per unit length) are calculated as:

\begin{equation}
    \left({\Phi_p}/{L}\right)^{\mathrm{LFF}} = \left( 1/x_{01} \right) B_{0} \, R_{c} \; ,
\end{equation}
\begin{equation}
    \left({\Phi_p}/{L}\right)^{\mathrm{GH}} = \ln{\left[1 + \tau^2R_c^2\right]} \, B_{0}  \left(1/2\tau\right) \; ,
\end{equation}
\begin{equation}
    \left({\Phi_p}/{L}\right)^{\mathrm{CCS}} =  (1/4) \left( \mu_0 \,  j_{z}^{\, 0} R_c \right) \, R_c \; .
\end{equation}
Here we note the general form of $\Phi_p / L \sim$ constant~$\times \, B \, R$.
The uncertainties for the in-situ flux rope model magnetic flux quantities are obtained via the usual fractional error addition: $\sigma_{t}/\Phi_t \approx 22\%$ and $\sigma_{p}/(\Phi_p/L) \approx 14\%$.

To evaluate the in-situ flux rope model fluxes, we calculate a $\Phi_t$ and $\Phi_p/L$ from the MHD simulation data cube at each time indicated in Figures~\ref{fig:fit0}, \ref{fig:fit1}, \ref{fig:fit2}, and \ref{fig:fit3}. The toroidal/axial flux is determined by integrating $B_\phi \, dA$ for $B_\phi \ge 0$ over a spherical wedge in the $r$--$\theta$ plane encompassing the flux rope cross-section. The poloidal/twist flux (per unit length) is obtained by integrating $|B_\theta| \, dr$ in our spherical wedge for the whole range of $\theta$ values and averaging the positive ($B_\theta > 0$) and negative ($B_\theta < 0$) values. The MHD flux estimates are shown as black squares in Figure~\ref{fig:flux}. The mean $\Phi_t$ and $\Phi_p/L$ values for the Type~1 and Type~2 events are $\langle \Phi_t^{\rm MHD} \rangle = 1.95 \pm 0.16 \times 10^{22}$~Mx and $\langle (\Phi_p/L)^{\rm MHD} \rangle = 1.13 \pm 0.22 \times 10^{10}$~Mx~cm$^{-1}$. These averages are shown in the corresponding left Figure~\ref{fig:flux} panels as the solid horizontal lines with $\pm1\sigma$ shown as the dashed lines. The yellow horizontal line shows the \citet{Lynch2019} reconnection flux estimate $\Phi_{\mathrm{rxn}}/L = 1.64 \times 10^{10}$~Mx~cm$^{-1}$ from the flare ribbon area at $t=152$~hr and approximating the flux rope axis length as $L=2 \pi R_{\mathrm{CME}}$ with $R_{\mathrm{CME}}=25R_\odot$.

We note that the Type~3 and 4 MHD flux values that deviate the most from the Type~1 and 2 average values in one or both quantities, i.e. S3, P6, S4, and P2, have their estimates at the largest simulation times (155.67, 154.50, 155.67, and 153.0 hr, respectively). As seen in Figures~\ref{fig:fit1} and \ref{fig:fit3}, for each of these synthetic observers, the leading edge or a nearby section of the magnetic ejecta has passed through the $r=30R_{\odot}$ outer boundary, so it seems reasonable to expect both the toroidal/axial and poloidal/twist fluxes to be a bit lower in these cases.

The right panel of Figure~\ref{fig:flux} shows the same data (with the same colors and plot symbols) as the left panels, but now as a two-dimensional distribution of $\left( \Phi_t, \Phi_p/L \right)$. The $\langle \Phi_t^{\rm MHD} \rangle$ and $\langle (\Phi_p/L)^{\rm MHD} \rangle$ values are the vertical and horizontal black lines. In order to assess the overlap of the in-situ flux rope model flux values with the corresponding MHD flux estimates, we fit a 2D Gaussian to the smoothed, 2D histogram of the in-situ flux rope model points using the standard IDL function {\tt gauss2Dfit.pro}. The red contours show the resulting Gaussian distribution at 0.5-$\sigma$ intervals between 0.5 and 3.0 standard deviations. The 1-$\sigma$ width of the distribution function in the $\Phi_t$ direction is $\sigma_t = 0.525 \times 10^{21}$~Mx whereas in the $\Phi_p/L$ direction it is $\sigma_p = 0.325 \times 10^{10}$~Mx~cm$^{-1}$. The tilt of the contour ellipses show that the variables $\left( \Phi_t, \Phi_p/L \right)$ are not independent (as expected given their dependence on the same model parameters). We have also manually constructed a similar 2D Gaussian from the classic-profile MHD flux averages with their respective standard deviations given above. The MHD-average Gaussian distribution is shown as the black elliptical contours (at 0.5-$\sigma$ intervals) centered on $\left( \langle \Phi_t^{\rm MHD} \rangle, \langle (\Phi_p/L)^{\rm MHD} \rangle \right)$.

The overlap between the in-situ model fit and MHD-average distributions presented in Figure~\ref{fig:flux} provides another way to visualize the flux rope model performance. The MHD distribution contours (black) fall almost entirely within the 1$\sigma$ and 2$\sigma$ contours (red) of the in-situ flux rope model distribution. The peak of the in-situ flux rope model distribution lies below the peak of the MHD distribution in both toroidal and poloidal flux coordinates (also seen in the clustering of the individual model fit points in the lower-left quadrant rather than a more uniform distribution across each quadrant). This suggests that the in-situ flux rope models are, at least statistically, underestimating both components of the CME flux content: the toroidal/axial flux by $\sim$60\%, i.e., $\langle \Phi_t^{\rm MHD}  \rangle \approx 1.63 \, \langle \Phi_t^{\rm FR}  \rangle$, and the poloidal/twist flux by $\sim$25\%, i.e., $\langle (\Phi_p / L)^{\rm MHD} \rangle \approx 1.26 \, \langle (\Phi_p / L)^{\rm FR} \rangle$. However, the MHD flux estimates for the problematic Type 3 and Type 4 events are also lower than the classic-profile averages, in roughly the same direction and proportion as the in-situ flux rope model points.


\section{Discussion and Conclusions} \label{sec:conclusions}

In summary, we conclude from this numerical experiment that the in-situ flux rope models can be used to infer the large-scale orientation and estimate some of the physical properties such as size and flux content. Since our MHD CME is a complex, 3D structure, we do expect some level of variation in profiles from different observers at different locations, and we certainly see that. There is also at least as much---and probably more---variation introduced by the limitations, assumptions, and simplifications of the flux rope models. In other words, the differences in quality of the fits with a single model between synthetic observers and their event types is comparable to the differences between different model fits to the same event. The optimistic interpretation is, therefore, that each of the flux rope models we have examined does a reasonable job, most of the time. The pessimistic interpretation, however, is that every flux rope model fit to the MHD profiles does an equally poor job of determining the flux rope (axis) orientation and CME flux content. Regardless of personal worldview on the merits of in-situ flux rope models, our results do contribute to a number of ongoing research questions that we will discuss below.

There are open questions about the specific details of \emph{when} and \emph{how} magnetic cloud/flux-rope CMEs become force-free or sufficiently relaxed enough that there are no longer any drastic changes to the internal magnetic structure of the CME. \citet{Lynch2004} found that in an axisymmetric (2.5D) MHD simulation, the CME flux rope structure at ${\sim}15\,R_\odot$ was reasonably well-described by the LFF model. This provides an important constraint on the internal dissipation of ``excess'' magnetic energy via restructuring or relaxation of the magnetic fields within the flux rope CME. Several researchers have examined the possibility that this dissipation process acts as a source of localized plasma heating within the CME \citep[e.g.][]{Kumar1996, Rakowski2007, LeeJY2009, Landi2010}. The results presented here show that reasonable fits are obtained for the magnetic structure of the 3D MHD flux rope ejecta between 10--30\,$R_\odot$ using both force-free (LFF, GH) and non-force free (CCS) in-situ flux rope models. \textcolor{red}{Having examined the internal structure of the $\boldsymbol{j} \times \boldsymbol{B}$ misalignment angle $\vartheta$, we conclude that our MHD ejecta intervals, while not force-free, appear to be \emph{more} force-free than the best-fit CCS flux rope solutions (at least statistically) and \emph{less} force free than the \citet{Kumar1996} assumption that the magnetic field structure is only a small departure from the LFF cylinder model. Given the typical values we have obtained for the percentage of the MHD flux rope intervals that fall below the ``approximately force free'' threshold of $\vartheta \approx 17^{\circ}$ \citep{Moestl2009b, Nieves-Chinchilla2016}, one may characterize the MHD ejecta as having dissipated somewhere between $\sim$30--60\% of the excess magnetic energy responsible for the non-zero Lorentz force. In principle, this reinforces the \citet{Lynch2004} suggestion that a significant amount of the excess magnetic energy \emph{within} the CME flux rope can be dissipated by $\sim$15\,$R_\odot$.}

Another open question is regarding the impact of a moving observer on the inferred structure of the magnetic flux rope ejecta. The spatial scales of ICME flux ropes at 1~AU are large enough ($\sim$0.1~AU) that treating the spacecraft position as fixed with respect to the ICME ejecta passage is a valid approximation. However, since PSP is moving significantly faster than previous spacecraft and could be so close to the Sun that the CME spatial scales are on the order of a few $R_\odot$, it is unclear whether this will significantly alter the observed magnetic structure. In our simulation data, the synthetic observers on PSP-like trajectories through various portions of the MHD CME flux rope see essentially the same profiles as the stationary observers, despite their ${\gtrsim}100$~km/s speed. \textcolor{red}{The longitudinal change in our example P5 observer of Section~\ref{sec:orbits:psp} was $\Delta \phi_{\rm obs} \sim 5^{\circ}$ which is much smaller than, e.g., the mean observed angular width of 45$^{\circ}$--60$^{\circ}$ in the LASCO CME catalog \citep{Yashiro2004} or the mean angular width of $\sim$70$^{\circ}$ for streamer blowout CMEs \citep{Vourlidas2018}.  Therefore, a PSP-like trajectory that intersects the ``nose'' (or apex) of the CME should not see a significant difference compared to a similarly situated stationary observer ($\Delta \phi_{\rm obs} = 0^{\circ}$), especially if the flux rope can be considered translationally-invariant along its axis.}

\textcolor{red}{However, there are multi-spacecraft observations \citep[e.g.][]{Kilpua2011,Lugaz2018} that show longitudinal differences between spacecraft of only 1$^{\circ}$--10$^{\circ}$ can and do result in significant and substantial differences in the plasma and magnetic field measurements. This may reflect a limitation of our simulation configuration; the specific geometry of our global, 360$^{\circ}$-wide ejecta in the numerical simulation means almost every one of our synthetic spacecraft encounters can be considered a ``nose''/apex encounter (with the exception of the Type 3 events that represent a more skewed intersection). 
For real CME events with typical angular widths, we might expect to observe at least some longitudinal variation \citep[][]{Kilpua2009a,Mulligan2013} and the occasional flank- or leg-encounter where a few degrees change in the spacecraft longitude over the event duration would cause the spacecraft to leave the coherent flux rope portion of the ejecta or, e.g., even potential dual encounters through different portions of the ejecta \citep[][]{Moestl2020}.}

A somewhat related issue is that of the impacts or observable consequences of CME flux ropes ``aging'' \citep{Demoulin2020}. The proximity of PSP measurements to the Sun could mean that any evolution of the magnetic structure of the CME (e.g. expansion, rotation, distortion, etc) would appear much more dramatic than is typically observed at greater heliocentric distances. Specifically, the duration of the PSP-like spacecraft encounters (${\sim}$3~hours) is a significant portion of the CME's lifetime (i.e., the CME eruption starts at $t\approx145$~hours, so roughly 5--10 hours by Figures~\ref{fig:fit0}, \ref{fig:fit1}, \ref{fig:fit2}, and \ref{fig:fit3}). However, despite our simulated CME being relatively ``young,'' there does not appear to be any identifiable evolutionary aspects of the synthetic time series that sufficiently distort or complicate the flux rope structure beyond the capabilities of the in-situ cylindrical flux rope models to approximate its large-scale coherent field rotation.

In our synthetic time series, the large-scale coherent field rotations are largely confined to the $B_T$--$B_N$ plane. Since magnetic field hodograms are increasingly being used in the analysis of in-situ ICME observations \citep[e.g.][]{Nieves-Chinchilla2019,Scolini2022,DaviesEE2021b}, including recent PSP events \citep[e.g.][]{Nieves-Chinchilla2020, Winslow2021}, we constructed hodogram visualizations of both the MHD simulation data and the in-situ flux rope model fits to further evaluate the quality of the flux rope model reconstructions. We find that reasonably good fits to the $\boldsymbol{B}_{\rm RTN}(t)$ component time series do not necessarily mean an equally good fit in the $B_T$--$B_N$ hodogram representation. In general, the classic Type 1 and 2 profiles are better fit than the problematic Type 3 and 4 profiles (cf. Figures~\ref{fig:hodor}, \ref{fig:hodor2}), but not universally, and there is not an individual model with a clearly superior performance for each event.

Not surprisingly, the in-situ flux rope model estimates of the toroidal/axial and poloidal/twist fluxes in the CME ejecta also tend to be a bit better for the classic profiles than the problematic profiles. Although, again, not universally and not even consistently within the same event. For example, in the S2 classic unipolar (Type 2) event, every in-situ model had an axial flux estimate $\Phi_t$ that was close to the MHD value, whereas for the twist flux density $(\Phi_p/L)$, every flux rope model overestimated the twist flux by $\gtrsim$50\% compared to the MHD value. On the other hand, all the flux rope model underestimated the axial fluxes of the Type 1 events while doing relatively well with the poloidal flux for those cases. The 2D distribution of the MHD values and in-situ flux rope model fitting results in toroidal--poloidal flux space (right panel of Figure~\ref{fig:flux}) summarizes the overall performance of our entire set of model fits in determining the CME flux content. At least on average, the flux rope models tend to underestimate the poloidal/twist flux component by $\sim$25\% and the toroidal/axial flux component by $\sim$60\%, but in general, the distribution of the MHD flux estimates for the classic bipolar and unipolar orientations are not inconsistent with the much broader distribution of in-situ model fit values, i.e., almost every MHD data point lies within the 2$\sigma$ contour of the distribution of in-situ model fitting results.

\textcolor{red}{We expect the results of this study to be generally applicable to previous and future PSP CME encounters. However, there are some limitations in our approach, largely resulting from the idealized, global-scale nature of the CME ejecta in our MHD simulation. Here we have treated our set of synthetic spacecraft trajectories over a broad longitudinal distribution as independent CME events, but it will be important to perform similar numerical experiments on simulations where the CME flux rope structures are more realistic, i.e., have smaller angular widths and/or originate from active regions. Likewise, it will be important to perform analyses where MHD ejecta are sampled with synthetic spacecraft trajectories that correspond to a range of multi-spacecraft observational geometries with varying radial and longitudinal separation.} In conclusion, while we certainly acknowledge that there is room for improvement on the in-situ flux rope modeling front, it is encouraging that the idealized structures of relatively simple magnetic flux rope models are apparently as good of an approximation of the internal magnetic configuration of our simulated MHD ejecta during local encounters at $\lesssim$30\,$R_\odot$ as they are to in-situ observations of ICME flux ropes further out in the inner heliosphere---with all of the usual caveats intact. We are looking forward to testing these modeling results with future PSP measurements of CMEs in the extended corona.

\section{Acknowledgments}

B.J.L. acknowledges support from NSF AGS-1622495 and NASA LWS 80NSSC19K0088, as well as helpful discussion within the \emph{HEliospheRic Magnetic Energy Storage and conversion} (HERMES) DRIVE Science Center.
N.A. and W.Y. acknowledge support from NSF AGS-1954983 and NASA ECIP 80NSSC21K0463.
N.A. and E.P. acknowledge support from NASA PSP-GI 80NSSC22K0349, N.A. also acknowledges support from NSF AGS-2027322, and E.P. also acknowledges support from NASA HTMS 80NSSC20K1274.
N.L. acknowledges support from NASA HSR 80NSSC19K0831 and HGI 80NSSC20K0700.

%
%

\appendix
\counterwithin{figure}{section}
\counterwithin{table}{section}

\section{In-situ Flux Rope Models}
\renewcommand{\thesection}{A}
\label{sec:frmodels}


We examine the performance of three representative flux rope models used to characterize the in-situ magnetic field profiles of magnetic cloud/flux rope ICMEs (as in \citealt{Palmerio2021c}, but see also, e.g., \citealt{Dasso2006}, \citealt{Al-Haddad2013}). For each of the following flux rope models, the 3D spatial orientation of the flux rope requires three free parameters. These are typically written as $\phi_0$, $\theta_0$, and $p_0$ where the symmetry axis of the flux rope (given by unit vector $\boldsymbol{\hat{z}}$) makes an angle $\phi_0$ within the $R$--$T$ plane, makes an angle $\theta_0$ out of the $R$--$T$ plane, and the relative distance between the spacecraft trajectory and the flux rope axis is represented by $p_0$, usually in units of flux rope radius $R_c$. 

\begin{table*}[t]
 \centering
 \begin{threeparttable}[T]
 \caption{In-situ flux rope model fit parameters and MHD estimates for Type 1 (classic bipolar) and Type 2 (classic unipolar) encounters.}
 \label{tab:fits12}
 \begin{tabular}{|c|c|c|rrrrrc|rc|c|}
\hline
Type/Fig. & Obs. & Model & $\phi_0$ [$^\circ$] & $\theta_0$ [$^\circ$] & $p_0/R_{c}$ & $R_{c}$ [AU] & $H$ & $B_0$ [$\mu$T] & $\Phi_t$ [Mx] & $\Phi_p/L$ [Mx/cm] & $\chi^2$ \\
\hline
\hline
& & LFF & 74.6 & 14.5 & 0.001 & 0.0245 & $+1$ & 6.46 & 1.188e+22 & 0.989e+10 & 0.228  \\
Type 1 & {S1} & GH  & 85.1 & 15.2 & 0.101 & 0.0254 & $+1$ & 6.81 & 1.358e+22 & 1.039e+10 &  0.220  \\
Fig.~\ref{fig:fit0}& & CCS & 70.5 & 12.7 & -0.058 & 0.0240 & $+1$ & 7.16 & 0.967e+22 & 1.015e+10 &  0.303   \\
 & & {MHD} & {80.2} & {-9.5} & {0.187} & {0.0315} & ${+1}$ & {5.95} & {2.033e+22} & {0.934e+10} & {---} \\
\hline
 &  & LFF & 74.9 & 1.1 & 0.043 & 0.0212 & $+1$ & 6.69 & 0.911e+22 & 0.881e+10 &  0.056   \\
Type 1 & {S7} & GH  & 65.6 & 1.0 & -0.055 & 0.0200 & $+1$ & 6.94 & 0.882e+22 & 0.833e+10 &  0.117   \\
Fig.~\ref{fig:fit2} & & CCS & 77.5 & 0.8 & 0.088 & 0.0215 & $+1$ & 7.59 & 0.822e+22 & 0.935e+10 &  0.088  \\
 & & {MHD} & {83.6} & {-1.1} & {0.133} & {0.0283} & ${+1}$ & {7.82} & {1.967e+22} & {1.202e+10} & {---} \\
\hline
 &  & LFF & 76.5 & 6.5 & -0.031 & 0.0270 & $+1$ & 5.36 & 1.197e+22 & 0.904e+10 &  0.118   \\
Type 1 & {P1} & GH  & 82.3 & 7.0 & 0.032 & 0.0275 & $+1$ & 5.61 & 1.341e+22 & 0.926e+10 &  0.139   \\
Fig.~\ref{fig:fit2} & & CCS & 73.5 & 5.7 & -0.090 & 0.0267 & $+1$ & 6.29 & 1.051e+22 & 0.864e+10 &  0.188   \\
 & & {MHD} & {74.9} & {-8.8} & {0.119} & {0.0271} & ${+1}$ & {6.78} & {1.825e+22} & {0.875e+10} & {---} \\
\hline
& & LFF & 77.8 & 0.7 & 0.091 & 0.0208 & $+1$ & 6.17 & 0.818e+22 & 0.802e+10 & 0.089  \\
Type 1 & {P5} & GH  & 79.6 & 0.3 & 0.091 & 0.0209 & $+1$ & 6.43 & 0.895e+22 & 0.806e+10 & 0.142   \\
Fig.~\ref{fig:fit0} & & CCS & 79.1 & 0.4 & 0.133 & 0.0210 & $+1$ & 7.33 & 0.758e+22 & 0.793e+10 & 0.181 \\
 & & {MHD} & {88.4} & {-5.0} & {0.216} & {0.0278} & ${+1}$ & {5.82} & {1.875e+22} & {0.918e+10} & {---} \\
\hline
\hline
 & & LFF & 270.8 & 73.4 & 0.138 & 0.0290 & $+1$ & 9.01 & 2.295e+22 & 1.623e+10 & 0.428  \\
Type 2 & {S2} & {GH}  & {265.6} & {70.5} & {0.031} & {0.0288} & $+1$ & {11.77} & 1.868e+22 & 1.968e+10 & 0.303   \\
Fig.~\ref{fig:fit2} & & CCS & 276.6 & 73.3 & 0.021 & 0.0288 & $+1$ & 10.14 & 1.972e+22 & 1.532e+10 & 0.818   \\
 & & {MHD} & {302.2} & {70.2} & {0.178} & {0.0334} & ${+1}$ & {8.30} & {2.132e+22} & {1.153e+10} & {---} \\
\hline
& & LFF & 304.9 & 82.5 & 0.015 & 0.0200 & $+1$ & 11.63 & 1.409e+22 & 1.445e+10 & 0.083   \\
Type  2 & {S6} & GH  & 197.7 & 72.1 & 0.231 & 0.0196 & $+1$ & 13.24 & 1.515e+22 & 1.498e+10 & 0.118   \\
Fig.~\ref{fig:fit0} & & CCS & 329.9 & 79.1 & -0.056 & 0.0196 & $+1$ & 13.24 & 1.192e+22 & 1.446e+10 & 0.119  \\
 & & {MHD} & {327.8} & {82.2} & {0.127} & {0.0220} & ${+1}$ & {12.22} & {1.963e+22} & {1.526e+10} & {---} \\
\hline
& & LFF & 232.8 & 44.8 & -0.024 & 0.0240 & $+1$ & 5.76 & 1.005e+22 & 0.859e+10 & 0.148   \\
Type 2 & {P3} & {GH}  & {228.5} & {41.5} & {0.021} & {0.0230} & $+1$ & {6.98} & 0.819e+22 & 0.954e+10 & 0.124   \\
Fig.~\ref{fig:fit2} & & CCS & 232.1 & 45.5 & -0.024 & 0.0240 & $+1$ & 6.31 & 0.852e+22 & 0.901e+10 & 0.247   \\
 & & {MHD} & {297.7} & {55.3} & {-0.267} & {0.0342} & ${+1}$ & {5.62} & {2.163e+22} & {1.163e+10} & {---} \\
\hline
& & LFF & 24.9 & 78.8 & 0.214 & 0.0196 & $+1$ & 17.94 & 2.088e+22 & 2.185e+10 & 0.154   \\
Type 2  & {P7}  & GH  & 4.9 & 78.8 & 0.180 & 0.0194 & $+1$ & 18.41 & 2.393e+22 & 2.124e+10 & 0.186   \\
Fig.~\ref{fig:fit0} & & CCS & 4.9 & 68.8 & 0.163 & 0.0184 & $+1$ & 23.98 & 1.904e+22 & 1.542e+10 & 0.180   \\
 & & {MHD} & {348.9} & {72.4} & {0.248} & {0.0163} & ${+1}$ & {17.60} & {1.6785e+22} & {1.2856e+10} & {---} \\
\hline
\end{tabular}
\end{threeparttable}
\end{table*}

\begin{table*}[t]
 \centering
 \begin{threeparttable}[T]
 \caption{Flux rope model fit parameters and MHD estimates for Type 3 (problematic orientation) and Type 4 (problematic impact parameter) encounters.}
 \label{tab:fits34}
 \begin{tabular}{|c|c|c|rrrrrc|rc|c|}
\hline
Type/Fig. & Obs. & Model & $\phi_0$ [$^\circ$] & $\theta_0$ [$^\circ$] & $p_0/R_{c}$ & $R_{c}$ [AU] & $H$ & $B_0$ [$\mu$T] & $\Phi_t$ [Mx] & $\Phi_p/L$ [Mx/cm]  & $\chi^2$ \\
\hline
\hline
& & LFF & 111.0 & 13.9 & -0.587 & 0.0450 & $+1$ & 5.06 & 3.104e+22 & 1.415e+10 &  0.099  \\
Type 3  & {S3}  & GH  & 160.4 & 3.2 & -0.001 & 0.0132 & $+1$ & 3.94 & 0.317e+22 & 0.280e+10 & 0.120 \\
Fig.~\ref{fig:fit1} & & CCS & 129.2 & 12.9 & -0.536 & 0.0362 & $+1$ & 8.08 & 2.481e+22 & 0.850e+10 & 0.113  \\
 & & {MHD} & {125.6} & {0.3} & {-0.219} & {0.0398} & ${+1}$ & {5.11} & {1.651e+22} & {0.601e+10} & {---} \\
\hline
& & LFF & 71.7 & -10.2 & 0.082 & 0.0188 & $+1$ & 24.13 & 2.583e+22 & 2.819e+10 & 0.123   \\
Type 3  & {S5} & GH  & 101.0 & -9.6 & 0.169 & 0.0197 & $+1$ & 30.13 & 2.639e+22 & 3.524e+10 & 0.100   \\
Fig.~\ref{fig:fit3} & & CCS & 73.8 & -11.1 & 0.075 & 0.0190 & $+1$ & 28.84 & 2.440e+22 & 2.537e+10 & 0.203   \\
 & & {MHD} & {60.5} & {-13.6} & {0.086} & {0.0137} & ${+1}$ & {43.20} & {2.334e+22} & {2.376e+10} & {---} \\
\hline
 &  & LFF & 95.0 & -10.1 & -0.557 & 0.0372 & $+1$ & 5.18 & 2.171e+22 & 1.197e+10 & 0.129   \\
Type 3  & {P6} & GH  & 145.3 & -7.8 & 0.039 & 0.0180 & $+1$ & 4.19 & 0.606e+22 & 0.414e+10 & 0.137   \\
Fig.~\ref{fig:fit1} & & CCS & 116.4 & -8.3 & -0.449 & 0.0312 & $+1$ & 7.56 & 1.725e+22 & 0.741e+10 & 0.152   \\
 & & {MHD} & {124.2} & {-22.4} & {0.000} & {0.0289} & ${+1}$ & {6.47} & {1.142e+22} & {0.642e+10} & {---} \\
\hline
& & LFF & 142.3 & 12.3 & -0.327 & 0.0217 & $+1$ & 9.33 & 1.331e+22 & 1.258e+10 & 0.193  \\
Type  3 & {P8} & GH  & 146.3 & 10.3 & -0.258 & 0.0190 & $+1$ & 9.37 & 1.182e+22 & 1.057e+10 & 0.287  \\
Fig.~\ref{fig:fit3} & & CCS & 153.6 & 9.3 & -0.331 & 0.0160 & $+1$ & 14.09 & 0.845e+22 & 0.657e+10 & 0.210  \\
 & & {MHD} & {50.4} & {10.8} & {-0.327} & {0.0178} & ${+1}$ & {16.80} & {1.574e+22} & {1.054e+10} & {---} \\
\hline
\hline
 & & LFF & 15.9 & -13.3 & -0.816 & 0.0233 & $+1$ & 4.33 & 0.720e+22 & 0.630e+10 & 0.201   \\
 Type 4 & {S4} & GH  & 109.2 & -39.3 & 0.330 & 0.0394 & $+1$ & 2.68 & 2.484e+22 & 0.412e+10 & 0.247   \\
Fig.~\ref{fig:fit3} & & CCS & 35.5 & -27.0 & -0.843 & 0.0490 & $+1$ & 14.87 & {8.732e+22} & 0.541e+10 & 0.164   \\
 & & {MHD} & {83.3} & {-36.7} & {-0.782} & {0.0260} & ${+1}$ & {4.04} & {0.874e+22} & {0.347e+10} & {---} \\
\hline
 & & LFF & 277.9 & 67.4 & 0.546 & 0.0359 & $+1$ & 7.34 & 2.899e+22 & 1.647e+10 & 0.191  \\
Type 4  & {S8} & {GH}  & {293.0} & {66.4} & {0.448} & {0.0332} & $+1$ & {7.77} & 2.547e+22 & 1.552e+10 & 0.274  \\
Fig.~\ref{fig:fit1} & & CCS & 302.4 & 65.3 & 0.590 & 0.0363 & $+1$ & 12.73 & 3.931e+22 & 1.300e+10 & 0.183   \\
 & & {MHD} & {311.3} & {89.8} & {0.460} & {0.0187} & ${+1}$ & {13.07} & {1.640e+22} & {0.778e+10} & {---} \\
\hline
& & {LFF} & {80.7} & {-37.1} & {-0.939} & {0.1650} & $+1$ & {4.21} & 35.12e+22 & 4.341e+10 & 0.055   \\
Type 4 & {P2} & GH  & 164.1 & -13.9 & -0.421 & 0.0228 & $+1$ & 2.32 & 0.808e+22 & 0.121e+10 & 0.064   \\
Fig.~\ref{fig:fit3} & & {CCS} & {100.3} & {-31.0} & {-0.956} & {0.1934} & $+1$ & {27.95} & 245.1e+22 & 2.106e+10 & 0.036  \\
 & & {MHD} & {244.2} & {-56.0} & {-0.835} & {0.0328} & ${+1}$ & {3.65} & {0.729e+22} & {0.484e+10} & {---} \\
\hline
 & & LFF & 102.1 & 2.5 & 0.401 & 0.0337 & $+1$ & 4.28 & 1.489e+22 & 0.901e+10 & 0.120   \\
Type 4 & {P4} & GH  & 75.3 & 2.7 & 0.078 & 0.0307 & $+1$ & 3.88 & 1.394e+22 & 0.695e+10 & 0.168  \\
Fig.~\ref{fig:fit1} & & CCS & 88.8 & 2.2 & 0.372 & 0.0341 & $+1$ & 6.33 & 1.726e+22 & 0.689e+10 & 0.123  \\
 & & {MHD} & {105.7} & {-2.2} & {0.492} & {0.0341} & ${+1}$ & {5.05} & {1.904e+22} & {0.660e+10} & {---} \\
 \hline
\end{tabular}
\end{threeparttable}
\end{table*}

\subsection*{Lundquist, Constant-$\alpha$ Linear Force-free Model (LFF).}

The first model is the constant-$\alpha$, linear force-free cylinder model based on the \citet{Lundquist1950} Bessel function solution \citep[e.g.][]{Lepping1990}. In local, cylindrical flux rope coordinates $(\, \boldsymbol{\hat{\rho}}, \, \boldsymbol{\hat{\varphi}}, \, \boldsymbol{\hat{z}} \, )$, the magnetic field structure is given by

\begin{equation}
B_{\rho} = 0 \;, \;\;\; B_{\varphi} = H B_0 J_1(\alpha \rho) \; , \;\;\; B_{z} = B_0 J_0 (\alpha \rho) \; ,
\end{equation}
where $H=\pm1$ is the sign of the magnetic helicity, $B_0$ is the field strength at the center of the flux rope (at the axis) and $\alpha$ is determined by setting $\alpha R_c = x_{01} \simeq 2.405$ to the first zero of $J_0$. Since the model flux rope size $R_c$ is obtained from the radial velocity and the identified flux rope boundaries in time (e.g. see Equation 17 in \citealt{Lepping2003a} or Equation A4 in \citealt{Lynch2005}), the Lundquist solution model has a total of five free parameters, $\{ \, \phi_0, \, \theta_0, \, p_0, \, H, \, B_0 \,\}$.

\subsection*{Gold-Hoyle, Uniform-Twist Model (GH).}

The second is the force-free, uniform-twist cylinder model based on the \citet{Gold1960} solution \citep[e.g.][]{Farrugia1999}. This gives a field distribution of

\begin{equation}
B_{\rho} = 0 \;, \;\;\; B_{\varphi} = \frac{\omega \rho B_0}{1+\omega^2\rho^2} \; , \;\;\; B_{z} = \frac{B_0}{1+\omega^2\rho^2} \; ,
\end{equation}
where $\omega = 2\pi \tau$ is the number of field line turns (in radians) about the cylinder axis per unit length and $B_0$ is again the field strength at the cylinder axis. The uniform-twist model also has a total of five free parameters, $\{ \, \phi_0, \, \theta_0, \, p_0, \, \tau, \, B_0 \,\}$. The sign of the flux rope helicity is contained in the sign of the twist, $\tau$.

\subsection*{Non-Force Free, Circular Cross-Section Model (CCS).}

The third model is a non-force free flux rope solution with a circular cross-section \citep{Hidalgo2000,Hidalgo2002a,Nieves-Chinchilla2016}. Here, the model field profiles are given in terms of the electric current density $\boldsymbol{j} = j_\varphi^{\, 0} \, \boldsymbol{\hat{\varphi}} + j_z^{\, 0} \, \boldsymbol{\hat{z}}$ in the flux rope.
For the CCS model these current density components are taken as constant, yielding a field distribution throughout the flux rope of

\begin{equation}
\label{eq:ccs}
    B_{\rho} = 0 \;, \;\;\; B_{\varphi} = \mu_0 j_{z}^{\, 0} \, \frac{\rho}{2} \; , \;\;\;  B_{z} = \mu_0 j_{\varphi}^{\, 0} \left( R_c - {\rho} \right) \, .
\end{equation}
The circular cross-section model, again, has five free parameters, $\{ \, \phi_0, \, \theta_0, \, p_0, \, j_{z}^{\, 0}, \, j_{\varphi}^{\, 0} \,\}$, with the sign of the helicity contained in the (relative) sign of the current density components, $j_{\varphi}^{\, 0}$ and $j_{z}^{\, 0}$. The field strength on the cylinder axis, $B_0$, is obtained from $B_0 = B_z(\rho=0) = \mu_0 \, j_{\varphi}^{\, 0} \, R_c$. 

\textcolor{red}{As discussed in Section~\ref{subsec:ffness}, a quantitative measure of how non-force free the CCS in-situ model fits are can be evaluated by constructing the Lorentz force magnitude $|\, \boldsymbol{j} \times \boldsymbol{B}\,| = |\boldsymbol{j}|\,|\boldsymbol{B}|\sin{\vartheta}$ where the angle $\vartheta$ is the misalignment between the current density and magnetic field vectors. The Lorentz force vector for the CCS model is
\begin{equation}
    \boldsymbol{j} \times \boldsymbol{B} = \mu_0 \left[ \, \left(j_{\varphi}^{\, 0}\right)^2 (R_c - \rho) - \left( j_{z}^{\, 0} \right)^2 \left( \frac{\rho}{2} \right) \, \right] \, \boldsymbol{\hat{\rho}} \; .
\end{equation}
where we have substituted the analytic formulas above for the $\boldsymbol{j}$ and $\boldsymbol{B}$ components.
The misalignment angle, $\vartheta$, is therefore
\begin{equation}
\label{eq:angle}
    \vartheta(\, \rho\, ) = \sin^{-1} \frac{ \big| \left(j_{\varphi}^{\, 0}\right)^2 (R_c - \rho) - \left( j_{z}^{\, 0} \right)^2 (\rho/2) \, \big| }{ \sqrt{ \left(j_{\varphi}^{\, 0}\right)^2 + \left(j_{z}^{\, 0}\right)^2 } \sqrt{ \left( j_{z}^{\, 0} \right)^2 (\rho^2/4) + \left(j_{\varphi}^{\, 0}\right)^2 (R_c - \rho)^2 } }\; .
\end{equation}
The time-dependence of the above expression is captured in the spacecraft's trajectory through the model flux rope, $\rho\,(t)$. The P5 example encounter's CCS $\vartheta(\, \rho\,(t) \, )$ profile is shown in Figures~\ref{fig:nff}(a) as the orange W-shaped curve, while the ejecta duration-averaged values, $\langle \, \vartheta \, \rangle^{\rm CCS}$, are shown in Figure~\ref{fig:nff}(b) for each of the synthetic observers. }

\subsection*{Parameter Optimization and Error Minimization Procedure.}

For each flux rope model, the values of the free parameters are obtained via a least-squares minimization procedure, similar to that originally described by \citet{Lepping1990}. Each of the mean-square error calculations between the ``model'' and ``observational'' values are performed in RTN coordinates. The in-situ flux rope model fields,  $\boldsymbol{B}^{\rm FR}(t)$ (projected into the RTN system) are taken as the ``model'' values whereas the MHD simulation data, $\boldsymbol{B}^{\rm MHD}(t)$ (also projected into the RTN system) are taken as the ``observational'' values.

In the current study, we use a chi-squared error norm, defined as

\begin{equation}
\label{eq:chi2}
\chi^2 = \frac{1}{ N } \sum_{i=1}^N \, \frac{\left( \boldsymbol{B}^{\rm FR}_{j}(t_i) - \boldsymbol{B}^{\rm MHD}_{j}(t_i) \right)^2}{ |B^{\rm MHD}(t_i)|^2 } \; ,
\end{equation}
where $N$ is the total number of data points during the synthetic observer's passage through the CME (typically a $\sim$3 hour interval at 12 samples per hour for a total of 30--40 points in each time series), and we sum over the index $j$ from 1--3 representing the individual $\boldsymbol{\hat{r}}$, $\boldsymbol{\hat{t}}$, and $\boldsymbol{\hat{n}}$ components of each vector field measurement.

\subsection*{Flux Rope Model Fit Parameters}

Tables~\ref{tab:fits12} and ~\ref{tab:fits34} present the best fit parameters associated with the LFF, GH, and CCS in-situ flux rope models applied to each of our 16 synthetic time series through the MHD simulation data, organized by encounter type.

Table~\ref{tab:fits12} lists each of the model fits to the classic bipolar profiles (Type 1; S1, P5, S7, and P1) and unipolar profiles (Type 2; S6, P7, S2, P3). Table~\ref{tab:fits34} lists the same quantities for the problematic orientation profiles (Type 3; S3, P6, S5, P8) and problematic impact parameter profiles (Type 4; S8, P4, S4, P2). The first column identifies the synthetic observer's encounter Type and lists the Figure where the corresponding MHD time series and in-situ flux rope model fits are shown (Figures~\ref{fig:fit0}, \ref{fig:fit1}, \ref{fig:fit2}, or \ref{fig:fit3}).

The parameters for each of the flux rope model fits include the symmetry axis orientation with respect to the synthetic observer (i.e. $\phi_0$, $\theta_0$, and $p_0$), the flux rope radius $R_c$, the flux rope chirality $H$, and the magnitude of the axial field at the flux rope center, $B_0$. We note that these are the LFF model's free parameters and the other models have either slightly different parameters (e.g. the field line twist, current density component amplitudes, etc) or additional parameters that are not listed. We have focused on these quantities as they are the most easily compared across models. Also listed are the toroidal/axial magnetic flux and the poloidal/twist flux per unit length that are derived from each model (and given in Section~\ref{subsec:fluxes} as functions of the model fit parameters). In the last column, we present the numerical value of the $\chi^2$ error norm for each entry.

\renewcommand{\thesection}{Appendix B}
\section{Additional Flux Rope Profiles and Model Fits}
\renewcommand{\thesection}{B}
\label{sec:insitu2}

\begin{figure*}[t!]
    \centering
    \includegraphics[width=0.85\textwidth]{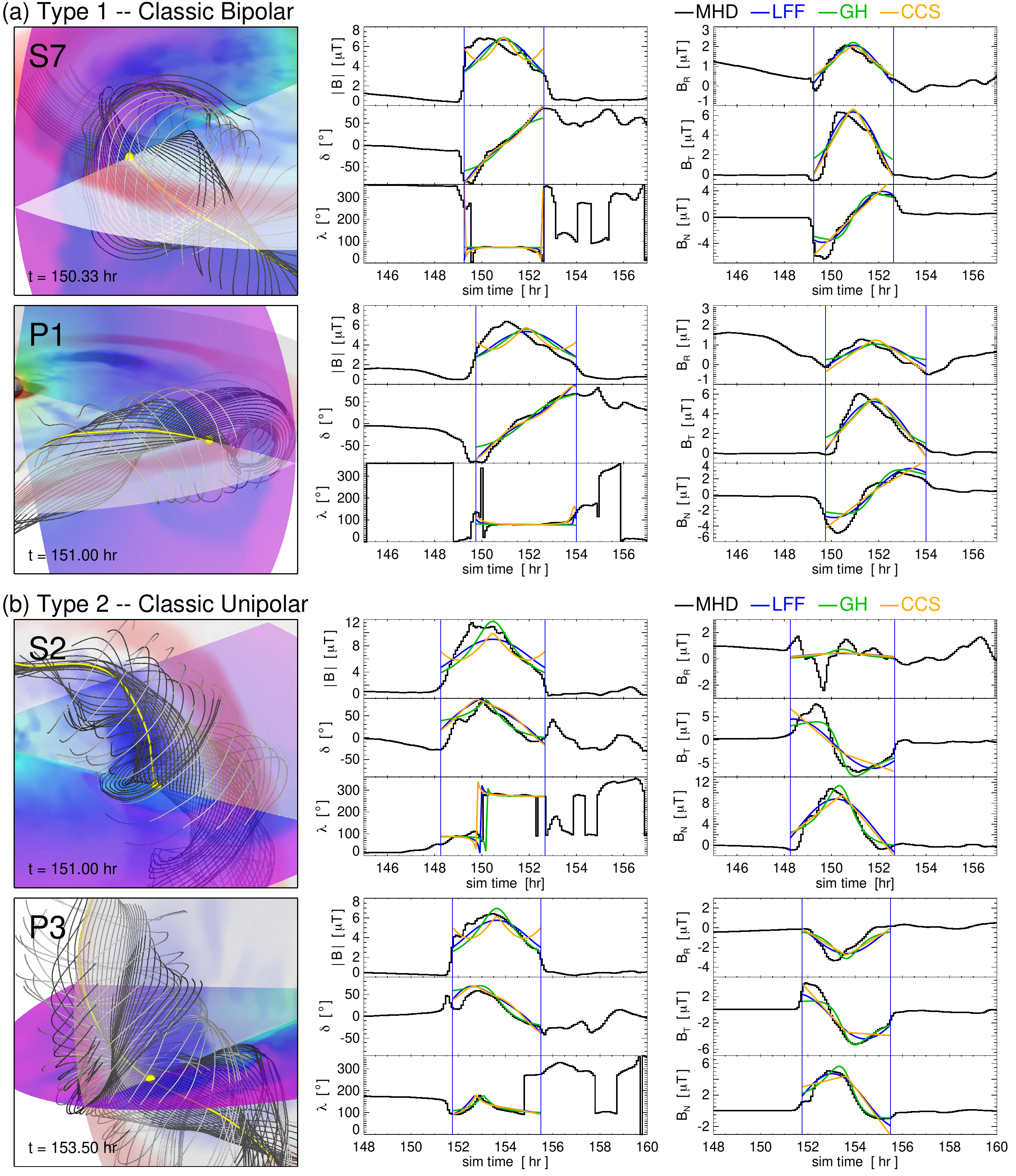}
    \caption{Remaining classic cases in the same format as Figure~\ref{fig:fit0}. (a) Type 1 magnetic field profiles representing classic bipolar MC/ICME orientations from synthetic observers S7 and P1. 
    (b) Type 2 magnetic field profiles representing classic unipolar MC/ICME orientations from synthetic observers S2 and P3.
    }
    \label{fig:fit2}
\end{figure*}
\begin{figure*}[t!]
    \centering
    \includegraphics[width=0.85\textwidth]{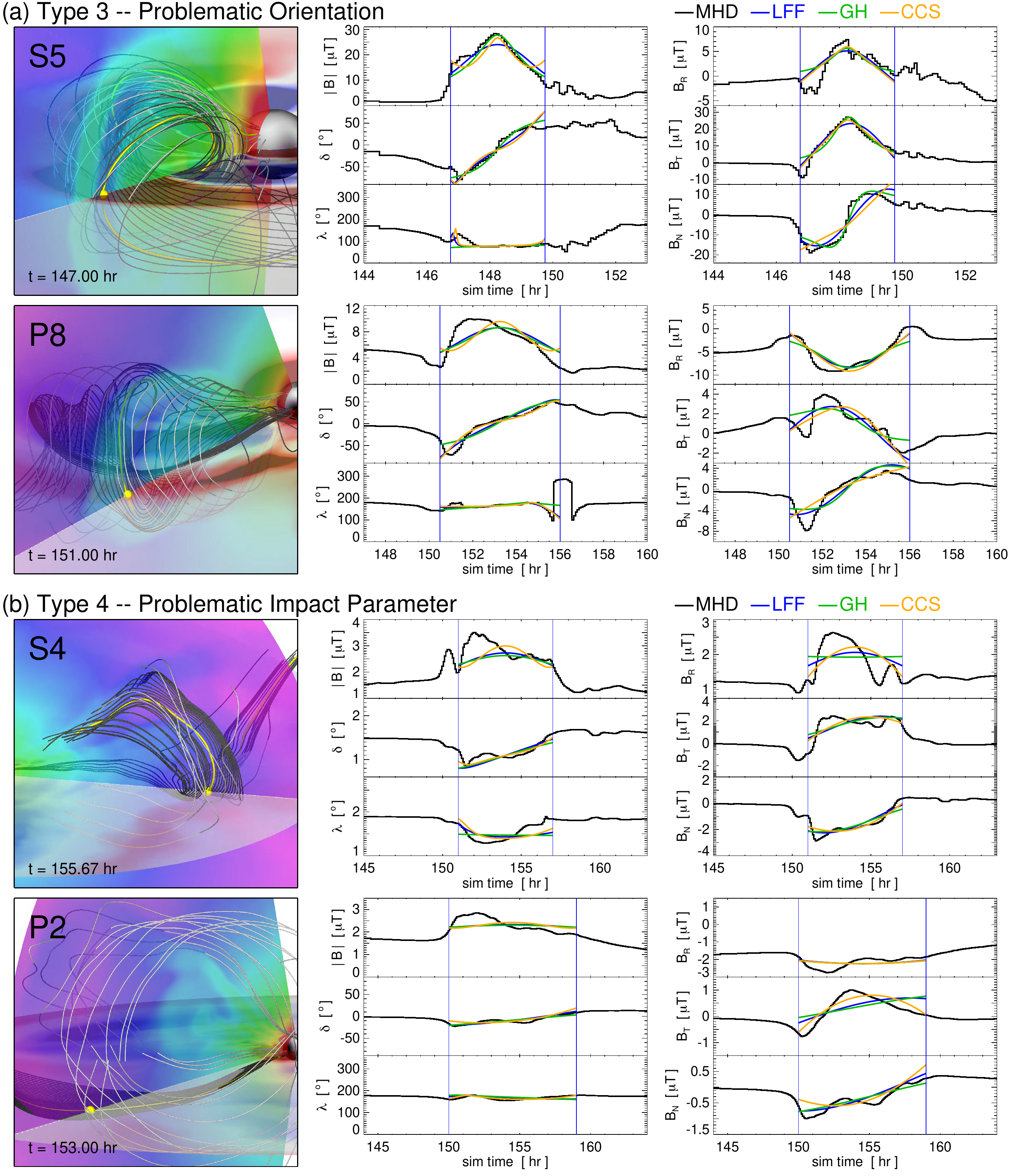}
    \caption{Remaining problematic cases in the same format as Figure~\ref{fig:fit1}. (a) Type 3 magnetic field profiles for problematic orientation MC/ICME events from synthetic observers S5 and P8. 
    (b) Type 4 magnetic field profiles for problematic impact parameter MC/ICME events from synthetic observers S4 and P2.}
    \label{fig:fit3}
\end{figure*}
\begin{figure*}[t!]
    {\includegraphics[width=0.48\textwidth]{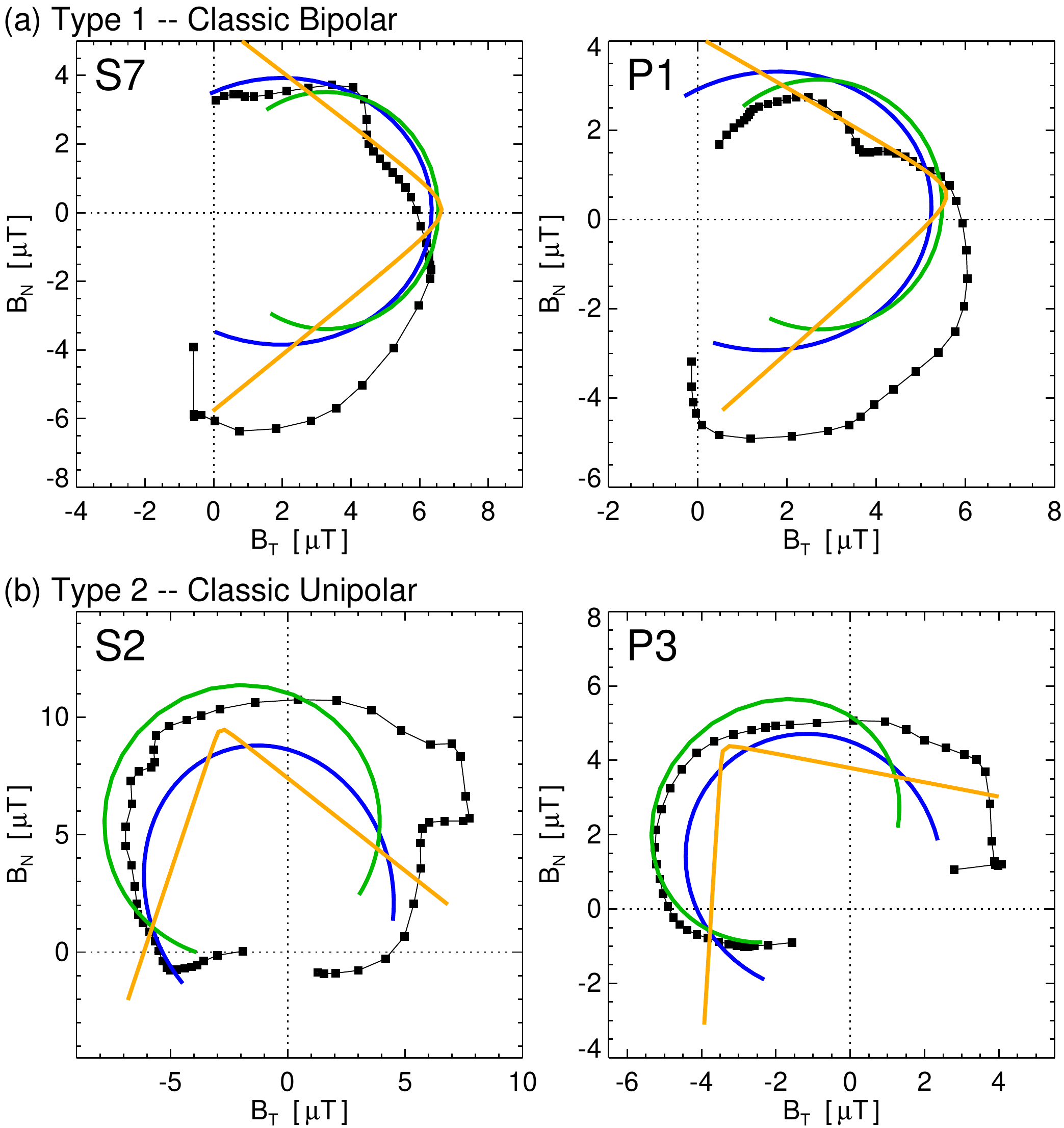}}
    \xspace\xspace
    {\includegraphics[width=0.48\textwidth]{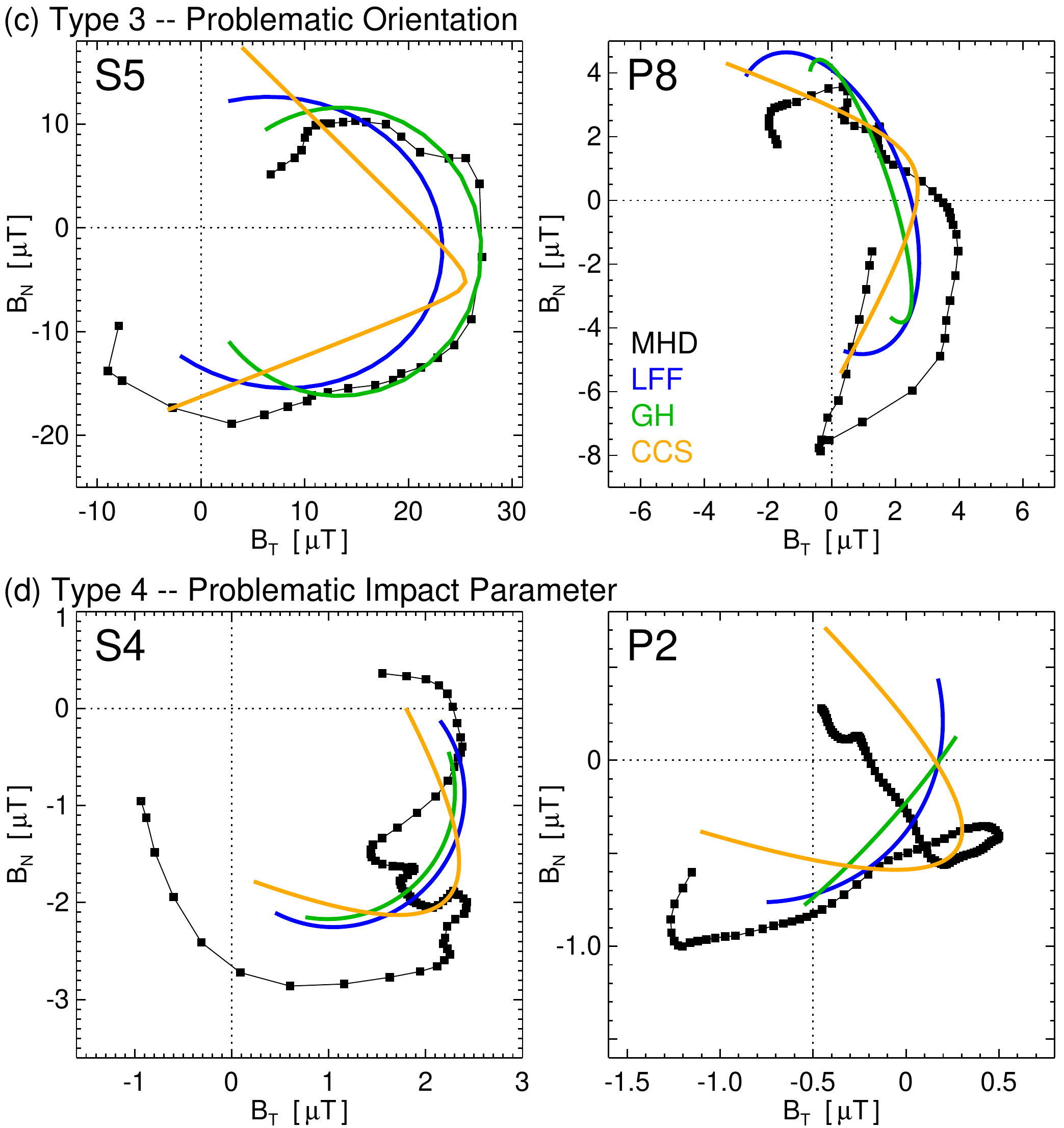}}
    \caption{Remaining $B_T$--$B_N$ hodograms during each observers' CME ejecta encounter organized by profile type in the same format as Figure~\ref{fig:hodor}. (a) Type~1: S7, P1; (b) Type~2: S2, P3; (c) Type~3: S5, P8; (d) Type~4: S4, P2.}
    \label{fig:hodor2}
\end{figure*}

For completeness, here we present the remaining synthetic observer profiles and model fits that were not presented in the main text. These are S2, S4, S5, and S7 for the stationary observer cases and P1, P2, P3, and P8 for the PSP trajectory cases.

Figure~\ref{fig:fit2}(a) shows the Type 1 classic bipolar cases S7, P1 and Figure~\ref{fig:fit2}(b) shows the Type 2 classic unipolar cases S2, P3, in the same format as Figure~\ref{fig:fit0}. Likewise, Figure~\ref{fig:fit3}(a) shows the Type 3 problematic orientation cases S5, P8 and Figure~\ref{fig:fit3}(b) shows the Type 4 problematic impact parameter cases S4, P2, in the same format as Figure~\ref{fig:fit1}.

The $B_N$--$B_T$ hodograms for each of these cases are shown in Figure~\ref{fig:hodor2} where the synthetic observer cases are again organized by type: panel~\ref{fig:hodor2}(a) shows Type~1, \ref{fig:hodor2}(b) Type~2, \ref{fig:hodor2}(c) Type~3, and \ref{fig:hodor2}(d) Type~4.

\input{journal_macros.sty}
\bibliographystyle{model5-names}
\biboptions{authoryear}
\bibliography{master}

\begin{thebibliography}{101}
\expandafter\ifx\csname natexlab\endcsname\relax\def\natexlab#1{#1}\fi
\providecommand{\url}[1]{\texttt{#1}}
\providecommand{\href}[2]{#2}
\providecommand{\path}[1]{#1}
\providecommand{\DOIprefix}{doi:}
\providecommand{\ArXivprefix}{arXiv:}
\providecommand{\URLprefix}{URL: }
\providecommand{\Pubmedprefix}{pmid:}
\providecommand{\doi}[1]{\href{http://dx.doi.org/#1}{\path{#1}}}
\providecommand{\Pubmed}[1]{\href{pmid:#1}{\path{#1}}}
\providecommand{\bibinfo}[2]{#2}
\ifx\xfnm\relax \def\xfnm[#1]{\unskip,\space#1}\fi
\bibitem[{{Al-Haddad} et~al.(2019{\natexlab{a}}){Al-Haddad}, {Lugaz}, {Poedts},
  {Farrugia}, {Nieves-Chinchilla} \& {Roussev}}]{Al-Haddad2019b}
\bibinfo{author}{{Al-Haddad}, N.}, \bibinfo{author}{{Lugaz}, N.},
  \bibinfo{author}{{Poedts}, S.}, \bibinfo{author}{{Farrugia}, C.~J.},
  \bibinfo{author}{{Nieves-Chinchilla}, T.}, \& \bibinfo{author}{{Roussev},
  I.~I.} (\bibinfo{year}{2019}{\natexlab{a}}).
\newblock \bibinfo{title}{{Evolution of Coronal Mass Ejection Properties in the
  Inner Heliosphere: Prediction for the Solar Orbiter and Parker Solar Probe}}.
\newblock {\it \bibinfo{journal}{\apj}\/},  {\it
  \bibinfo{volume}{884}\/}\bibinfo{issue}{(2)}, \bibinfo{pages}{179}.
  \DOIprefix\doi{10.3847/1538-4357/ab4126}.
  \href{http://arxiv.org/abs/1910.04811}{\tt arXiv:1910.04811}.
\bibitem[{{Al-Haddad} et~al.(2018){Al-Haddad}, {Nieves-Chinchilla}, {Savani},
  {Lugaz} \& {Roussev}}]{Al-Haddad2018}
\bibinfo{author}{{Al-Haddad}, N.}, \bibinfo{author}{{Nieves-Chinchilla}, T.},
  \bibinfo{author}{{Savani}, N.~P.}, \bibinfo{author}{{Lugaz}, N.}, \&
  \bibinfo{author}{{Roussev}, I.~I.} (\bibinfo{year}{2018}).
\newblock \bibinfo{title}{{Fitting and Reconstruction of Thirteen Simple
  Coronal Mass Ejections}}.
\newblock {\it \bibinfo{journal}{\solphys}\/},  {\it
  \bibinfo{volume}{293}\/}\bibinfo{issue}{(5)}, \bibinfo{pages}{73}.
  \DOIprefix\doi{10.1007/s11207-018-1288-3}.
  \href{http://arxiv.org/abs/1804.02359}{\tt arXiv:1804.02359}.
\bibitem[{{Al-Haddad} et~al.(2013){Al-Haddad}, {Nieves-Chinchilla}, {Savani},
  {M{\"o}stl}, {Marubashi}, {Hidalgo}, {Roussev}, {Poedts} \&
  {Farrugia}}]{Al-Haddad2013}
\bibinfo{author}{{Al-Haddad}, N.}, \bibinfo{author}{{Nieves-Chinchilla}, T.},
  \bibinfo{author}{{Savani}, N.~P.}, \bibinfo{author}{{M{\"o}stl}, C.},
  \bibinfo{author}{{Marubashi}, K.}, \bibinfo{author}{{Hidalgo}, M.~A.},
  \bibinfo{author}{{Roussev}, I.~I.}, \bibinfo{author}{{Poedts}, S.}, \&
  \bibinfo{author}{{Farrugia}, C.~J.} (\bibinfo{year}{2013}).
\newblock \bibinfo{title}{{Magnetic Field Configuration Models and
  Reconstruction Methods for Interplanetary Coronal Mass Ejections}}.
\newblock {\it \bibinfo{journal}{\solphys}\/},  {\it \bibinfo{volume}{284}\/},
  \bibinfo{pages}{129--149}. \DOIprefix\doi{10.1007/s11207-013-0244-5}.
  \href{http://arxiv.org/abs/1209.6394}{\tt arXiv:1209.6394}.
\bibitem[{{Al-Haddad} et~al.(2019{\natexlab{b}}){Al-Haddad}, {Poedts},
  {Roussev}, {Farrugia}, {Yu} \& {Lugaz}}]{Al-Haddad2019a}
\bibinfo{author}{{Al-Haddad}, N.}, \bibinfo{author}{{Poedts}, S.},
  \bibinfo{author}{{Roussev}, I.}, \bibinfo{author}{{Farrugia}, C.~J.},
  \bibinfo{author}{{Yu}, W.}, \& \bibinfo{author}{{Lugaz}, N.}
  (\bibinfo{year}{2019}{\natexlab{b}}).
\newblock \bibinfo{title}{{The Magnetic Morphology of Magnetic Clouds:
  Multi-spacecraft Investigation of Twisted and Writhed Coronal Mass
  Ejections}}.
\newblock {\it \bibinfo{journal}{\apj}\/},  {\it
  \bibinfo{volume}{870}\/}\bibinfo{issue}{(2)}, \bibinfo{pages}{100}.
  \DOIprefix\doi{10.3847/1538-4357/aaf38d}.
\bibitem[{{Al-Haddad} et~al.(2011){Al-Haddad}, {Roussev}, {M{\"o}stl},
  {Jacobs}, {Lugaz}, {Poedts} \& {Farrugia}}]{Al-Haddad2011}
\bibinfo{author}{{Al-Haddad}, N.}, \bibinfo{author}{{Roussev}, I.~I.},
  \bibinfo{author}{{M{\"o}stl}, C.}, \bibinfo{author}{{Jacobs}, C.},
  \bibinfo{author}{{Lugaz}, N.}, \bibinfo{author}{{Poedts}, S.}, \&
  \bibinfo{author}{{Farrugia}, C.~J.} (\bibinfo{year}{2011}).
\newblock \bibinfo{title}{{On the Internal Structure of the Magnetic Field in
  Magnetic Clouds and Interplanetary Coronal Mass Ejections: Writhe versus
  Twist}}.
\newblock {\it \bibinfo{journal}{\apjl}\/},  {\it
  \bibinfo{volume}{738}\/}\bibinfo{issue}{(2)}, \bibinfo{pages}{L18}.
  \DOIprefix\doi{10.1088/2041-8205/738/2/L18}.
\bibitem[{{Berchem} \& {Russell}(1982)}]{Berchem1982}
\bibinfo{author}{{Berchem}, J.}, \& \bibinfo{author}{{Russell}, C.~T.}
  (\bibinfo{year}{1982}).
\newblock \bibinfo{title}{{Magnetic field rotation through the magnetopause:
  ISEE 1 and 2 observations}}.
\newblock {\it \bibinfo{journal}{\jgr}\/},  {\it
  \bibinfo{volume}{87}\/}\bibinfo{issue}{(A10)}, \bibinfo{pages}{8139--8148}.
  \DOIprefix\doi{10.1029/JA087iA10p08139}.
\bibitem[{{Bothmer} \& {Schwenn}(1998)}]{Bothmer1998}
\bibinfo{author}{{Bothmer}, V.}, \& \bibinfo{author}{{Schwenn}, R.}
  (\bibinfo{year}{1998}).
\newblock \bibinfo{title}{{The structure and origin of magnetic clouds in the
  solar wind}}.
\newblock {\it \bibinfo{journal}{\angeo}\/},  {\it \bibinfo{volume}{16}\/},
  \bibinfo{pages}{1--24}. \DOIprefix\doi{10.1007/s00585-997-0001-x}.
\bibitem[{{Burlaga} et~al.(1981){Burlaga}, {Sittler}, {Mariani} \&
  {Schwenn}}]{Burlaga1981}
\bibinfo{author}{{Burlaga}, L.}, \bibinfo{author}{{Sittler}, E.},
  \bibinfo{author}{{Mariani}, F.}, \& \bibinfo{author}{{Schwenn}, R.}
  (\bibinfo{year}{1981}).
\newblock \bibinfo{title}{{Magnetic loop behind an interplanetary shock -
  Voyager, Helios, and IMP 8 observations}}.
\newblock {\it \bibinfo{journal}{\jgr}\/},  {\it \bibinfo{volume}{86}\/},
  \bibinfo{pages}{6673--6684}. \DOIprefix\doi{10.1029/JA086iA08p06673}.
\bibitem[{{Burlaga}(1988)}]{Burlaga1988}
\bibinfo{author}{{Burlaga}, L.~F.} (\bibinfo{year}{1988}).
\newblock \bibinfo{title}{{Magnetic clouds and force-free fields with constant
  alpha}}.
\newblock {\it \bibinfo{journal}{\jgr}\/},  {\it \bibinfo{volume}{93}\/},
  \bibinfo{pages}{7217--7224}. \DOIprefix\doi{10.1029/JA093iA07p07217}.
\bibitem[{{Dasso} et~al.(2006){Dasso}, {Mandrini}, {D{\'e}moulin} \&
  {Luoni}}]{Dasso2006}
\bibinfo{author}{{Dasso}, S.}, \bibinfo{author}{{Mandrini}, C.~H.},
  \bibinfo{author}{{D{\'e}moulin}, P.}, \& \bibinfo{author}{{Luoni}, M.~L.}
  (\bibinfo{year}{2006}).
\newblock \bibinfo{title}{{A new model-independent method to compute magnetic
  helicity in magnetic clouds}}.
\newblock {\it \bibinfo{journal}{\aap}\/},  {\it \bibinfo{volume}{455}\/},
  \bibinfo{pages}{349--359}. \DOIprefix\doi{10.1051/0004-6361:20064806}.
\bibitem[{{Davies} et~al.(2021){Davies}, {Forsyth}, {Winslow}, {M{\"o}stl} \&
  {Lugaz}}]{DaviesEE2021b}
\bibinfo{author}{{Davies}, E.~E.}, \bibinfo{author}{{Forsyth}, R.~J.},
  \bibinfo{author}{{Winslow}, R.~M.}, \bibinfo{author}{{M{\"o}stl}, C.}, \&
  \bibinfo{author}{{Lugaz}, N.} (\bibinfo{year}{2021}).
\newblock \bibinfo{title}{{A Catalog of Interplanetary Coronal Mass Ejections
  Observed by Juno between 1 and 5.4 au}}.
\newblock {\it \bibinfo{journal}{\apj}\/},  {\it
  \bibinfo{volume}{923}\/}\bibinfo{issue}{(2)}, \bibinfo{pages}{136}.
  \DOIprefix\doi{10.3847/1538-4357/ac2ccb}.
  \href{http://arxiv.org/abs/2111.11336}{\tt arXiv:2111.11336}.
\bibitem[{{D\'{e}moulin}(2008)}]{Demoulin2008}
\bibinfo{author}{{D\'{e}moulin}, P.} (\bibinfo{year}{2008}).
\newblock \bibinfo{title}{{A review of the quantitative links between CMEs and
  magnetic clouds}}.
\newblock {\it \bibinfo{journal}{Annales Geophysicae}\/},  {\it
  \bibinfo{volume}{26}\/}, \bibinfo{pages}{3113--3125}.
  \DOIprefix\doi{10.5194/angeo-26-3113-2008}.
\bibitem[{{D{\'e}moulin} et~al.(2020){D{\'e}moulin}, {Dasso}, {Lanabere} \&
  {Janvier}}]{Demoulin2020}
\bibinfo{author}{{D{\'e}moulin}, P.}, \bibinfo{author}{{Dasso}, S.},
  \bibinfo{author}{{Lanabere}, V.}, \& \bibinfo{author}{{Janvier}, M.}
  (\bibinfo{year}{2020}).
\newblock \bibinfo{title}{{Contribution of the ageing effect to the observed
  asymmetry of interplanetary magnetic clouds}}.
\newblock {\it \bibinfo{journal}{\aap}\/},  {\it \bibinfo{volume}{639}\/},
  \bibinfo{pages}{A6}. \DOIprefix\doi{10.1051/0004-6361/202038077}.
  \href{http://arxiv.org/abs/2005.05049}{\tt arXiv:2005.05049}.
\bibitem[{{DeVore}(1991)}]{DeVore1991}
\bibinfo{author}{{DeVore}, C.~R.} (\bibinfo{year}{1991}).
\newblock \bibinfo{title}{{Flux-corrected transport techniques for
  multidimensional compressible magnetohydrodynamics}}.
\newblock {\it \bibinfo{journal}{J. Comp. Phys.}\/},  {\it
  \bibinfo{volume}{92}\/}, \bibinfo{pages}{142--160}.
  \DOIprefix\doi{10.1016/0021-9991(91)90295-V}.
\bibitem[{{DeVore} \& {Antiochos}(2008)}]{DeVore2008}
\bibinfo{author}{{DeVore}, C.~R.}, \& \bibinfo{author}{{Antiochos}, S.~K.}
  (\bibinfo{year}{2008}).
\newblock \bibinfo{title}{{Homologous Confined Filament Eruptions via Magnetic
  Breakout}}.
\newblock {\it \bibinfo{journal}{\apj}\/},  {\it \bibinfo{volume}{680}\/},
  \bibinfo{pages}{740--756}. \DOIprefix\doi{10.1086/588011}.
\bibitem[{{Farrugia} et~al.(1999){Farrugia}, {Janoo}, {Torbert}, {Quinn},
  {Ogilvie}, {Lepping}, {Fitzenreiter}, {Steinberg}, {Lazarus}, {Lin},
  {Larson}, {Dasso}, {Gratton}, {Lin} \& {Berdichevsky}}]{Farrugia1999}
\bibinfo{author}{{Farrugia}, C.~J.}, \bibinfo{author}{{Janoo}, L.~A.},
  \bibinfo{author}{{Torbert}, R.~B.}, \bibinfo{author}{{Quinn}, J.~M.},
  \bibinfo{author}{{Ogilvie}, K.~W.}, \bibinfo{author}{{Lepping}, R.~P.},
  \bibinfo{author}{{Fitzenreiter}, R.~J.}, \bibinfo{author}{{Steinberg},
  J.~T.}, \bibinfo{author}{{Lazarus}, A.~J.}, \bibinfo{author}{{Lin}, R.~P.},
  \bibinfo{author}{{Larson}, D.}, \bibinfo{author}{{Dasso}, S.},
  \bibinfo{author}{{Gratton}, F.~T.}, \bibinfo{author}{{Lin}, Y.}, \&
  \bibinfo{author}{{Berdichevsky}, D.} (\bibinfo{year}{1999}).
\newblock \bibinfo{title}{{A uniform-twist magnetic flux rope in the solar
  wind}}.
\newblock In \bibinfo{editor}{S.~R. {Habbal}}, \bibinfo{editor}{R.~{Esser}},
  \bibinfo{editor}{J.~V. {Hollweg}}, \& \bibinfo{editor}{P.~A. {Isenberg}}
  (Eds.), {\it \bibinfo{booktitle}{Solar Wind Nine}\/} (pp.
  \bibinfo{pages}{745--748}).
\newblock volume \bibinfo{volume}{471} of {\it \bibinfo{series}{American
  Institute of Physics Conference Series}\/}.
\newblock \DOIprefix\doi{10.1063/1.58724}.
\bibitem[{{Forbes} \& {Priest}(1983)}]{Forbes1983}
\bibinfo{author}{{Forbes}, T.~G.}, \& \bibinfo{author}{{Priest}, E.~R.}
  (\bibinfo{year}{1983}).
\newblock \bibinfo{title}{{A numerical experiment relevant to line-tied
  reconnection in two-ribbon flares}}.
\newblock {\it \bibinfo{journal}{\solphys}\/},  {\it \bibinfo{volume}{84}\/},
  \bibinfo{pages}{169--188}. \DOIprefix\doi{10.1007/BF00157455}.
\bibitem[{{Fox} et~al.(2016){Fox}, {Velli}, {Bale}, {Decker}, {Driesman},
  {Howard}, {Kasper}, {Kinnison}, {Kusterer}, {Lario}, {Lockwood}, {McComas},
  {Raouafi} \& {Szabo}}]{Fox2016}
\bibinfo{author}{{Fox}, N.~J.}, \bibinfo{author}{{Velli}, M.~C.},
  \bibinfo{author}{{Bale}, S.~D.}, \bibinfo{author}{{Decker}, R.},
  \bibinfo{author}{{Driesman}, A.}, \bibinfo{author}{{Howard}, R.~A.},
  \bibinfo{author}{{Kasper}, J.~C.}, \bibinfo{author}{{Kinnison}, J.},
  \bibinfo{author}{{Kusterer}, M.}, \bibinfo{author}{{Lario}, D.},
  \bibinfo{author}{{Lockwood}, M.~K.}, \bibinfo{author}{{McComas}, D.~J.},
  \bibinfo{author}{{Raouafi}, N.~E.}, \& \bibinfo{author}{{Szabo}, A.}
  (\bibinfo{year}{2016}).
\newblock \bibinfo{title}{{The Solar Probe Plus Mission: Humanity's First Visit
  to Our Star}}.
\newblock {\it \bibinfo{journal}{\ssr}\/},  {\it
  \bibinfo{volume}{204}\/}\bibinfo{issue}{(1-4)}, \bibinfo{pages}{7--48}.
  \DOIprefix\doi{10.1007/s11214-015-0211-6}.
\bibitem[{{Gold} \& {Hoyle}(1960)}]{Gold1960}
\bibinfo{author}{{Gold}, T.}, \& \bibinfo{author}{{Hoyle}, F.}
  (\bibinfo{year}{1960}).
\newblock \bibinfo{title}{{On the origin of solar flares}}.
\newblock {\it \bibinfo{journal}{\mnras}\/},  {\it \bibinfo{volume}{120}\/},
  \bibinfo{pages}{89}. \DOIprefix\doi{10.1093/mnras/120.2.89}.
\bibitem[{{Gopalswamy} et~al.(2018){Gopalswamy}, {Akiyama}, {Yashiro} \&
  {Xie}}]{Gopalswamy2018}
\bibinfo{author}{{Gopalswamy}, N.}, \bibinfo{author}{{Akiyama}, S.},
  \bibinfo{author}{{Yashiro}, S.}, \& \bibinfo{author}{{Xie}, H.}
  (\bibinfo{year}{2018}).
\newblock \bibinfo{title}{{A New Technique to Provide Realistic Input to CME
  Forecasting Models}}.
\newblock In \bibinfo{editor}{C.~{Foullon}}, \& \bibinfo{editor}{O.~E.
  {Malandraki}} (Eds.), {\it \bibinfo{booktitle}{Space Weather of the
  Heliosphere: Processes and Forecasts}\/} (pp. \bibinfo{pages}{258--262}).
\newblock volume \bibinfo{volume}{335} of {\it \bibinfo{series}{IAU
  Symposium}\/}.
\newblock \DOIprefix\doi{10.1017/S1743921317011048}.
  \href{http://arxiv.org/abs/1709.03160}{\tt arXiv:1709.03160}.
\bibitem[{{Gopalswamy} et~al.(2017){Gopalswamy}, {Yashiro}, {Akiyama} \&
  {Xie}}]{Gopalswamy2017}
\bibinfo{author}{{Gopalswamy}, N.}, \bibinfo{author}{{Yashiro}, S.},
  \bibinfo{author}{{Akiyama}, S.}, \& \bibinfo{author}{{Xie}, H.}
  (\bibinfo{year}{2017}).
\newblock \bibinfo{title}{{Estimation of Reconnection Flux Using Post-eruption
  Arcades and Its Relevance to Magnetic Clouds at 1 AU}}.
\newblock {\it \bibinfo{journal}{\solphys}\/},  {\it \bibinfo{volume}{292}\/},
  \bibinfo{pages}{65}. \DOIprefix\doi{10.1007/s11207-017-1080-9}.
  \href{http://arxiv.org/abs/1701.01943}{\tt arXiv:1701.01943}.
\bibitem[{{Hadid} et~al.(2021){Hadid}, {G{\'e}not}, {Aizawa}, {Milillo},
  {Zender}, {Murakami}, {Benkhoff}, {Zouganelis}, {Alberti}, {Andr{\'e}},
  {Bebesi}, {Califano}, {Dimmock}, {Dosa}, {Escoubet}, {Griton}, {Ho},
  {Horbury}, {Iwai}, {Janvier}, {Kilpua}, {Lavraud}, {Madar}, {Miyoshi},
  {M{\"u}ller}, {Pinto}, {Rouillard}, {Raines}, {Raouafi}, {Sahraoui},
  {S{\'a}nchez-Cano}, {Shiota}, {Vainio} \& {Walsh}}]{Hadid2021}
\bibinfo{author}{{Hadid}, L.~Z.}, \bibinfo{author}{{G{\'e}not}, V.},
  \bibinfo{author}{{Aizawa}, S.}, \bibinfo{author}{{Milillo}, A.},
  \bibinfo{author}{{Zender}, J.}, \bibinfo{author}{{Murakami}, G.},
  \bibinfo{author}{{Benkhoff}, J.}, \bibinfo{author}{{Zouganelis}, I.},
  \bibinfo{author}{{Alberti}, T.}, \bibinfo{author}{{Andr{\'e}}, N.},
  \bibinfo{author}{{Bebesi}, Z.}, \bibinfo{author}{{Califano}, F.},
  \bibinfo{author}{{Dimmock}, A.~P.}, \bibinfo{author}{{Dosa}, M.},
  \bibinfo{author}{{Escoubet}, C.~P.}, \bibinfo{author}{{Griton}, L.},
  \bibinfo{author}{{Ho}, G.~C.}, \bibinfo{author}{{Horbury}, T.~S.},
  \bibinfo{author}{{Iwai}, K.}, \bibinfo{author}{{Janvier}, M.},
  \bibinfo{author}{{Kilpua}, E.}, \bibinfo{author}{{Lavraud}, B.},
  \bibinfo{author}{{Madar}, A.}, \bibinfo{author}{{Miyoshi}, Y.},
  \bibinfo{author}{{M{\"u}ller}, D.}, \bibinfo{author}{{Pinto}, R.~F.},
  \bibinfo{author}{{Rouillard}, A.~P.}, \bibinfo{author}{{Raines}, J.~M.},
  \bibinfo{author}{{Raouafi}, N.}, \bibinfo{author}{{Sahraoui}, F.},
  \bibinfo{author}{{S{\'a}nchez-Cano}, B.}, \bibinfo{author}{{Shiota}, D.},
  \bibinfo{author}{{Vainio}, R.}, \& \bibinfo{author}{{Walsh}, A.}
  (\bibinfo{year}{2021}).
\newblock \bibinfo{title}{{BepiColombo's cruise phase: unique opportunity for
  synergistic observations}}.
\newblock {\it \bibinfo{journal}{\frass}\/},  {\it \bibinfo{volume}{8}\/},
  \bibinfo{pages}{154}. \DOIprefix\doi{10.3389/fspas.2021.718024}.
\bibitem[{{Hess} et~al.(2020){Hess}, {Rouillard}, {Kouloumvakos}, {Liewer},
  {Zhang}, {Dhakal}, {Stenborg}, {Colaninno} \& {Howard}}]{Hess2020}
\bibinfo{author}{{Hess}, P.}, \bibinfo{author}{{Rouillard}, A.~P.},
  \bibinfo{author}{{Kouloumvakos}, A.}, \bibinfo{author}{{Liewer}, P.~C.},
  \bibinfo{author}{{Zhang}, J.}, \bibinfo{author}{{Dhakal}, S.},
  \bibinfo{author}{{Stenborg}, G.}, \bibinfo{author}{{Colaninno}, R.~C.}, \&
  \bibinfo{author}{{Howard}, R.~A.} (\bibinfo{year}{2020}).
\newblock \bibinfo{title}{{WISPR Imaging of a Pristine CME}}.
\newblock {\it \bibinfo{journal}{\apjs}\/},  {\it
  \bibinfo{volume}{246}\/}\bibinfo{issue}{(2)}, \bibinfo{pages}{25}.
  \DOIprefix\doi{10.3847/1538-4365/ab4ff0}.
  \href{http://arxiv.org/abs/1912.02255}{\tt arXiv:1912.02255}.
\bibitem[{{Hidalgo} et~al.(2000){Hidalgo}, {Cid}, {Medina} \&
  {Vi{\~n}as}}]{Hidalgo2000}
\bibinfo{author}{{Hidalgo}, M.~A.}, \bibinfo{author}{{Cid}, C.},
  \bibinfo{author}{{Medina}, J.}, \& \bibinfo{author}{{Vi{\~n}as}, A.~F.}
  (\bibinfo{year}{2000}).
\newblock \bibinfo{title}{{A new model for the topology of magnetic clouds in
  the solar wind}}.
\newblock {\it \bibinfo{journal}{\solphys}\/},  {\it
  \bibinfo{volume}{194}\/}\bibinfo{issue}{(1)}, \bibinfo{pages}{165--174}.
  \DOIprefix\doi{10.1023/A:1005206107017}.
\bibitem[{{Hidalgo} et~al.(2002){Hidalgo}, {Cid}, {Vinas} \&
  {Sequeiros}}]{Hidalgo2002a}
\bibinfo{author}{{Hidalgo}, M.~A.}, \bibinfo{author}{{Cid}, C.},
  \bibinfo{author}{{Vinas}, A.~F.}, \& \bibinfo{author}{{Sequeiros}, J.}
  (\bibinfo{year}{2002}).
\newblock \bibinfo{title}{{A non-force-free approach to the topology of
  magnetic clouds in the solar wind}}.
\newblock {\it \bibinfo{journal}{Journal of Geophysical Research (Space
  Physics)}\/},  {\it \bibinfo{volume}{107}\/}\bibinfo{issue}{(A1)},
  \bibinfo{pages}{1002}. \DOIprefix\doi{10.1029/2001JA900100}.
\bibitem[{{Howard} et~al.(2019){Howard}, {Vourlidas}, {Bothmer}, {Colaninno},
  {DeForest}, {Gallagher}, {Hall}, {Hess}, {Higginson}, {Korendyke},
  {Kouloumvakos}, {Lamy}, {Liewer}, {Linker}, {Linton}, {Penteado}, {Plunkett},
  {Poirier}, {Raouafi}, {Rich}, {Rochus}, {Rouillard}, {Socker}, {Stenborg},
  {Thernisien} \& {Viall}}]{HowardR2019}
\bibinfo{author}{{Howard}, R.~A.}, \bibinfo{author}{{Vourlidas}, A.},
  \bibinfo{author}{{Bothmer}, V.}, \bibinfo{author}{{Colaninno}, R.~C.},
  \bibinfo{author}{{DeForest}, C.~E.}, \bibinfo{author}{{Gallagher}, B.},
  \bibinfo{author}{{Hall}, J.~R.}, \bibinfo{author}{{Hess}, P.},
  \bibinfo{author}{{Higginson}, A.~K.}, \bibinfo{author}{{Korendyke}, C.~M.},
  \bibinfo{author}{{Kouloumvakos}, A.}, \bibinfo{author}{{Lamy}, P.~L.},
  \bibinfo{author}{{Liewer}, P.~C.}, \bibinfo{author}{{Linker}, J.},
  \bibinfo{author}{{Linton}, M.}, \bibinfo{author}{{Penteado}, P.},
  \bibinfo{author}{{Plunkett}, S.~P.}, \bibinfo{author}{{Poirier}, N.},
  \bibinfo{author}{{Raouafi}, N.~E.}, \bibinfo{author}{{Rich}, N.},
  \bibinfo{author}{{Rochus}, P.}, \bibinfo{author}{{Rouillard}, A.~P.},
  \bibinfo{author}{{Socker}, D.~G.}, \bibinfo{author}{{Stenborg}, G.},
  \bibinfo{author}{{Thernisien}, A.~F.}, \& \bibinfo{author}{{Viall}, N.~M.}
  (\bibinfo{year}{2019}).
\newblock \bibinfo{title}{{Near-Sun observations of an F-corona decrease and
  K-corona fine structure}}.
\newblock {\it \bibinfo{journal}{\nat}\/},  {\it
  \bibinfo{volume}{576}\/}\bibinfo{issue}{(7786)}, \bibinfo{pages}{232--236}.
  \DOIprefix\doi{10.1038/s41586-019-1807-x}.
\bibitem[{{Hu} et~al.(2014){Hu}, {Qiu}, {Dasgupta}, {Khare} \& {Webb}}]{Hu2014}
\bibinfo{author}{{Hu}, Q.}, \bibinfo{author}{{Qiu}, J.},
  \bibinfo{author}{{Dasgupta}, B.}, \bibinfo{author}{{Khare}, A.}, \&
  \bibinfo{author}{{Webb}, G.~M.} (\bibinfo{year}{2014}).
\newblock \bibinfo{title}{{Structures of Interplanetary Magnetic Flux Ropes and
  Comparison with Their Solar Sources}}.
\newblock {\it \bibinfo{journal}{\apj}\/},  {\it \bibinfo{volume}{793}\/},
  \bibinfo{pages}{53}. \DOIprefix\doi{10.1088/0004-637X/793/1/53}.
  \href{http://arxiv.org/abs/1408.1470}{\tt arXiv:1408.1470}.
\bibitem[{{Illing} \& {Hundhausen}(1985)}]{Illing1985}
\bibinfo{author}{{Illing}, R.~M.~E.}, \& \bibinfo{author}{{Hundhausen}, A.~J.}
  (\bibinfo{year}{1985}).
\newblock \bibinfo{title}{{Observation of a coronal transient from 1.2 to 6
  solar radii}}.
\newblock {\it \bibinfo{journal}{\jgr}\/},  {\it
  \bibinfo{volume}{90}\/}\bibinfo{issue}{(A1)}, \bibinfo{pages}{275--282}.
  \DOIprefix\doi{10.1029/JA090iA01p00275}.
\bibitem[{{Janvier} et~al.(2014){Janvier}, {D{\'e}moulin} \&
  {Dasso}}]{Janvier2014b}
\bibinfo{author}{{Janvier}, M.}, \bibinfo{author}{{D{\'e}moulin}, P.}, \&
  \bibinfo{author}{{Dasso}, S.} (\bibinfo{year}{2014}).
\newblock \bibinfo{title}{{In situ properties of small and large flux ropes in
  the solar wind}}.
\newblock {\it \bibinfo{journal}{\jgr}\/},  {\it \bibinfo{volume}{119}\/},
  \bibinfo{pages}{7088--7107}. \DOIprefix\doi{10.1002/2014JA020218}.
  \href{http://arxiv.org/abs/1408.5520}{\tt arXiv:1408.5520}.
\bibitem[{{Jian} et~al.(2018){Jian}, {Russell}, {Luhmann} \&
  {Galvin}}]{Jian2018}
\bibinfo{author}{{Jian}, L.~K.}, \bibinfo{author}{{Russell}, C.~T.},
  \bibinfo{author}{{Luhmann}, J.~G.}, \& \bibinfo{author}{{Galvin}, A.~B.}
  (\bibinfo{year}{2018}).
\newblock \bibinfo{title}{{STEREO Observations of Interplanetary Coronal Mass
  Ejections in 2007-2016}}.
\newblock {\it \bibinfo{journal}{\apj}\/},  {\it
  \bibinfo{volume}{855}\/}\bibinfo{issue}{(2)}, \bibinfo{pages}{114}.
  \DOIprefix\doi{10.3847/1538-4357/aab189}.
\bibitem[{{Jing} et~al.(2005){Jing}, {Qiu}, {Lin}, {Qu}, {Xu} \&
  {Wang}}]{Jing2005}
\bibinfo{author}{{Jing}, J.}, \bibinfo{author}{{Qiu}, J.},
  \bibinfo{author}{{Lin}, J.}, \bibinfo{author}{{Qu}, M.},
  \bibinfo{author}{{Xu}, Y.}, \& \bibinfo{author}{{Wang}, H.}
  (\bibinfo{year}{2005}).
\newblock \bibinfo{title}{{Magnetic Reconnection Rate and Flux-Rope
  Acceleration of Two-Ribbon Flares}}.
\newblock {\it \bibinfo{journal}{\apj}\/},  {\it \bibinfo{volume}{620}\/},
  \bibinfo{pages}{1085--1091}. \DOIprefix\doi{10.1086/427165}.
\bibitem[{{Kazachenko} et~al.(2017){Kazachenko}, {Lynch}, {Welsch} \&
  {Sun}}]{Kazachenko2017}
\bibinfo{author}{{Kazachenko}, M.~D.}, \bibinfo{author}{{Lynch}, B.~J.},
  \bibinfo{author}{{Welsch}, B.~T.}, \& \bibinfo{author}{{Sun}, X.}
  (\bibinfo{year}{2017}).
\newblock \bibinfo{title}{{A Database of Flare Ribbon Properties from the Solar
  Dynamics Observatory. I. Reconnection Flux}}.
\newblock {\it \bibinfo{journal}{\apj}\/},  {\it \bibinfo{volume}{845}\/},
  \bibinfo{pages}{49}. \DOIprefix\doi{10.3847/1538-4357/aa7ed6}.
  \href{http://arxiv.org/abs/1704.05097}{\tt arXiv:1704.05097}.
\bibitem[{{Kilpua} et~al.(2017){Kilpua}, {Koskinen} \&
  {Pulkkinen}}]{Kilpua2017}
\bibinfo{author}{{Kilpua}, E.}, \bibinfo{author}{{Koskinen}, H. E.~J.}, \&
  \bibinfo{author}{{Pulkkinen}, T.~I.} (\bibinfo{year}{2017}).
\newblock \bibinfo{title}{{Coronal mass ejections and their sheath regions in
  interplanetary space}}.
\newblock {\it \bibinfo{journal}{\lrsp}\/},  {\it
  \bibinfo{volume}{14}\/}\bibinfo{issue}{(1)}, \bibinfo{pages}{5}.
  \DOIprefix\doi{10.1007/s41116-017-0009-6}.
\bibitem[{{Kilpua} et~al.(2011){Kilpua}, {Jian}, {Li}, {Luhmann} \&
  {Russell}}]{Kilpua2011}
\bibinfo{author}{{Kilpua}, E.~K.~J.}, \bibinfo{author}{{Jian}, L.~K.},
  \bibinfo{author}{{Li}, Y.}, \bibinfo{author}{{Luhmann}, J.~G.}, \&
  \bibinfo{author}{{Russell}, C.~T.} (\bibinfo{year}{2011}).
\newblock \bibinfo{title}{{Multipoint ICME encounters: Pre-STEREO and STEREO
  observations}}.
\newblock {\it \bibinfo{journal}{J. Atmos. Sol. Terr. Phys.}\/},  {\it
  \bibinfo{volume}{73}\/}, \bibinfo{pages}{1228--1241}.
  \DOIprefix\doi{10.1016/j.jastp.2010.10.012}.
\bibitem[{{Kilpua} et~al.(2009){Kilpua}, {Liewer}, {Farrugia}, {Luhmann},
  {M{\"o}stl}, {Li}, {Liu}, {Lynch}, {Russell}, {Vourlidas}, {Acuna}, {Galvin},
  {Larson} \& {Sauvaud}}]{Kilpua2009a}
\bibinfo{author}{{Kilpua}, E.~K.~J.}, \bibinfo{author}{{Liewer}, P.~C.},
  \bibinfo{author}{{Farrugia}, C.}, \bibinfo{author}{{Luhmann}, J.~G.},
  \bibinfo{author}{{M{\"o}stl}, C.}, \bibinfo{author}{{Li}, Y.},
  \bibinfo{author}{{Liu}, Y.}, \bibinfo{author}{{Lynch}, B.~J.},
  \bibinfo{author}{{Russell}, C.~T.}, \bibinfo{author}{{Vourlidas}, A.},
  \bibinfo{author}{{Acuna}, M.~H.}, \bibinfo{author}{{Galvin}, A.~B.},
  \bibinfo{author}{{Larson}, D.}, \& \bibinfo{author}{{Sauvaud}, J.~A.}
  (\bibinfo{year}{2009}).
\newblock \bibinfo{title}{{Multispacecraft Observations of Magnetic Clouds and
  Their Solar Origins between 19 and 23 May 2007}}.
\newblock {\it \bibinfo{journal}{\solphys}\/},  {\it \bibinfo{volume}{254}\/},
  \bibinfo{pages}{325--344}. \DOIprefix\doi{10.1007/s11207-008-9300-y}.
\bibitem[{{Klein} \& {Burlaga}(1982)}]{Klein1982}
\bibinfo{author}{{Klein}, L.~W.}, \& \bibinfo{author}{{Burlaga}, L.~F.}
  (\bibinfo{year}{1982}).
\newblock \bibinfo{title}{{Interplanetary magnetic clouds at 1 AU}}.
\newblock {\it \bibinfo{journal}{\jgr}\/},  {\it
  \bibinfo{volume}{87}\/}\bibinfo{issue}{(A2)}, \bibinfo{pages}{613--624}.
  \DOIprefix\doi{10.1029/JA087iA02p00613}.
\bibitem[{{Knizhnik} et~al.(2015){Knizhnik}, {Antiochos} \&
  {DeVore}}]{knizhnik2015}
\bibinfo{author}{{Knizhnik}, K.~J.}, \bibinfo{author}{{Antiochos}, S.~K.}, \&
  \bibinfo{author}{{DeVore}, C.~R.} (\bibinfo{year}{2015}).
\newblock \bibinfo{title}{{Filament Channel Formation via Magnetic Helicity
  Condensation}}.
\newblock {\it \bibinfo{journal}{\apj}\/},  {\it \bibinfo{volume}{809}\/},
  \bibinfo{pages}{137}. \DOIprefix\doi{10.1088/0004-637X/809/2/137}.
  \href{http://arxiv.org/abs/1411.5396}{\tt arXiv:1411.5396}.
\bibitem[{{Knizhnik} et~al.(2017){Knizhnik}, {Antiochos} \&
  {DeVore}}]{knizhnik2017}
\bibinfo{author}{{Knizhnik}, K.~J.}, \bibinfo{author}{{Antiochos}, S.~K.}, \&
  \bibinfo{author}{{DeVore}, C.~R.} (\bibinfo{year}{2017}).
\newblock \bibinfo{title}{{The Role of Magnetic Helicity in Structuring the
  Solar Corona}}.
\newblock {\it \bibinfo{journal}{\apj}\/},  {\it \bibinfo{volume}{835}\/},
  \bibinfo{pages}{85}. \DOIprefix\doi{10.3847/1538-4357/835/1/85}.
  \href{http://arxiv.org/abs/1607.06756}{\tt arXiv:1607.06756}.
\bibitem[{{Korreck} et~al.(2020){Korreck}, {Szabo}, {Chinchilla}, {Lavraud},
  {Luhmann}, {Niembro}, {Higginson}, {Alzate}, {Wallace}, {Paulson},
  {Rouillard}, {Kouloumvakos}, {Poirier}, {Kasper}, {Case}, {Stevens}, {Bale},
  {Pulupa}, {Whittlesey}, {Livi}, {Goetz}, {Larson}, {Malaspina}, {Morgan},
  {Narock}, {Schwadron}, {Bonnell}, {Harvey} \& {Wygant}}]{Korreck2020}
\bibinfo{author}{{Korreck}, K.~E.}, \bibinfo{author}{{Szabo}, A.},
  \bibinfo{author}{{Chinchilla}, T.~N.}, \bibinfo{author}{{Lavraud}, B.},
  \bibinfo{author}{{Luhmann}, J.}, \bibinfo{author}{{Niembro}, T.},
  \bibinfo{author}{{Higginson}, A.}, \bibinfo{author}{{Alzate}, N.},
  \bibinfo{author}{{Wallace}, S.}, \bibinfo{author}{{Paulson}, K.},
  \bibinfo{author}{{Rouillard}, A.}, \bibinfo{author}{{Kouloumvakos}, A.},
  \bibinfo{author}{{Poirier}, N.}, \bibinfo{author}{{Kasper}, J.~C.},
  \bibinfo{author}{{Case}, A.~W.}, \bibinfo{author}{{Stevens}, M.~L.},
  \bibinfo{author}{{Bale}, S.~D.}, \bibinfo{author}{{Pulupa}, M.},
  \bibinfo{author}{{Whittlesey}, P.}, \bibinfo{author}{{Livi}, R.},
  \bibinfo{author}{{Goetz}, K.}, \bibinfo{author}{{Larson}, D.},
  \bibinfo{author}{{Malaspina}, D.~M.}, \bibinfo{author}{{Morgan}, H.},
  \bibinfo{author}{{Narock}, A.~A.}, \bibinfo{author}{{Schwadron}, N.~A.},
  \bibinfo{author}{{Bonnell}, J.}, \bibinfo{author}{{Harvey}, P.}, \&
  \bibinfo{author}{{Wygant}, J.} (\bibinfo{year}{2020}).
\newblock \bibinfo{title}{{Source and Propagation of a Streamer Blowout Coronal
  Mass Ejection Observed by the Parker Solar Probe}}.
\newblock {\it \bibinfo{journal}{\apjs}\/},  {\it
  \bibinfo{volume}{246}\/}\bibinfo{issue}{(2)}, \bibinfo{pages}{69}.
  \DOIprefix\doi{10.3847/1538-4365/ab6ff9}.
\bibitem[{{Kumar} \& {Rust}(1996)}]{Kumar1996}
\bibinfo{author}{{Kumar}, A.}, \& \bibinfo{author}{{Rust}, D.~M.}
  (\bibinfo{year}{1996}).
\newblock \bibinfo{title}{{Interplanetary magnetic clouds, helicity
  conservation, and current-core flux-ropes}}.
\newblock {\it \bibinfo{journal}{\jgr}\/},  {\it \bibinfo{volume}{101}\/},
  \bibinfo{pages}{15667--15684}. \DOIprefix\doi{10.1029/96JA00544}.
\bibitem[{{Landi} et~al.(2010){Landi}, {Raymond}, {Miralles} \&
  {Hara}}]{Landi2010}
\bibinfo{author}{{Landi}, E.}, \bibinfo{author}{{Raymond}, J.~C.},
  \bibinfo{author}{{Miralles}, M.~P.}, \& \bibinfo{author}{{Hara}, H.}
  (\bibinfo{year}{2010}).
\newblock \bibinfo{title}{{Physical Conditions in a Coronal Mass Ejection from
  Hinode, Stereo, and SOHO Observations}}.
\newblock {\it \bibinfo{journal}{\apj}\/},  {\it \bibinfo{volume}{711}\/},
  \bibinfo{pages}{75--98}. \DOIprefix\doi{10.1088/0004-637X/711/1/75}.
\bibitem[{{Lario} et~al.(2020){Lario}, {Balmaceda}, {Alzate}, {Mays},
  {Richardson}, {Allen}, {Florido-Llinas}, {Nieves-Chinchilla}, {Koval},
  {Lugaz}, {Jian}, {Arge}, {Macneice}, {Odstrcil}, {Morgan}, {Szabo}, {Desai},
  {Whittlesey}, {Stevens}, {Ho} \& {Luhmann}}]{Lario2020}
\bibinfo{author}{{Lario}, D.}, \bibinfo{author}{{Balmaceda}, L.},
  \bibinfo{author}{{Alzate}, N.}, \bibinfo{author}{{Mays}, M.~L.},
  \bibinfo{author}{{Richardson}, I.~G.}, \bibinfo{author}{{Allen}, R.~C.},
  \bibinfo{author}{{Florido-Llinas}, M.}, \bibinfo{author}{{Nieves-Chinchilla},
  T.}, \bibinfo{author}{{Koval}, A.}, \bibinfo{author}{{Lugaz}, N.},
  \bibinfo{author}{{Jian}, L.~K.}, \bibinfo{author}{{Arge}, C.~N.},
  \bibinfo{author}{{Macneice}, P.~J.}, \bibinfo{author}{{Odstrcil}, D.},
  \bibinfo{author}{{Morgan}, H.}, \bibinfo{author}{{Szabo}, A.},
  \bibinfo{author}{{Desai}, M.~I.}, \bibinfo{author}{{Whittlesey}, P.~L.},
  \bibinfo{author}{{Stevens}, M.~L.}, \bibinfo{author}{{Ho}, G.~C.}, \&
  \bibinfo{author}{{Luhmann}, J.~G.} (\bibinfo{year}{2020}).
\newblock \bibinfo{title}{{The Streamer Blowout Origin of a Flux Rope and
  Energetic Particle Event Observed by Parker Solar Probe at 0.5 au}}.
\newblock {\it \bibinfo{journal}{\apj}\/},  {\it
  \bibinfo{volume}{897}\/}\bibinfo{issue}{(2)}, \bibinfo{pages}{134}.
  \DOIprefix\doi{10.3847/1538-4357/ab9942}.
\bibitem[{{Leamon} et~al.(2004){Leamon}, {Canfield}, {Jones}, {Lambkin},
  {Lundberg} \& {Pevtsov}}]{Leamon2004}
\bibinfo{author}{{Leamon}, R.~J.}, \bibinfo{author}{{Canfield}, R.~C.},
  \bibinfo{author}{{Jones}, S.~L.}, \bibinfo{author}{{Lambkin}, K.},
  \bibinfo{author}{{Lundberg}, B.~J.}, \& \bibinfo{author}{{Pevtsov}, A.~A.}
  (\bibinfo{year}{2004}).
\newblock \bibinfo{title}{{Helicity of magnetic clouds and their associated
  active regions}}.
\newblock {\it \bibinfo{journal}{\jgr}\/},  {\it \bibinfo{volume}{109}\/},
  \bibinfo{pages}{5106}. \DOIprefix\doi{10.1029/2003JA010324}.
\bibitem[{{Lee} et~al.(2009){Lee}, {Raymond}, {Ko} \& {Kim}}]{LeeJY2009}
\bibinfo{author}{{Lee}, J.~Y.}, \bibinfo{author}{{Raymond}, J.~C.},
  \bibinfo{author}{{Ko}, Y.~K.}, \& \bibinfo{author}{{Kim}, K.~S.}
  (\bibinfo{year}{2009}).
\newblock \bibinfo{title}{{Three-Dimensional Structure and Energy Balance of a
  Coronal Mass Ejection}}.
\newblock {\it \bibinfo{journal}{\apj}\/},  {\it
  \bibinfo{volume}{692}\/}\bibinfo{issue}{(2)}, \bibinfo{pages}{1271--1286}.
  \DOIprefix\doi{10.1088/0004-637X/692/2/1271}.
  \href{http://arxiv.org/abs/0810.4950}{\tt arXiv:0810.4950}.
\bibitem[{{Lepping} et~al.(2003){Lepping}, {Berdichevsky} \&
  {Ferguson}}]{Lepping2003a}
\bibinfo{author}{{Lepping}, R.~P.}, \bibinfo{author}{{Berdichevsky}, D.~B.}, \&
  \bibinfo{author}{{Ferguson}, T.~J.} (\bibinfo{year}{2003}).
\newblock \bibinfo{title}{{Estimated errors in magnetic cloud model fit
  parameters with force-free cylindrically symmetric assumptions}}.
\newblock {\it \bibinfo{journal}{\jgr}\/},  {\it \bibinfo{volume}{108}\/},
  \bibinfo{pages}{1356}. \DOIprefix\doi{10.1029/2002JA009657}.
\bibitem[{{Lepping} et~al.(1990){Lepping}, {Burlaga} \& {Jones}}]{Lepping1990}
\bibinfo{author}{{Lepping}, R.~P.}, \bibinfo{author}{{Burlaga}, L.~F.}, \&
  \bibinfo{author}{{Jones}, J.~A.} (\bibinfo{year}{1990}).
\newblock \bibinfo{title}{{Magnetic field structure of interplanetary magnetic
  clouds at 1 AU}}.
\newblock {\it \bibinfo{journal}{\jgr}\/},  {\it \bibinfo{volume}{95}\/},
  \bibinfo{pages}{11957--11965}. \DOIprefix\doi{10.1029/JA095iA08p11957}.
\bibitem[{{Li} et~al.(2011){Li}, {Luhmann}, {Lynch} \& {Kilpua}}]{Li2011}
\bibinfo{author}{{Li}, Y.}, \bibinfo{author}{{Luhmann}, J.~G.},
  \bibinfo{author}{{Lynch}, B.~J.}, \& \bibinfo{author}{{Kilpua}, E.~K.~J.}
  (\bibinfo{year}{2011}).
\newblock \bibinfo{title}{{Cyclic Reversal of Magnetic Cloud Poloidal Field}}.
\newblock {\it \bibinfo{journal}{\solphys}\/},  {\it \bibinfo{volume}{270}\/},
  \bibinfo{pages}{331--346}. \DOIprefix\doi{10.1007/s11207-011-9722-9}.
\bibitem[{{Li} et~al.(2014){Li}, {Luhmann}, {Lynch} \& {Kilpua}}]{Li2014}
\bibinfo{author}{{Li}, Y.}, \bibinfo{author}{{Luhmann}, J.~G.},
  \bibinfo{author}{{Lynch}, B.~J.}, \& \bibinfo{author}{{Kilpua}, E.~K.~J.}
  (\bibinfo{year}{2014}).
\newblock \bibinfo{title}{{Magnetic clouds and origins in STEREO era}}.
\newblock {\it \bibinfo{journal}{\jgr}\/},  {\it \bibinfo{volume}{119}\/},
  \bibinfo{pages}{3237--3246}. \DOIprefix\doi{10.1002/2013JA019538}.
\bibitem[{{Liewer} et~al.(2021){Liewer}, {Qiu}, {Vourlidas}, {Hall} \&
  {Penteado}}]{Liewer2021}
\bibinfo{author}{{Liewer}, P.~C.}, \bibinfo{author}{{Qiu}, J.},
  \bibinfo{author}{{Vourlidas}, A.}, \bibinfo{author}{{Hall}, J.~R.}, \&
  \bibinfo{author}{{Penteado}, P.} (\bibinfo{year}{2021}).
\newblock \bibinfo{title}{{Evolution of a streamer-blowout CME as observed by
  imagers on Parker Solar Probe and the Solar Terrestrial Relations
  Observatory}}.
\newblock {\it \bibinfo{journal}{\aap}\/},  {\it \bibinfo{volume}{650}\/},
  \bibinfo{pages}{A32}. \DOIprefix\doi{10.1051/0004-6361/202039641}.
  \href{http://arxiv.org/abs/2012.05174}{\tt arXiv:2012.05174}.
\bibitem[{{Lugaz} et~al.(2018){Lugaz}, {Farrugia}, {Winslow}, {Al-Haddad},
  {Galvin}, {Nieves-Chinchilla}, {Lee} \& {Janvier}}]{Lugaz2018}
\bibinfo{author}{{Lugaz}, N.}, \bibinfo{author}{{Farrugia}, C.~J.},
  \bibinfo{author}{{Winslow}, R.~M.}, \bibinfo{author}{{Al-Haddad}, N.},
  \bibinfo{author}{{Galvin}, A.~B.}, \bibinfo{author}{{Nieves-Chinchilla}, T.},
  \bibinfo{author}{{Lee}, C.~O.}, \& \bibinfo{author}{{Janvier}, M.}
  (\bibinfo{year}{2018}).
\newblock \bibinfo{title}{{On the Spatial Coherence of Magnetic Ejecta:
  Measurements of Coronal Mass Ejections by Multiple Spacecraft Longitudinally
  Separated by 0.01 au}}.
\newblock {\it \bibinfo{journal}{\apjl}\/},  {\it
  \bibinfo{volume}{864}\/}\bibinfo{issue}{(1)}, \bibinfo{pages}{L7}.
  \DOIprefix\doi{10.3847/2041-8213/aad9f4}.
  \href{http://arxiv.org/abs/1812.00911}{\tt arXiv:1812.00911}.
\bibitem[{{Lugaz} et~al.(2020){Lugaz}, {Winslow} \& {Farrugia}}]{Lugaz2020}
\bibinfo{author}{{Lugaz}, N.}, \bibinfo{author}{{Winslow}, R.~M.}, \&
  \bibinfo{author}{{Farrugia}, C.~J.} (\bibinfo{year}{2020}).
\newblock \bibinfo{title}{{Evolution of a Long-Duration Coronal Mass Ejection
  and Its Sheath Region Between Mercury and Earth on 9-14 July 2013}}.
\newblock {\it \bibinfo{journal}{\jgr}\/},  {\it
  \bibinfo{volume}{125}\/}\bibinfo{issue}{(1)}, \bibinfo{pages}{e27213}.
  \DOIprefix\doi{10.1029/2019JA027213}.
  \href{http://arxiv.org/abs/1912.05446}{\tt arXiv:1912.05446}.
\bibitem[{{Luhmann} et~al.(2020){Luhmann}, {Gopalswamy}, {Jian} \&
  {Lugaz}}]{Luhmann2020}
\bibinfo{author}{{Luhmann}, J.~G.}, \bibinfo{author}{{Gopalswamy}, N.},
  \bibinfo{author}{{Jian}, L.~K.}, \& \bibinfo{author}{{Lugaz}, N.}
  (\bibinfo{year}{2020}).
\newblock \bibinfo{title}{{ICME Evolution in the Inner Heliosphere}}.
\newblock {\it \bibinfo{journal}{\solphys}\/},  {\it
  \bibinfo{volume}{295}\/}\bibinfo{issue}{(4)}, \bibinfo{pages}{61}.
  \DOIprefix\doi{10.1007/s11207-020-01624-0}.
\bibitem[{{Lundquist}(1950)}]{Lundquist1950}
\bibinfo{author}{{Lundquist}, S.} (\bibinfo{year}{1950}).
\newblock \bibinfo{title}{Magnetohydrostatic fields}.
\newblock {\it \bibinfo{journal}{Ark. Fys.}\/},  {\it \bibinfo{volume}{2}\/},
  \bibinfo{pages}{361}.
\bibitem[{{Lynch} et~al.(2019){Lynch}, {Airapetian}, {DeVore}, {Kazachenko},
  {L{\"u}ftinger}, {Kochukhov}, {Ros{\'e}n} \& {Abbett}}]{Lynch2019}
\bibinfo{author}{{Lynch}, B.~J.}, \bibinfo{author}{{Airapetian}, V.~S.},
  \bibinfo{author}{{DeVore}, C.~R.}, \bibinfo{author}{{Kazachenko}, M.~D.},
  \bibinfo{author}{{L{\"u}ftinger}, T.}, \bibinfo{author}{{Kochukhov}, O.},
  \bibinfo{author}{{Ros{\'e}n}, L.}, \& \bibinfo{author}{{Abbett}, W.~P.}
  (\bibinfo{year}{2019}).
\newblock \bibinfo{title}{{Modeling a Carrington-scale Stellar Superflare and
  Coronal Mass Ejection from $\kappa^{1}{Cet}$}}.
\newblock {\it \bibinfo{journal}{\apj}\/},  {\it
  \bibinfo{volume}{880}\/}\bibinfo{issue}{(2)}, \bibinfo{pages}{97}.
  \DOIprefix\doi{10.3847/1538-4357/ab287e}.
  \href{http://arxiv.org/abs/1906.03189}{\tt arXiv:1906.03189}.
\bibitem[{{Lynch} et~al.(2004){Lynch}, {Antiochos}, {MacNeice}, {Zurbuchen} \&
  {Fisk}}]{Lynch2004}
\bibinfo{author}{{Lynch}, B.~J.}, \bibinfo{author}{{Antiochos}, S.~K.},
  \bibinfo{author}{{MacNeice}, P.~J.}, \bibinfo{author}{{Zurbuchen}, T.~H.}, \&
  \bibinfo{author}{{Fisk}, L.~A.} (\bibinfo{year}{2004}).
\newblock \bibinfo{title}{{Observable Properties of the Breakout Model for
  Coronal Mass Ejections}}.
\newblock {\it \bibinfo{journal}{\apj}\/},  {\it \bibinfo{volume}{617}\/},
  \bibinfo{pages}{589--599}. \DOIprefix\doi{10.1086/424564}.
\bibitem[{{Lynch} et~al.(2005){Lynch}, {Gruesbeck}, {Zurbuchen} \&
  {Antiochos}}]{Lynch2005}
\bibinfo{author}{{Lynch}, B.~J.}, \bibinfo{author}{{Gruesbeck}, J.~R.},
  \bibinfo{author}{{Zurbuchen}, T.~H.}, \& \bibinfo{author}{{Antiochos}, S.~K.}
  (\bibinfo{year}{2005}).
\newblock \bibinfo{title}{{Solar cycle-dependent helicity transport by magnetic
  clouds}}.
\newblock {\it \bibinfo{journal}{\jgr}\/},  {\it \bibinfo{volume}{110}\/},
  \bibinfo{pages}{8107}. \DOIprefix\doi{10.1029/2005JA011137}.
\bibitem[{{Lynch} et~al.(2016){Lynch}, {Masson}, {Li}, {Devore}, {Luhmann},
  {Antiochos} \& {Fisher}}]{lynch2016b}
\bibinfo{author}{{Lynch}, B.~J.}, \bibinfo{author}{{Masson}, S.},
  \bibinfo{author}{{Li}, Y.}, \bibinfo{author}{{Devore}, C.~R.},
  \bibinfo{author}{{Luhmann}, J.~G.}, \bibinfo{author}{{Antiochos}, S.~K.}, \&
  \bibinfo{author}{{Fisher}, G.~H.} (\bibinfo{year}{2016}).
\newblock \bibinfo{title}{{A model for stealth coronal mass ejections}}.
\newblock {\it \bibinfo{journal}{\jgr}\/},  {\it \bibinfo{volume}{121}\/},
  \bibinfo{pages}{10677}. \DOIprefix\doi{10.1002/2016JA023432}.
\bibitem[{{Lynch} et~al.(2021){Lynch}, {Palmerio}, {DeVore}, {Kazachenko},
  {Dahlin}, {Pomoell} \& {Kilpua}}]{lynch2021}
\bibinfo{author}{{Lynch}, B.~J.}, \bibinfo{author}{{Palmerio}, E.},
  \bibinfo{author}{{DeVore}, C.~R.}, \bibinfo{author}{{Kazachenko}, M.~D.},
  \bibinfo{author}{{Dahlin}, J.~T.}, \bibinfo{author}{{Pomoell}, J.}, \&
  \bibinfo{author}{{Kilpua}, E. K.~J.} (\bibinfo{year}{2021}).
\newblock \bibinfo{title}{{Modeling a Coronal Mass Ejection from an Extended
  Filament Channel. I. Eruption and Early Evolution}}.
\newblock {\it \bibinfo{journal}{\apj}\/},  {\it
  \bibinfo{volume}{914}\/}\bibinfo{issue}{(1)}, \bibinfo{pages}{39}.
  \DOIprefix\doi{10.3847/1538-4357/abf9a9}.
  \href{http://arxiv.org/abs/2104.08643}{\tt arXiv:2104.08643}.
\bibitem[{{Lynch} et~al.(2003){Lynch}, {Zurbuchen}, {Fisk} \&
  {Antiochos}}]{Lynch2003}
\bibinfo{author}{{Lynch}, B.~J.}, \bibinfo{author}{{Zurbuchen}, T.~H.},
  \bibinfo{author}{{Fisk}, L.~A.}, \& \bibinfo{author}{{Antiochos}, S.~K.}
  (\bibinfo{year}{2003}).
\newblock \bibinfo{title}{{Internal structure of magnetic clouds: Plasma and
  composition}}.
\newblock {\it \bibinfo{journal}{\jgr}\/},  {\it \bibinfo{volume}{108}\/},
  \bibinfo{pages}{1239}. \DOIprefix\doi{10.1029/2002JA009591}.
\bibitem[{{MacNeice} et~al.(2000){MacNeice}, {Olson}, {Mobarry}, {de
  Fainchtein} \& {Packer}}]{MacNeice2000}
\bibinfo{author}{{MacNeice}, P.}, \bibinfo{author}{{Olson}, K.~M.},
  \bibinfo{author}{{Mobarry}, C.}, \bibinfo{author}{{de Fainchtein}, R.}, \&
  \bibinfo{author}{{Packer}, C.} (\bibinfo{year}{2000}).
\newblock \bibinfo{title}{{PARAMESH: A parallel adaptive mesh refinement
  community toolkit}}.
\newblock {\it \bibinfo{journal}{Comp. Phys. Comm.}\/},  {\it
  \bibinfo{volume}{126}\/}, \bibinfo{pages}{330--354}.
  \DOIprefix\doi{10.1016/S0010-4655(99)00501-9}.
\bibitem[{{Manchester} et~al.(2017){Manchester}, {Kilpua}, {Liu}, {Lugaz},
  {Riley}, {T{\"o}r{\"o}k} \& {Vr{\v s}nak}}]{Manchester2017}
\bibinfo{author}{{Manchester}, W.}, \bibinfo{author}{{Kilpua}, E.~K.~J.},
  \bibinfo{author}{{Liu}, Y.~D.}, \bibinfo{author}{{Lugaz}, N.},
  \bibinfo{author}{{Riley}, P.}, \bibinfo{author}{{T{\"o}r{\"o}k}, T.}, \&
  \bibinfo{author}{{Vr{\v s}nak}, B.} (\bibinfo{year}{2017}).
\newblock \bibinfo{title}{{The Physical Processes of CME/ICME Evolution}}.
\newblock {\it \bibinfo{journal}{\ssr}\/},  {\it \bibinfo{volume}{212}\/},
  \bibinfo{pages}{1159--1219}. \DOIprefix\doi{10.1007/s11214-017-0394-0}.
\bibitem[{{Marubashi}(1986)}]{Marubashi1986}
\bibinfo{author}{{Marubashi}, K.} (\bibinfo{year}{1986}).
\newblock \bibinfo{title}{{Structure of the interplanetary magnetic clouds and
  their solar origins}}.
\newblock {\it \bibinfo{journal}{\adv}\/},  {\it \bibinfo{volume}{6}\/},
  \bibinfo{pages}{335--338}. \DOIprefix\doi{10.1016/0273-1177(86)90172-9}.
\bibitem[{{Marubashi}(1997)}]{Marubashi1997}
\bibinfo{author}{{Marubashi}, K.} (\bibinfo{year}{1997}).
\newblock \bibinfo{title}{{Interplanetary magnetic flux ropes and solar
  filaments}}.
\newblock {\it \bibinfo{journal}{Washington DC American Geophysical Union
  Geophysical Monograph Series}\/},  {\it \bibinfo{volume}{99}\/},
  \bibinfo{pages}{147--156}. \DOIprefix\doi{10.1029/GM099p0147}.
\bibitem[{{M{\"o}stl} et~al.(2009){M{\"o}stl}, {Farrugia}, {Biernat},
  {Leitner}, {Kilpua}, {Galvin} \& {Luhmann}}]{Moestl2009b}
\bibinfo{author}{{M{\"o}stl}, C.}, \bibinfo{author}{{Farrugia}, C.~J.},
  \bibinfo{author}{{Biernat}, H.~K.}, \bibinfo{author}{{Leitner}, M.},
  \bibinfo{author}{{Kilpua}, E.~K.~J.}, \bibinfo{author}{{Galvin}, A.~B.}, \&
  \bibinfo{author}{{Luhmann}, J.~G.} (\bibinfo{year}{2009}).
\newblock \bibinfo{title}{{Optimized Grad - Shafranov Reconstruction of a
  Magnetic Cloud Using STEREO- Wind Observations}}.
\newblock {\it \bibinfo{journal}{\solphys}\/},  {\it
  \bibinfo{volume}{256}\/}\bibinfo{issue}{(1-2)}, \bibinfo{pages}{427--441}.
  \DOIprefix\doi{10.1007/s11207-009-9360-7}.
\bibitem[{{M{\"o}stl} et~al.(2008){M{\"o}stl}, {Miklenic}, {Farrugia},
  {Temmer}, {Veronig}, {Galvin}, {Vr{\v{s}}nak} \& {Biernat}}]{Moestl2008}
\bibinfo{author}{{M{\"o}stl}, C.}, \bibinfo{author}{{Miklenic}, C.},
  \bibinfo{author}{{Farrugia}, C.~J.}, \bibinfo{author}{{Temmer}, M.},
  \bibinfo{author}{{Veronig}, A.}, \bibinfo{author}{{Galvin}, A.~B.},
  \bibinfo{author}{{Vr{\v{s}}nak}, B.}, \& \bibinfo{author}{{Biernat}, H.~K.}
  (\bibinfo{year}{2008}).
\newblock \bibinfo{title}{{Two-spacecraft reconstruction of a magnetic cloud
  and comparison to its solar source}}.
\newblock {\it \bibinfo{journal}{\angeo}\/},  {\it
  \bibinfo{volume}{26}\/}\bibinfo{issue}{(10)}, \bibinfo{pages}{3139--3152}.
  \DOIprefix\doi{10.5194/angeo-26-3139-2008}.
\bibitem[{{M{\"o}stl} et~al.(2020){M{\"o}stl}, {Weiss}, {Bailey}, {Reiss},
  {Amerstorfer}, {Hinterreiter}, {Bauer}, {McIntosh}, {Lugaz} \&
  {Stansby}}]{Moestl2020}
\bibinfo{author}{{M{\"o}stl}, C.}, \bibinfo{author}{{Weiss}, A.~J.},
  \bibinfo{author}{{Bailey}, R.~L.}, \bibinfo{author}{{Reiss}, M.~A.},
  \bibinfo{author}{{Amerstorfer}, T.}, \bibinfo{author}{{Hinterreiter}, J.},
  \bibinfo{author}{{Bauer}, M.}, \bibinfo{author}{{McIntosh}, S.~W.},
  \bibinfo{author}{{Lugaz}, N.}, \& \bibinfo{author}{{Stansby}, D.}
  (\bibinfo{year}{2020}).
\newblock \bibinfo{title}{{Prediction of the In Situ Coronal Mass Ejection Rate
  for Solar Cycle 25: Implications for Parker Solar Probe In Situ
  Observations}}.
\newblock {\it \bibinfo{journal}{\apj}\/},  {\it
  \bibinfo{volume}{903}\/}\bibinfo{issue}{(2)}, \bibinfo{pages}{92}.
  \DOIprefix\doi{10.3847/1538-4357/abb9a1}.
  \href{http://arxiv.org/abs/2007.14743}{\tt arXiv:2007.14743}.
\bibitem[{{M{\"o}stl} et~al.(2022){M{\"o}stl}, {Weiss}, {Reiss}, {Amerstorfer},
  {Bailey}, {Hinterreiter}, {Bauer}, {Barnes}, {Davies}, {Harrison}, {Freiherr
  von Forstner}, {Davies}, {Heyner}, {Horbury} \& {Bale}}]{Moestl2022}
\bibinfo{author}{{M{\"o}stl}, C.}, \bibinfo{author}{{Weiss}, A.~J.},
  \bibinfo{author}{{Reiss}, M.~A.}, \bibinfo{author}{{Amerstorfer}, T.},
  \bibinfo{author}{{Bailey}, R.~L.}, \bibinfo{author}{{Hinterreiter}, J.},
  \bibinfo{author}{{Bauer}, M.}, \bibinfo{author}{{Barnes}, D.},
  \bibinfo{author}{{Davies}, J.~A.}, \bibinfo{author}{{Harrison}, R.~A.},
  \bibinfo{author}{{Freiherr von Forstner}, J.~L.}, \bibinfo{author}{{Davies},
  E.~E.}, \bibinfo{author}{{Heyner}, D.}, \bibinfo{author}{{Horbury}, T.}, \&
  \bibinfo{author}{{Bale}, S.~D.} (\bibinfo{year}{2022}).
\newblock \bibinfo{title}{{Multipoint Interplanetary Coronal Mass Ejections
  Observed with Solar Orbiter, BepiColombo, Parker Solar Probe, Wind, and
  STEREO-A}}.
\newblock {\it \bibinfo{journal}{\apjl}\/},  {\it
  \bibinfo{volume}{924}\/}\bibinfo{issue}{(1)}, \bibinfo{pages}{L6}.
  \DOIprefix\doi{10.3847/2041-8213/ac42d0}.
  \href{http://arxiv.org/abs/2109.07200}{\tt arXiv:2109.07200}.
\bibitem[{{Mulligan} et~al.(2013){Mulligan}, {Reinard} \&
  {Lynch}}]{Mulligan2013}
\bibinfo{author}{{Mulligan}, T.}, \bibinfo{author}{{Reinard}, A.~A.}, \&
  \bibinfo{author}{{Lynch}, B.~J.} (\bibinfo{year}{2013}).
\newblock \bibinfo{title}{{Advancing in situ modeling of ICMEs: New techniques
  for new observations}}.
\newblock {\it \bibinfo{journal}{\jgr}\/},  {\it \bibinfo{volume}{118}\/},
  \bibinfo{pages}{1410--1427}. \DOIprefix\doi{10.1002/jgra.50101}.
  \href{http://arxiv.org/abs/1208.5107}{\tt arXiv:1208.5107}.
\bibitem[{{Mulligan} et~al.(1998){Mulligan}, {Russell} \&
  {Luhmann}}]{Mulligan1998}
\bibinfo{author}{{Mulligan}, T.}, \bibinfo{author}{{Russell}, C.~T.}, \&
  \bibinfo{author}{{Luhmann}, J.~G.} (\bibinfo{year}{1998}).
\newblock \bibinfo{title}{{Solar cycle evolution of the structure of magnetic
  clouds in the inner heliosphere}}.
\newblock {\it \bibinfo{journal}{\grl}\/},  {\it \bibinfo{volume}{25}\/},
  \bibinfo{pages}{2959--2962}. \DOIprefix\doi{10.1029/98GL01302}.
\bibitem[{{Nieves-Chinchilla} et~al.(2019){Nieves-Chinchilla}, {Jian},
  {Balmaceda}, {Vourlidas}, {dos Santos} \& {Szabo}}]{Nieves-Chinchilla2019}
\bibinfo{author}{{Nieves-Chinchilla}, T.}, \bibinfo{author}{{Jian}, L.~K.},
  \bibinfo{author}{{Balmaceda}, L.}, \bibinfo{author}{{Vourlidas}, A.},
  \bibinfo{author}{{dos Santos}, L. F.~G.}, \& \bibinfo{author}{{Szabo}, A.}
  (\bibinfo{year}{2019}).
\newblock \bibinfo{title}{{Unraveling the Internal Magnetic Field Structure of
  the Earth-directed Interplanetary Coronal Mass Ejections During 1995 -
  2015}}.
\newblock {\it \bibinfo{journal}{\solphys}\/},  {\it
  \bibinfo{volume}{294}\/}\bibinfo{issue}{(7)}, \bibinfo{pages}{89}.
  \DOIprefix\doi{10.1007/s11207-019-1477-8}.
\bibitem[{{Nieves-Chinchilla} et~al.(2016){Nieves-Chinchilla}, {Linton},
  {Hidalgo}, {Vourlidas}, {Savani}, {Szabo}, {Farrugia} \&
  {Yu}}]{Nieves-Chinchilla2016}
\bibinfo{author}{{Nieves-Chinchilla}, T.}, \bibinfo{author}{{Linton}, M.~G.},
  \bibinfo{author}{{Hidalgo}, M.~A.}, \bibinfo{author}{{Vourlidas}, A.},
  \bibinfo{author}{{Savani}, N.~P.}, \bibinfo{author}{{Szabo}, A.},
  \bibinfo{author}{{Farrugia}, C.}, \& \bibinfo{author}{{Yu}, W.}
  (\bibinfo{year}{2016}).
\newblock \bibinfo{title}{{A Circular-cylindrical Flux-rope Analytical Model
  for Magnetic Clouds}}.
\newblock {\it \bibinfo{journal}{\apj}\/},  {\it
  \bibinfo{volume}{823}\/}\bibinfo{issue}{(1)}, \bibinfo{pages}{27}.
  \DOIprefix\doi{10.3847/0004-637X/823/1/27}.
\bibitem[{{Nieves-Chinchilla} et~al.(2020){Nieves-Chinchilla}, {Szabo},
  {Korreck}, {Alzate}, {Balmaceda}, {Lavraud}, {Paulson}, {Narock}, {Wallace},
  {Jian}, {Luhmann}, {Morgan}, {Higginson}, {Arge}, {Bale}, {Case}, {Wit},
  {Giacalone}, {Goetz}, {Harvey}, {Jones-Melosky}, {Kasper}, {Larson}, {Livi},
  {McComas}, {MacDowall}, {Malaspina}, {Pulupa}, {Raouafi}, {Schwadron},
  {Stevens} \& {Whittlesey}}]{Nieves-Chinchilla2020}
\bibinfo{author}{{Nieves-Chinchilla}, T.}, \bibinfo{author}{{Szabo}, A.},
  \bibinfo{author}{{Korreck}, K.~E.}, \bibinfo{author}{{Alzate}, N.},
  \bibinfo{author}{{Balmaceda}, L.~A.}, \bibinfo{author}{{Lavraud}, B.},
  \bibinfo{author}{{Paulson}, K.}, \bibinfo{author}{{Narock}, A.~A.},
  \bibinfo{author}{{Wallace}, S.}, \bibinfo{author}{{Jian}, L.~K.},
  \bibinfo{author}{{Luhmann}, J.~G.}, \bibinfo{author}{{Morgan}, H.},
  \bibinfo{author}{{Higginson}, A.}, \bibinfo{author}{{Arge}, C.~N.},
  \bibinfo{author}{{Bale}, S.~D.}, \bibinfo{author}{{Case}, A.~W.},
  \bibinfo{author}{{Wit}, T. D.~d.}, \bibinfo{author}{{Giacalone}, J.},
  \bibinfo{author}{{Goetz}, K.}, \bibinfo{author}{{Harvey}, P.~R.},
  \bibinfo{author}{{Jones-Melosky}, S.~I.}, \bibinfo{author}{{Kasper}, J.~C.},
  \bibinfo{author}{{Larson}, D.~E.}, \bibinfo{author}{{Livi}, R.},
  \bibinfo{author}{{McComas}, D.~J.}, \bibinfo{author}{{MacDowall}, R.~J.},
  \bibinfo{author}{{Malaspina}, D.~M.}, \bibinfo{author}{{Pulupa}, M.},
  \bibinfo{author}{{Raouafi}, N.~E.}, \bibinfo{author}{{Schwadron}, N.},
  \bibinfo{author}{{Stevens}, M.~L.}, \& \bibinfo{author}{{Whittlesey}, P.~L.}
  (\bibinfo{year}{2020}).
\newblock \bibinfo{title}{{Analysis of the Internal Structure of the Streamer
  Blowout Observed by the Parker Solar Probe During the First Solar
  Encounter}}.
\newblock {\it \bibinfo{journal}{\apjs}\/},  {\it
  \bibinfo{volume}{246}\/}\bibinfo{issue}{(2)}, \bibinfo{pages}{63}.
  \DOIprefix\doi{10.3847/1538-4365/ab61f5}.
\bibitem[{{Nieves-Chinchilla} et~al.(2018){Nieves-Chinchilla}, {Vourlidas},
  {Raymond}, {Linton}, {Al-haddad}, {Savani}, {Szabo} \&
  {Hidalgo}}]{Nieves-Chinchilla2018a}
\bibinfo{author}{{Nieves-Chinchilla}, T.}, \bibinfo{author}{{Vourlidas}, A.},
  \bibinfo{author}{{Raymond}, J.~C.}, \bibinfo{author}{{Linton}, M.~G.},
  \bibinfo{author}{{Al-haddad}, N.}, \bibinfo{author}{{Savani}, N.~P.},
  \bibinfo{author}{{Szabo}, A.}, \& \bibinfo{author}{{Hidalgo}, M.~A.}
  (\bibinfo{year}{2018}).
\newblock \bibinfo{title}{{Understanding the Internal Magnetic Field
  Configurations of ICMEs Using More than 20 Years of Wind Observations}}.
\newblock {\it \bibinfo{journal}{\solphys}\/},  {\it \bibinfo{volume}{293}\/},
  \bibinfo{pages}{25}. \DOIprefix\doi{10.1007/s11207-018-1247-z}.
\bibitem[{{Owens}(2008)}]{Owens2008}
\bibinfo{author}{{Owens}, M.~J.} (\bibinfo{year}{2008}).
\newblock \bibinfo{title}{{Combining remote and in situ observations of coronal
  mass ejections to better constrain magnetic cloud reconstruction}}.
\newblock {\it \bibinfo{journal}{\jgr}\/},  {\it
  \bibinfo{volume}{113}\/}\bibinfo{issue}{(A12)}, \bibinfo{pages}{A12102}.
  \DOIprefix\doi{10.1029/2008JA013589}.
\bibitem[{{Owens} et~al.(2012){Owens}, {D{\'e}moulin}, {Savani}, {Lavraud} \&
  {Ruffenach}}]{Owens2012}
\bibinfo{author}{{Owens}, M.~J.}, \bibinfo{author}{{D{\'e}moulin}, P.},
  \bibinfo{author}{{Savani}, N.~P.}, \bibinfo{author}{{Lavraud}, B.}, \&
  \bibinfo{author}{{Ruffenach}, A.} (\bibinfo{year}{2012}).
\newblock \bibinfo{title}{{Implications of Non-cylindrical Flux Ropes for
  Magnetic Cloud Reconstruction Techniques and the Interpretation of Double
  Flux Rope Events}}.
\newblock {\it \bibinfo{journal}{\solphys}\/},  {\it
  \bibinfo{volume}{278}\/}\bibinfo{issue}{(2)}, \bibinfo{pages}{435--446}.
  \DOIprefix\doi{10.1007/s11207-012-9939-2}.
\bibitem[{{Owens} et~al.(2006){Owens}, {Merkin} \& {Riley}}]{Owens2006a}
\bibinfo{author}{{Owens}, M.~J.}, \bibinfo{author}{{Merkin}, V.~G.}, \&
  \bibinfo{author}{{Riley}, P.} (\bibinfo{year}{2006}).
\newblock \bibinfo{title}{{A kinematically distorted flux rope model for
  magnetic clouds}}.
\newblock {\it \bibinfo{journal}{\jgr}\/},  {\it \bibinfo{volume}{111}\/},
  \bibinfo{pages}{3104}. \DOIprefix\doi{10.1029/2005JA011460}.
\bibitem[{{Pal} et~al.(2021){Pal}, {Kilpua}, {Good}, {Pomoell} \&
  {Price}}]{Pal2021}
\bibinfo{author}{{Pal}, S.}, \bibinfo{author}{{Kilpua}, E.},
  \bibinfo{author}{{Good}, S.}, \bibinfo{author}{{Pomoell}, J.}, \&
  \bibinfo{author}{{Price}, D.~J.} (\bibinfo{year}{2021}).
\newblock \bibinfo{title}{{Uncovering erosion effects on magnetic flux rope
  twist}}.
\newblock {\it \bibinfo{journal}{\aap}\/},  {\it \bibinfo{volume}{650}\/},
  \bibinfo{pages}{A176}. \DOIprefix\doi{10.1051/0004-6361/202040070}.
  \href{http://arxiv.org/abs/2104.03569}{\tt arXiv:2104.03569}.
\bibitem[{{Palmerio} et~al.(2021){Palmerio}, {Kay}, {Al-Haddad}, {Lynch}, {Yu},
  {Stevens}, {Pal} \& {Lee}}]{Palmerio2021c}
\bibinfo{author}{{Palmerio}, E.}, \bibinfo{author}{{Kay}, C.},
  \bibinfo{author}{{Al-Haddad}, N.}, \bibinfo{author}{{Lynch}, B.~J.},
  \bibinfo{author}{{Yu}, W.}, \bibinfo{author}{{Stevens}, M.~L.},
  \bibinfo{author}{{Pal}, S.}, \& \bibinfo{author}{{Lee}, C.~O.}
  (\bibinfo{year}{2021}).
\newblock \bibinfo{title}{{Predicting the Magnetic Fields of a Stealth CME
  Detected by Parker Solar Probe at 0.5 au}}.
\newblock {\it \bibinfo{journal}{\apj}\/},  {\it
  \bibinfo{volume}{920}\/}\bibinfo{issue}{(2)}, \bibinfo{pages}{65}.
  \DOIprefix\doi{10.3847/1538-4357/ac25f4}.
  \href{http://arxiv.org/abs/2109.04933}{\tt arXiv:2109.04933}.
\bibitem[{{Palmerio} et~al.(2017){Palmerio}, {Kilpua}, {James}, {Green},
  {Pomoell}, {Isavnin} \& {Valori}}]{Palmerio2017}
\bibinfo{author}{{Palmerio}, E.}, \bibinfo{author}{{Kilpua}, E.~K.~J.},
  \bibinfo{author}{{James}, A.~W.}, \bibinfo{author}{{Green}, L.~M.},
  \bibinfo{author}{{Pomoell}, J.}, \bibinfo{author}{{Isavnin}, A.}, \&
  \bibinfo{author}{{Valori}, G.} (\bibinfo{year}{2017}).
\newblock \bibinfo{title}{{Determining the Intrinsic CME Flux Rope Type Using
  Remote-sensing Solar Disk Observations}}.
\newblock {\it \bibinfo{journal}{\solphys}\/},  {\it \bibinfo{volume}{292}\/},
  \bibinfo{pages}{39}. \DOIprefix\doi{10.1007/s11207-017-1063-x}.
  \href{http://arxiv.org/abs/1701.08595}{\tt arXiv:1701.08595}.
\bibitem[{{Palmerio} et~al.(2018){Palmerio}, {Kilpua}, {M{\"o}stl}, {Bothmer},
  {James}, {Green}, {Isavnin}, {Davies} \& {Harrison}}]{Palmerio2018}
\bibinfo{author}{{Palmerio}, E.}, \bibinfo{author}{{Kilpua}, E.~K.~J.},
  \bibinfo{author}{{M{\"o}stl}, C.}, \bibinfo{author}{{Bothmer}, V.},
  \bibinfo{author}{{James}, A.~W.}, \bibinfo{author}{{Green}, L.~M.},
  \bibinfo{author}{{Isavnin}, A.}, \bibinfo{author}{{Davies}, J.~A.}, \&
  \bibinfo{author}{{Harrison}, R.~A.} (\bibinfo{year}{2018}).
\newblock \bibinfo{title}{{Coronal Magnetic Structure of Earthbound CMEs and In
  Situ Comparison}}.
\newblock {\it \bibinfo{journal}{Space Weather}\/},  {\it
  \bibinfo{volume}{16}\/}, \bibinfo{pages}{442--460}.
  \DOIprefix\doi{10.1002/2017SW001767}.
  \href{http://arxiv.org/abs/1803.04769}{\tt arXiv:1803.04769}.
\bibitem[{{Pariat} et~al.(2015){Pariat}, {Dalmasse}, {DeVore}, {Antiochos} \&
  {Karpen}}]{pariat2015}
\bibinfo{author}{{Pariat}, E.}, \bibinfo{author}{{Dalmasse}, K.},
  \bibinfo{author}{{DeVore}, C.~R.}, \bibinfo{author}{{Antiochos}, S.~K.}, \&
  \bibinfo{author}{{Karpen}, J.~T.} (\bibinfo{year}{2015}).
\newblock \bibinfo{title}{{Model for straight and helical solar jets. I.
  Parametric studies of the magnetic field geometry}}.
\newblock {\it \bibinfo{journal}{\aap}\/},  {\it \bibinfo{volume}{573}\/},
  \bibinfo{pages}{A130}. \DOIprefix\doi{10.1051/0004-6361/201424209}.
\bibitem[{{Qiu} et~al.(2007){Qiu}, {Hu}, {Howard} \& {Yurchyshyn}}]{Qiu2007}
\bibinfo{author}{{Qiu}, J.}, \bibinfo{author}{{Hu}, Q.},
  \bibinfo{author}{{Howard}, T.~A.}, \& \bibinfo{author}{{Yurchyshyn}, V.~B.}
  (\bibinfo{year}{2007}).
\newblock \bibinfo{title}{{On the Magnetic Flux Budget in Low-Corona Magnetic
  Reconnection and Interplanetary Coronal Mass Ejections}}.
\newblock {\it \bibinfo{journal}{\apj}\/},  {\it \bibinfo{volume}{659}\/},
  \bibinfo{pages}{758--772}. \DOIprefix\doi{10.1086/512060}.
\bibitem[{{Qiu} et~al.(2010){Qiu}, {Liu}, {Hill} \& {Kazachenko}}]{Qiu2010}
\bibinfo{author}{{Qiu}, J.}, \bibinfo{author}{{Liu}, W.},
  \bibinfo{author}{{Hill}, N.}, \& \bibinfo{author}{{Kazachenko}, M.}
  (\bibinfo{year}{2010}).
\newblock \bibinfo{title}{{Reconnection and Energetics in Two-ribbon Flares: A
  Revisit of the Bastille-day Flare}}.
\newblock {\it \bibinfo{journal}{\apj}\/},  {\it \bibinfo{volume}{725}\/},
  \bibinfo{pages}{319--330}. \DOIprefix\doi{10.1088/0004-637X/725/1/319}.
\bibitem[{{Rakowski} et~al.(2007){Rakowski}, {Laming} \&
  {Lepri}}]{Rakowski2007}
\bibinfo{author}{{Rakowski}, C.~E.}, \bibinfo{author}{{Laming}, J.~M.}, \&
  \bibinfo{author}{{Lepri}, S.~T.} (\bibinfo{year}{2007}).
\newblock \bibinfo{title}{{Ion Charge States in Halo Coronal Mass Ejections:
  What Can We Learn about the Explosion?}}
\newblock {\it \bibinfo{journal}{\apj}\/},  {\it \bibinfo{volume}{667}\/},
  \bibinfo{pages}{602--609}. \DOIprefix\doi{10.1086/520914}.
  \href{http://arxiv.org/abs/0706.3395}{\tt arXiv:0706.3395}.
\bibitem[{{Richardson} \& {Cane}(2010)}]{Richardson2010}
\bibinfo{author}{{Richardson}, I.~G.}, \& \bibinfo{author}{{Cane}, H.~V.}
  (\bibinfo{year}{2010}).
\newblock \bibinfo{title}{{Near-Earth Interplanetary Coronal Mass Ejections
  During Solar Cycle 23 (1996 - 2009): Catalog and Summary of Properties}}.
\newblock {\it \bibinfo{journal}{\solphys}\/},  {\it \bibinfo{volume}{264}\/},
  \bibinfo{pages}{189--237}. \DOIprefix\doi{10.1007/s11207-010-9568-6}.
\bibitem[{{Riley} et~al.(2004){Riley}, {Linker}, {Lionello}, {Miki{\'c}},
  {Odstrcil}, {Hidalgo}, {Cid}, {Hu}, {Lepping}, {Lynch} \& {Rees}}]{Riley2004}
\bibinfo{author}{{Riley}, P.}, \bibinfo{author}{{Linker}, J.~A.},
  \bibinfo{author}{{Lionello}, R.}, \bibinfo{author}{{Miki{\'c}}, Z.},
  \bibinfo{author}{{Odstrcil}, D.}, \bibinfo{author}{{Hidalgo}, M.~A.},
  \bibinfo{author}{{Cid}, C.}, \bibinfo{author}{{Hu}, Q.},
  \bibinfo{author}{{Lepping}, R.~P.}, \bibinfo{author}{{Lynch}, B.~J.}, \&
  \bibinfo{author}{{Rees}, A.} (\bibinfo{year}{2004}).
\newblock \bibinfo{title}{{Fitting flux ropes to a global MHD solution: a
  comparison of techniques}}.
\newblock {\it \bibinfo{journal}{J. Atmos. Sol. Terr. Phys.}\/},  {\it
  \bibinfo{volume}{66}\/}, \bibinfo{pages}{1321--1331}.
  \DOIprefix\doi{10.1016/j.jastp.2004.03.019}.
\bibitem[{{Rosa Oliveira} et~al.(2020){Rosa Oliveira}, {da Silva Oliveira},
  {Ojeda-Gonz{\'a}lez} \& {De La Luz}}]{RosaOliveira2020}
\bibinfo{author}{{Rosa Oliveira}, R.~A.}, \bibinfo{author}{{da Silva Oliveira},
  M.~W.}, \bibinfo{author}{{Ojeda-Gonz{\'a}lez}, A.}, \& \bibinfo{author}{{De
  La Luz}, V.} (\bibinfo{year}{2020}).
\newblock \bibinfo{title}{{New Metric for Minimum Variance Analysis Validation
  in the Study of Interplanetary Magnetic Clouds}}.
\newblock {\it \bibinfo{journal}{\solphys}\/},  {\it
  \bibinfo{volume}{295}\/}\bibinfo{issue}{(3)}, \bibinfo{pages}{45}.
  \DOIprefix\doi{10.1007/s11207-020-01610-6}.
\bibitem[{{Rosa Oliveira} et~al.(2021){Rosa Oliveira}, {da Silva Oliveira},
  {Ojeda-Gonz{\'a}lez}, {Gil Pillat}, {Echer} \&
  {Nieves-Chinchilla}}]{RosaOliveira2021}
\bibinfo{author}{{Rosa Oliveira}, R.~A.}, \bibinfo{author}{{da Silva Oliveira},
  M.~W.}, \bibinfo{author}{{Ojeda-Gonz{\'a}lez}, A.}, \bibinfo{author}{{Gil
  Pillat}, V.}, \bibinfo{author}{{Echer}, E.}, \&
  \bibinfo{author}{{Nieves-Chinchilla}, T.} (\bibinfo{year}{2021}).
\newblock \bibinfo{title}{{Resolving the Ambiguity of a Magnetic Cloud's
  Orientation Caused by Minimum Variance Analysis Comparing it with a
  Force-Free Model}}.
\newblock {\it \bibinfo{journal}{\solphys}\/},  {\it
  \bibinfo{volume}{296}\/}\bibinfo{issue}{(12)}, \bibinfo{pages}{182}.
  \DOIprefix\doi{10.1007/s11207-021-01921-2}.
\bibitem[{{Ruffenach} et~al.(2012){Ruffenach}, {Lavraud}, {Owens}, {Sauvaud},
  {Savani}, {Rouillard}, {D{\'e}moulin}, {Foullon}, {Opitz}, {Fedorov},
  {Jacquey}, {G{\'e}not}, {Louarn}, {Luhmann}, {Russell}, {Farrugia} \&
  {Galvin}}]{Ruffenach2012}
\bibinfo{author}{{Ruffenach}, A.}, \bibinfo{author}{{Lavraud}, B.},
  \bibinfo{author}{{Owens}, M.~J.}, \bibinfo{author}{{Sauvaud}, J.~A.},
  \bibinfo{author}{{Savani}, N.~P.}, \bibinfo{author}{{Rouillard}, A.~P.},
  \bibinfo{author}{{D{\'e}moulin}, P.}, \bibinfo{author}{{Foullon}, C.},
  \bibinfo{author}{{Opitz}, A.}, \bibinfo{author}{{Fedorov}, A.},
  \bibinfo{author}{{Jacquey}, C.~J.}, \bibinfo{author}{{G{\'e}not}, V.},
  \bibinfo{author}{{Louarn}, P.}, \bibinfo{author}{{Luhmann}, J.~G.},
  \bibinfo{author}{{Russell}, C.~T.}, \bibinfo{author}{{Farrugia}, C.~J.}, \&
  \bibinfo{author}{{Galvin}, A.~B.} (\bibinfo{year}{2012}).
\newblock \bibinfo{title}{{Multispacecraft observation of magnetic cloud
  erosion by magnetic reconnection during propagation}}.
\newblock {\it \bibinfo{journal}{\jgr}\/},  {\it
  \bibinfo{volume}{117}\/}\bibinfo{issue}{(A9)}, \bibinfo{pages}{A09101}.
  \DOIprefix\doi{10.1029/2012JA017624}.
\bibitem[{{Scolini} et~al.(2022){Scolini}, {Winslow}, {Lugaz}, {Salman},
  {Davies} \& {Galvin}}]{Scolini2022}
\bibinfo{author}{{Scolini}, C.}, \bibinfo{author}{{Winslow}, R.~M.},
  \bibinfo{author}{{Lugaz}, N.}, \bibinfo{author}{{Salman}, T.~M.},
  \bibinfo{author}{{Davies}, E.~E.}, \& \bibinfo{author}{{Galvin}, A.~B.}
  (\bibinfo{year}{2022}).
\newblock \bibinfo{title}{{Causes and Consequences of Magnetic Complexity
  Changes within Interplanetary Coronal Mass Ejections: A Statistical Study}}.
\newblock {\it \bibinfo{journal}{\apj}\/},  {\it
  \bibinfo{volume}{927}\/}\bibinfo{issue}{(1)}, \bibinfo{pages}{102}.
  \DOIprefix\doi{10.3847/1538-4357/ac3e60}.
  \href{http://arxiv.org/abs/2111.12637}{\tt arXiv:2111.12637}.
\bibitem[{{Subramanian} et~al.(2014){Subramanian}, {Arunbabu}, {Vourlidas} \&
  {Mauriya}}]{Subramanian2014}
\bibinfo{author}{{Subramanian}, P.}, \bibinfo{author}{{Arunbabu}, K.~P.},
  \bibinfo{author}{{Vourlidas}, A.}, \& \bibinfo{author}{{Mauriya}, A.}
  (\bibinfo{year}{2014}).
\newblock \bibinfo{title}{{Self-similar Expansion of Solar Coronal Mass
  Ejections: Implications for Lorentz Self-force Driving}}.
\newblock {\it \bibinfo{journal}{\apj}\/},  {\it
  \bibinfo{volume}{790}\/}\bibinfo{issue}{(2)}, \bibinfo{pages}{125}.
  \DOIprefix\doi{10.1088/0004-637X/790/2/125}.
  \href{http://arxiv.org/abs/1406.0286}{\tt arXiv:1406.0286}.
\bibitem[{{Velli} et~al.(2020){Velli}, {Harra}, {Vourlidas}, {Schwadron},
  {Panasenco}, {Liewer}, {M{\"u}ller}, {Zouganelis}, {St Cyr}, {Gilbert},
  {Nieves-Chinchilla}, {Auch{\`e}re}, {Berghmans}, {Fludra}, {Horbury},
  {Howard}, {Krucker}, {Maksimovic}, {Owen}, {Rodr{\'\i}guez-Pacheco},
  {Romoli}, {Solanki}, {Wimmer-Schweingruber}, {Bale}, {Kasper}, {McComas},
  {Raouafi}, {Martinez-Pillet}, {Walsh}, {De Groof} \& {Williams}}]{Velli2020}
\bibinfo{author}{{Velli}, M.}, \bibinfo{author}{{Harra}, L.~K.},
  \bibinfo{author}{{Vourlidas}, A.}, \bibinfo{author}{{Schwadron}, N.},
  \bibinfo{author}{{Panasenco}, O.}, \bibinfo{author}{{Liewer}, P.~C.},
  \bibinfo{author}{{M{\"u}ller}, D.}, \bibinfo{author}{{Zouganelis}, I.},
  \bibinfo{author}{{St Cyr}, O.~C.}, \bibinfo{author}{{Gilbert}, H.},
  \bibinfo{author}{{Nieves-Chinchilla}, T.}, \bibinfo{author}{{Auch{\`e}re},
  F.}, \bibinfo{author}{{Berghmans}, D.}, \bibinfo{author}{{Fludra}, A.},
  \bibinfo{author}{{Horbury}, T.~S.}, \bibinfo{author}{{Howard}, R.~A.},
  \bibinfo{author}{{Krucker}, S.}, \bibinfo{author}{{Maksimovic}, M.},
  \bibinfo{author}{{Owen}, C.~J.}, \bibinfo{author}{{Rodr{\'\i}guez-Pacheco},
  J.}, \bibinfo{author}{{Romoli}, M.}, \bibinfo{author}{{Solanki}, S.~K.},
  \bibinfo{author}{{Wimmer-Schweingruber}, R.~F.}, \bibinfo{author}{{Bale},
  S.}, \bibinfo{author}{{Kasper}, J.}, \bibinfo{author}{{McComas}, D.~J.},
  \bibinfo{author}{{Raouafi}, N.}, \bibinfo{author}{{Martinez-Pillet}, V.},
  \bibinfo{author}{{Walsh}, A.~P.}, \bibinfo{author}{{De Groof}, A.}, \&
  \bibinfo{author}{{Williams}, D.} (\bibinfo{year}{2020}).
\newblock \bibinfo{title}{{Understanding the origins of the heliosphere:
  integrating observations and measurements from Parker Solar Probe, Solar
  Orbiter, and other space- and ground-based observatories}}.
\newblock {\it \bibinfo{journal}{\aap}\/},  {\it \bibinfo{volume}{642}\/},
  \bibinfo{pages}{A4}. \DOIprefix\doi{10.1051/0004-6361/202038245}.
\bibitem[{{Vourlidas} et~al.(2017){Vourlidas}, {Balmaceda}, {Stenborg} \& {Dal
  Lago}}]{Vourlidas2017}
\bibinfo{author}{{Vourlidas}, A.}, \bibinfo{author}{{Balmaceda}, L.~A.},
  \bibinfo{author}{{Stenborg}, G.}, \& \bibinfo{author}{{Dal Lago}, A.}
  (\bibinfo{year}{2017}).
\newblock \bibinfo{title}{{Multi-viewpoint Coronal Mass Ejection Catalog Based
  on STEREO COR2 Observations}}.
\newblock {\it \bibinfo{journal}{\apj}\/},  {\it
  \bibinfo{volume}{838}\/}\bibinfo{issue}{(2)}, \bibinfo{pages}{141}.
  \DOIprefix\doi{10.3847/1538-4357/aa67f0}.
\bibitem[{{Vourlidas} et~al.(2013){Vourlidas}, {Lynch}, {Howard} \&
  {Li}}]{Vourlidas2013}
\bibinfo{author}{{Vourlidas}, A.}, \bibinfo{author}{{Lynch}, B.~J.},
  \bibinfo{author}{{Howard}, R.~A.}, \& \bibinfo{author}{{Li}, Y.}
  (\bibinfo{year}{2013}).
\newblock \bibinfo{title}{{How Many CMEs Have Flux Ropes? Deciphering the
  Signatures of Shocks, Flux Ropes, and Prominences in Coronagraph Observations
  of CMEs}}.
\newblock {\it \bibinfo{journal}{\solphys}\/},  {\it \bibinfo{volume}{284}\/},
  \bibinfo{pages}{179--201}. \DOIprefix\doi{10.1007/s11207-012-0084-8}.
  \href{http://arxiv.org/abs/1207.1599}{\tt arXiv:1207.1599}.
\bibitem[{{Vourlidas} \& {Webb}(2018)}]{Vourlidas2018}
\bibinfo{author}{{Vourlidas}, A.}, \& \bibinfo{author}{{Webb}, D.~F.}
  (\bibinfo{year}{2018}).
\newblock \bibinfo{title}{{Streamer-blowout Coronal Mass Ejections: Their
  Properties and Relation to the Coronal Magnetic Field Structure}}.
\newblock {\it \bibinfo{journal}{\apj}\/},  {\it \bibinfo{volume}{861}\/},
  \bibinfo{pages}{103}. \DOIprefix\doi{10.3847/1538-4357/aaca3e}.
  \href{http://arxiv.org/abs/1806.00644}{\tt arXiv:1806.00644}.
\bibitem[{{Welsch}(2018)}]{Welsch2018}
\bibinfo{author}{{Welsch}, B.~T.} (\bibinfo{year}{2018}).
\newblock \bibinfo{title}{{Flux Accretion and Coronal Mass Ejection Dynamics}}.
\newblock {\it \bibinfo{journal}{\solphys}\/},  {\it \bibinfo{volume}{293}\/},
  \bibinfo{pages}{113}. \DOIprefix\doi{10.1007/s11207-018-1329-y}.
  \href{http://arxiv.org/abs/1701.09082}{\tt arXiv:1701.09082}.
\bibitem[{{Winslow} et~al.(2021){Winslow}, {Lugaz}, {Scolini} \&
  {Galvin}}]{Winslow2021}
\bibinfo{author}{{Winslow}, R.~M.}, \bibinfo{author}{{Lugaz}, N.},
  \bibinfo{author}{{Scolini}, C.}, \& \bibinfo{author}{{Galvin}, A.~B.}
  (\bibinfo{year}{2021}).
\newblock \bibinfo{title}{{First Simultaneous In Situ Measurements of a Coronal
  Mass Ejection by Parker Solar Probe and STEREO-A}}.
\newblock {\it \bibinfo{journal}{\apj}\/},  {\it
  \bibinfo{volume}{916}\/}\bibinfo{issue}{(2)}, \bibinfo{pages}{94}.
  \DOIprefix\doi{10.3847/1538-4357/ac0821}.
  \href{http://arxiv.org/abs/2106.04685}{\tt arXiv:2106.04685}.
\bibitem[{{Wood} et~al.(2020){Wood}, {Hess}, {Howard}, {Stenborg} \&
  {Wang}}]{Wood2020}
\bibinfo{author}{{Wood}, B.~E.}, \bibinfo{author}{{Hess}, P.},
  \bibinfo{author}{{Howard}, R.~A.}, \bibinfo{author}{{Stenborg}, G.}, \&
  \bibinfo{author}{{Wang}, Y.-M.} (\bibinfo{year}{2020}).
\newblock \bibinfo{title}{{Morphological Reconstruction of a Small Transient
  Observed by Parker Solar Probe on 2018 November 5}}.
\newblock {\it \bibinfo{journal}{\apjs}\/},  {\it
  \bibinfo{volume}{246}\/}\bibinfo{issue}{(2)}, \bibinfo{pages}{28}.
  \DOIprefix\doi{10.3847/1538-4365/ab5219}.
  \href{http://arxiv.org/abs/2001.08716}{\tt arXiv:2001.08716}.
\bibitem[{{Wood} et~al.(2017){Wood}, {Wu}, {Lepping}, {Nieves-Chinchilla},
  {Howard}, {Linton} \& {Socker}}]{Wood2017}
\bibinfo{author}{{Wood}, B.~E.}, \bibinfo{author}{{Wu}, C.-C.},
  \bibinfo{author}{{Lepping}, R.~P.}, \bibinfo{author}{{Nieves-Chinchilla},
  T.}, \bibinfo{author}{{Howard}, R.~A.}, \bibinfo{author}{{Linton}, M.~G.}, \&
  \bibinfo{author}{{Socker}, D.~G.} (\bibinfo{year}{2017}).
\newblock \bibinfo{title}{{A STEREO Survey of Magnetic Cloud Coronal Mass
  Ejections Observed at Earth in 2008-2012}}.
\newblock {\it \bibinfo{journal}{\apjs}\/},  {\it
  \bibinfo{volume}{229}\/}\bibinfo{issue}{(2)}, \bibinfo{pages}{29}.
  \DOIprefix\doi{10.3847/1538-4365/229/2/29}.
  \href{http://arxiv.org/abs/1701.01682}{\tt arXiv:1701.01682}.
\bibitem[{{Yashiro} et~al.(2004){Yashiro}, {Gopalswamy}, {Michalek}, {St.~Cyr},
  {Plunkett}, {Rich} \& {Howard}}]{Yashiro2004}
\bibinfo{author}{{Yashiro}, S.}, \bibinfo{author}{{Gopalswamy}, N.},
  \bibinfo{author}{{Michalek}, G.}, \bibinfo{author}{{St.~Cyr}, O.~C.},
  \bibinfo{author}{{Plunkett}, S.~P.}, \bibinfo{author}{{Rich}, N.~B.}, \&
  \bibinfo{author}{{Howard}, R.~A.} (\bibinfo{year}{2004}).
\newblock \bibinfo{title}{{A catalog of white light coronal mass ejections
  observed by the SOHO spacecraft}}.
\newblock {\it \bibinfo{journal}{\jgr}\/},  {\it \bibinfo{volume}{109}\/},
  \bibinfo{pages}{A07105}. \DOIprefix\doi{10.1029/2003JA010282}.
\bibitem[{{Zhang} et~al.(2004){Zhang}, {Liemohn}, {Kozyra}, {Lynch} \&
  {Zurbuchen}}]{Zhang2004}
\bibinfo{author}{{Zhang}, J.}, \bibinfo{author}{{Liemohn}, M.~W.},
  \bibinfo{author}{{Kozyra}, J.~U.}, \bibinfo{author}{{Lynch}, B.~J.}, \&
  \bibinfo{author}{{Zurbuchen}, T.~H.} (\bibinfo{year}{2004}).
\newblock \bibinfo{title}{{A statistical study of the geoeffectiveness of
  magnetic clouds during high solar activity years}}.
\newblock {\it \bibinfo{journal}{\jgr}\/},  {\it \bibinfo{volume}{109}\/},
  \bibinfo{pages}{9101}. \DOIprefix\doi{10.1029/2004JA010410}.

\end{thebibliography}



\end{document}